\documentclass[twocolumn]{aastex631}
\usepackage{amsmath}

\usepackage{xspace}
\def\m87{M87$^*$\xspace}
\def\sgra{Sgr~A$^*$\xspace}

\def\lsim{\mathrel{\raise.3ex\hbox{$<$\kern-.75em\lower1ex\hbox{$\sim$}}}}
\def\gsim{\mathrel{\raise.3ex\hbox{$>$\kern-.75em\lower1ex\hbox{$\sim$}}}}
\defcitealias{Issaoun_2022}{Issaoun, Wielgus et al. 2022}

\usepackage{comment}
\usepackage[encapsulated]{CJK}
\usepackage{ucs}
\newcommand{\cntext}[1]{\begin{CJK}{UTF8}{gbsn}#1\end{CJK}}

\shorttitle{Polarimetric geometric modeling for mm-VLBI observations of black holes}
\shortauthors{Roelofs et al.}
\graphicspath{{./}{}}

\defcitealias{M87PaperIX}{Paper IX}

\usepackage{multirow}

\begin{document}

\nocite{PaperI, PaperII, PaperIII, PaperIV, PaperV, PaperVI, PaperVII, PaperVIII, SgrAEHTCI, SgrAEHTCII, SgrAEHTCIII, SgrAEHTCIV, SgrAEHTCV, SgrAEHTCVI}

\title{Polarimetric geometric modeling for mm-VLBI observations of black holes}

\correspondingauthor{Freek Roelofs}
\email{freek.roelofs@cfa.harvard.edu}

\author[0000-0001-5461-3687]{Freek Roelofs}
\affiliation{Center for Astrophysics $|$ Harvard \& Smithsonian,  60 Garden Street, Cambridge, MA 02138, USA}
\affiliation{Black Hole Initiative at Harvard University, 20 Garden Street, Cambridge, MA 02138, USA}
\affiliation{Department of Astrophysics, Institute for Mathematics, Astrophysics and Particle Physics (IMAPP), Radboud University, P.O. Box 9010, 6500 GL Nijmegen, The Netherlands}

\author[0000-0002-4120-3029]{Michael D. Johnson}
\affiliation{Center for Astrophysics $|$ Harvard \& Smithsonian,  60 Garden Street, Cambridge, MA 02138, USA}
\affiliation{Black Hole Initiative, Harvard University, 20 Garden Street, Cambridge, MA 02138, USA}

\author[0000-0003-2966-6220]{Andrew Chael}
\affiliation{Princeton Gravity Initiative, Princeton University, Jadwin Hall, Princeton, NJ 08544, USA}

\author[0000-0001-8685-6544]{Michael Janssen}
\affiliation{Max-Planck-Institut f\"ur Radioastronomie, Auf dem H\"ugel 69, D-53121 Bonn, Germany}

\author[0000-0002-8635-4242]{Maciek Wielgus}
\affiliation{Max-Planck-Institut f\"ur Radioastronomie, Auf dem H\"ugel 69, D-53121 Bonn, Germany}

\author[0000-0002-3351-760X]{Avery E. Broderick}
\affiliation{Perimeter Institute for Theoretical Physics, 31 Caroline Street North, Waterloo, ON, N2L 2Y5, Canada}
\affiliation{Department of Physics and Astronomy, University of Waterloo, 200 University Avenue West, Waterloo, ON, N2L 3G1, Canada}
\affiliation{Waterloo Centre for Astrophysics, University of Waterloo, Waterloo, ON, N2L 3G1, Canada}

\author[0000-0002-9475-4254]{Kazunori Akiyama}
\affiliation{Massachusetts Institute of Technology Haystack Observatory, 99 Millstone Road, Westford, MA 01886, USA}
\affiliation{National Astronomical Observatory of Japan, 2-21-1 Osawa, Mitaka, Tokyo 181-8588, Japan}
\affiliation{Black Hole Initiative at Harvard University, 20 Garden Street, Cambridge, MA 02138, USA}

\author[0000-0002-9371-1033]{Antxon Alberdi}
\affiliation{Instituto de Astrofísica de Andalucía-CSIC, Glorieta de la Astronomía s/n, E-18008 Granada, Spain}

\author{Walter Alef}
\affiliation{Max-Planck-Institut für Radioastronomie, Auf dem Hügel 69, D-53121 Bonn, Germany}

\author[0000-0001-6993-1696]{Juan Carlos Algaba}
\affiliation{Department of Physics, Faculty of Science, Universiti Malaya, 50603 Kuala Lumpur, Malaysia}

% ...




\begin{abstract}
The Event Horizon Telescope (EHT) is a millimeter very-long-baseline interferometry (VLBI) array which has imaged the apparent shadows of the supermassive black holes \m87 and Sagittarius~A$^*$. Polarimetric data from these observations contain a wealth of information on the black hole and accretion flow properties. 
In this work, we develop polarimetric geometric modeling methods for mm-VLBI data, focusing on approaches that fit data products with differing degrees of invariance to broad classes of calibration errors. We establish a fitting procedure using a polarimetric ``m-ring'' model to approximate the image structure near a black hole. By fitting this model to synthetic EHT data from general relativistic magnetohydrodynamic (GRMHD) models, we show that the linear and circular polarization structure can be successfully approximated with relatively few model parameters. We then fit this model to EHT observations of \m87 taken in 2017. In total intensity and linear polarization, the m-ring fits are consistent with previous results from imaging methods. 
In circular polarization, the m-ring fits indicate the presence of event-horizon-scale circular polarization structure, with a persistent dipolar asymmetry and orientation across several days. The same structure was recovered independently of observing band, used data products, and model assumptions. 
Despite this broad agreement, imaging methods do not produce similarly consistent results. Our circular polarization results, which imposed additional assumptions on the source structure, should thus be interpreted with some caution. Polarimetric geometric modeling provides a useful and powerful method to constrain the properties of horizon-scale polarized emission, particularly for sparse arrays like the EHT.
\end{abstract}



\section{Introduction} \label{sec:intro}
The Event Horizon Telescope has imaged \m87, the $6.5 \times 10^9M_{\odot}$ supermassive black hole in the M87 galaxy, in both total intensity \citep{PaperI, PaperII, PaperIII, PaperIV, PaperV, PaperVI} and linear polarization \citep{PaperVII, PaperVIII} using data from its 2017 campaign. More recently, images have been produced in circular polarization; however, these show inconsistent structure among different imaging and calibration methods due to the weakness of the circular polarization signal, although some secure inferences on the structure could be made \citep[][hereafter \citetalias{M87PaperIX}]{M87PaperIX}. The presence of resolved circular polarization structure on event horizon scales was established unambiguously, and a $\sim 4\%$ upper limit on the image-averaged resolved circular polarization fraction was obtained. While imaging methods showed different circular polarization structure between, e.g., observing epochs and frequency bands, circular polarization geometric model fitting indicated a consistent dipolar asymmetry across the multi-day observing window. 

In this paper, we describe and test the polarimetric modeling methods used to obtain this circular polarization modeling result. Since our modeling methods solve for the total intensity and linear polarization structure as well, we also compare our fits to previously obtained results from EHT imaging and other geometric modeling methods. Before introducing the modeling methods, we will next briefly summarize the origin, utility, and previous measurements of linear polarization and circular polarization and the challenges of studying these quantities using very-long-baseline interferometry (VLBI). 

\subsection{Linear Polarization on Event Horizon Scales}

In radiatively inefficient accretion flows onto supermassive black holes, (sub)millimeter emission is produced near the event horizon as synchrotron radiation, which is intrinsically linearly polarized at a level of $\sim 70\%$ \citep[see, e.g.,][]{Yuan_Narayan_2014}. The electric vector polarization angle (EVPA) is orthogonal to the orientation of the magnetic field. Polarimetric imaging and modeling of this emission thus probes the magnetic field structure. 

However, on the photon trajectory towards the observer, the linear polarization fraction and EVPA may be altered by two effects. First, photon propagation along curved geodesics near the event horizon affects the EVPA and may lead to depolarization in the observed image \citep[e.g.,][]{Connors_Stark_1977,Moscibrodzka2017, Narayan2021, Palumbo2022, Ricarte2022}. Second, propagation through the magnetized accretion flow plasma results in Faraday rotation of the EVPA, with depolarizing effects as well. \citep[e.g.,][]{Moscibrodzka2017, Jimenez-Rosales2018, Ricarte2020}. A polarized image of the accretion flow hence probes the spacetime as well as the plasma properties. These effects may be difficult to untangle, although \citet{Palumbo2020} found that the black hole spin can be constrained from the twistiness of the polarization pattern.

Ray-traced general relativistic magnetohydrodynamic (GRMHD) simulations show that the appearance of the horizon-scale accretion flow may indeed depend strongly on black hole and accretion parameters. In particular, magnetically arrested disk \citep[MAD, ][]{Narayan2003} models, with large magnetic flux permeating the event horizon, produce significantly more ordered EVPA patterns than standard and normal evolution (SANE) models, due to the much larger Faraday depth and hence EVPA rotation and Faraday depolarization of the latter \citep[e.g.,][]{PaperVIII}. Also, a lower electron temperature results in a more disordered EVPA pattern in these simulations for both SANE and MAD, again due to a larger Faraday depth. \citet{Palumbo2020} found that, despite the plasma effects, the twistiness of the EVPA pattern (the $\beta_{\mathcal{P},2}$-mode, see also \autoref{sec:pol_mring}) is a proxy for black hole spin in GRMHD simulations, with larger black hole spins often resulting in more radial EVPA patterns. \citet{Ricarte2022handedness} showed that the handedness of the EVPA pattern switches sign as a function of radius for retrograde accretion flows, which have a black hole spin direction opposite from the large-scale plasma rotation direction.

While the total intensity EHT data of \m87 from 2017 ruled out a few models from the EHT GRMHD simulation library, the linear polarization data of \m87 provided stronger constraints, ruling out a significant fraction of the models in the library. These constraints favored MAD models over SANE models \citep{PaperVII, PaperVIII}. In particular, the constraint on $\beta_{\mathcal{P},2}$ was a strong discriminator for these results. In general, our theoretical models are increasingly challenged by EHT observations, even more so for \sgra \citep{SgrAEHTCV}, where no single GRMHD library model was able to fit all constraints on the (Stokes $\mathcal{I}$) source structure and light curve variability \citep{SgrAEHTCV, Wielgus2022}. Additional physics or parameter space may need to be explored.

\subsection{Circular Polarization on Event Horizon Scales}
Circular polarization in millimeter emission from black hole accretion flows may arise from two distinct physical processes \citep[e.g.,][]{Wardle2003, Moscibrodzka2021, Ricarte2021}. First, synchrotron radiation is intrinsically polarized depending on the observing frequency and magnetic field strength and configuration. 
The maximum intrinsic circular polarization from synchrotron radiation is of order $1/\gamma$, where $\gamma$ is the Lorentz factor of the (relativistic) synchrotron emitting electrons \citep[e.g.][]{Wardle1998, Wardle2003}. In numerical simulations of supermassive black hole accretion flows, the intrinsic circular polarization fraction comes down to $\sim1\%$, which is substantially lower than the intrinsic linear polarization fraction of $\sim70\%$ \citep{Ricarte2021}. Circular polarization is therefore more difficult to detect than linear polarization. The sign of intrinsic circular polarization (before any propagation effects occur) directly maps to the magnetic field orientation with respect to the emission direction, with a positive sign of the observed circular polarization corresponding to a magnetic field orientation pointing toward the observer.

Second, circular polarization may be produced from linear polarization through Faraday conversion. In the local plasma frame, Faraday conversion only operates on Stokes $\mathcal{U}$, while linear polarization in synchrotron emission is intrinsically produced only in Stokes $\mathcal{Q}$ (perpendicular to both the magnetic field and the photon propagation direction). Along the photon propagation path, part of the Stokes $\mathcal{Q}$ thus has to be recast into Stokes $\mathcal{U}$ for Faraday conversion to occur. Such an exchange between Stokes $\mathcal{Q}$ and $\mathcal{U}$ occurs in the case of Faraday rotation, where the linear polarization direction rotates depending on electron density and magnetic field component parallel to the photon propagation direction. Another pathway for Faraday conversion is a rotation of the magnetic field component perpendicular to the photon propagation direction due to a twist in the magnetic field along the photon propagation direction. Faraday conversion through Faraday rotation produces circular polarization in the same direction as the intrinsic emission (assuming a constant magnetic field). Faraday conversion through a positive twist of the magnetic field (counter-clockwise with respect to the photon propagation direction) produces negative circular polarization \citep[see, e.g., ][and references therein]{Ricarte2021}. 

In general, the fraction and direction of circular polarization depend on the magnetic field structure, the electron temperature and density, and plasma composition \citep{Jones1977, Kennett1998, Wardle1998}. These dependencies make circular polarization an excellent probe for constraining the plasma properties of black hole accretion flows. 

GRMHD simulations with resolved circular polarization on sub-event-horizon scales show that the Stokes~$\mathcal{V}$ image structure strongly depends on the inclination of the viewing angle with respect to the black hole spin axis \citep{Ricarte2021}. For high inclinations (edge-on view with respect to the spin axis), the circular polarization image has a clear quadrupolar structure---especially for MAD models---set by the helical magnetic field structure of the jet in combination with the viewing angle. 

For low inclinations (face-on view with respect to the spin axis), the relation between the magnetic field geometry and the Stokes~$\mathcal{V}$ image is less straightforward. The image contains contributions from features above and below the mid-plane, which have opposite sign. Stokes~$\mathcal{V}$ images from MAD models are visually dominated by the ($n=1$) photon ring (even though its contribution to the total Stokes~$\mathcal{V}$ emission is marginal), which has an opposite sign from the direct emission due to lensing effects (photons making a half orbit around the black hole) in combination with the magnetic field geometry \citep{Moscibrodzka2021}. Stokes~$\mathcal{V}$ images from SANE models are less clearly structured, with turbulent features depending on the details of the more turbulent magnetic field and Faraday effects \citep{Ricarte2021}.

In GRMHD and semi-analytic models, the circular polarization fraction generally increases at submillimeter wavelengths as a function of positron fraction due to an increase in Faraday conversion \citep{Emami2021, Anantua2020}. However, degeneracies with other simulation parameters exist \citep{Emami2022}.

\subsection{Polarimetric VLBI data}

Polarization poses particular challenges for the calibration of VLBI data. The polarization signal is often weak ($\lesssim 1-10\%$), especially for circular polarization. Most VLBI experiments, including the EHT, use orthogonal circular feeds, which are ideal for measuring linear polarization but require precise calibration of the right/left (R/L) gain ratios to measure circular polarization \citep[e.g.][see also \autoref{sec:m87cpolcasa}]{Homan1999,Homan2001,Homan2004,Homan2006,Gabuzda2008}. 
Besides systematics that also affect total intensity measurements, such as rapid atmospheric phase fluctuations, atmospheric opacity, and antenna pointing offsets, polarimetric VLBI data is particularly sensitive to ``leakage'' effects between orthogonal feeds (see \autoref{sec:data_products}), which often give spurious signals that exceed the true polarized signal, especially in linear polarization. The combination of weak signals and the additional systematics has often limited polarimetric measurements, especially for circular polarization. In this work, we therefore focus on constructing data products that are invariant to most calibration errors. In this process, we can directly explore the effects of calibration errors in both the bias and uncertainty of the measurements of polarized image structure.

\subsection{Outline}
In this work, we provide a framework for fitting geometric models to polarimetric VLBI data, and we apply these methods to synthetic data generated from GRMHD models and to EHT observations of \m87. We introduce our polarimetric ``m-ring'' model in \autoref{sec:mring}, and outline our procedure to fit this or other geometric models to VLBI data in \autoref{sec:fitting}. We test our model fitting framework on synthetic data from geometric models in \autoref{sec:synthdata_geomodels}, and apply it to synthetic data from GRMHD models in \autoref{sec:grmhd}. In \autoref{sec:m87}, we apply our methods to EHT observations of \m87 in 2017, estimating the source properties in both linear and circular polarization. We conclude and provide an outlook for future work in \autoref{sec:summary}.

\section{Polarimetric m-ring Modeling}
\label{sec:mring}
\subsection{Geometric Modeling of VLBI Data}
VLBI measurements, such as those made by the EHT, sample interferometric ``visibilities'' on each baseline connecting a pair of telescopes with mutual visibility of a target source. These visibilities are given by the correlation of the narrow-band complex electric fields sampled at the telescopes: $V_{12} \equiv \left \langle E_1 E_2^\ast \right \rangle$. Because each telescope can record two orthogonal polarization products (typically right and left circular, or orthogonal linear polarizations), each baseline can measure four correlations that can be easily mapped to the four Stokes parameters (see \autoref{sec:data_products}). By the van Cittert-Zernike theorem, these visibilities correspond to samples of the Fourier transform of the sky image, with the wavenumber given by the dimensionless baseline length (measured in wavelengths). For instance, the visibilities in total intensity, $\tilde{I}$ are given by \citep[see, e.g.,][]{TMS2017}
\begin{equation}
\begin{split}
\tilde{\mathcal{I}}(\vec{u}) &= \int d^2\vec{x}\, \mathcal{I}(\vec{x}) e^{-2\pi i \vec{u} \cdot \vec{x}}.
\end{split}
\end{equation}
Here, $\vec{u}$ is the dimensionless baseline vector projected orthogonally to the line of sight, $\vec{x}$ is the angular sky coordinate in radians, and $\mathcal{I}$ is the sky brightness distribution. The brightness distribution $\mathcal{I}(\vec{x})$ is real, so the corresponding visibilities have a conjugation symmetry: $\tilde{\mathcal{I}}(\vec{u}) = \tilde{\mathcal{I}}^\ast(-\vec{u})$.

Because an interferometer only sparsely samples the Fourier domain, geometric modeling of interferometric data provides a powerful alternative to imaging \citep[e.g.,][]{Pearson_1999,PaperVI,SgrAEHTCIV}. In particular, for sources with relatively simple morphology, geometric models may be parameterized using far fewer parameters than imaging. Geometric models are also flexible; multiple geometric models can be easily added to describe sources with complex morphologies because the Fourier transform is linear. Geometric modeling is an especially effective analysis strategy for VLBI arrays that have a small number of baselines (such as the EHT) or in cases where the signal-to-noise is low (as is frequently the case for polarimetric visibilities). The danger of using geometric models is that the choice of model may significantly affect the inferences, and model misspecification (i.e. using a model that does not fully describe the underlying structure) can result in parameter biases. 

Because EHT images of \m87 and \sgra are dominated by a prominent ring with an azimuthal brightness asymmetry \citep{PaperIV,SgrAEHTCIII}, both  ``crescent'' \citep{Kamruddin_2013,Benkevitch_2016,PaperVI,Wielgus2020,Lockhart_2022} and geometric ring models \citep{PaperVI,Johnson_2020,SgrAEHTCIII} provide good fits to EHT data from these sources. 

\subsection{m-ring Model}
In this paper, we will focus on using extensions of the ``m-ring'' model from \citet{Johnson_2020} to fit polarimetric EHT data. For our purposes, the principal benefits of this model are:

\begin{itemize}
\item Efficiency: the model has a simple analytic form in both the image and visibility domains, with analytic gradients.
\item Flexibility: the model can describe arbitrary azimuthal variations in brightness, ring shape asymmetry, and radial structure controlled by a Gaussian blurring kernel.
\item Polarimetry: the model naturally includes complex polarized structure in both linear and circular polarization. 
\item Interpretability: the m-ring model naturally describes key image features that are useful for physical interpretation, such as the ring's diameter, brightness asymmetry, shape asymmetry, and rotationally-invariant polarization. 
\end{itemize}

Specifically, the m-ring model is constructed from a thin ring with non-uniform brightness in azimuth expressed as a Fourier series. Written in polar image coordinates ($\rho$, $\varphi$), it takes the form

\begin{equation}
\begin{split}
\label{eq::Ring_Image}
\mathcal{I}(\rho,\varphi) &= \frac{F}{\pi d}\delta\left(\rho-\frac{d}{2}\right) \sum_{k=-m}^m \beta_k e^{ik\varphi},
\end{split}
\end{equation}
where $\delta$ is the Dirac delta distribution, $\beta_{-k} \equiv \beta_k^\ast$ since the image is real, and $\beta_0 \equiv 1$ so that $F>0$ gives the total flux density of the ring. By increasing the m-ring order $m$, increasingly complex azimuthal structures can be modeled. The corresponding visibility function in polar coordinates ($u$, $\phi$) is given by
\begin{equation}
\label{eq::Ring_Visibility}
\begin{split}
\tilde{\mathcal{I}}(u,\phi) &= F \sum_{k=-m}^m\beta_k J_k(\pi d u) e^{i k (\phi-\pi/2)},
\end{split}
\end{equation}

where $J_k$ denotes the $k^{\rm th}$ Bessel function of the first kind. Notably, the azimuthal Fourier coefficients in the image and visibility domains are identical up to a constant rescaling. 

Two natural extensions of this model are to include shape asymmetry and to introduce finite ring width. For the former, the m-ring can be stretched in any direction using the similarity property of the Fourier transform to compute the associated visibility function: if $\mathcal{I}(x,y)\to \mathcal{I}'(x,y)=\mathcal{I}(ax,by)$, where the arrow indicates the stretch transformation and $\mathcal{I}'(x,y)$ is the stretched m-ring, then $\tilde{\mathcal{I}}(u,v)\to \tilde{\mathcal{I}}'(u,v)=\left|ab\right|^{-1}\tilde{\mathcal{I}}(u/a,v/b)$. For the latter, the m-ring can be easily convolved with Gaussian of full width at half maximum $\alpha$. This blurred m-ring has visibility and image functions given by:

\begin{equation}
\begin{split}
\mathcal{I}(\rho,\varphi;\alpha) &= \frac{4 \ln 2}{\pi \alpha^2} F e^{-\frac{4\ln2}{\alpha^2}\left(\rho^2+d^2/4\right)}\\
\nonumber &\qquad \times \sum_{k=-m}^{m} \beta_k I_k\left(4 \ln2 \frac{\rho d}{\alpha^2} \right) e^{i k \varphi},\\
\tilde{\mathcal{I}}(u,\phi;\alpha) &= \tilde{\mathcal{I}}(u,\phi) e^{-\frac{ \left( \pi \alpha u \right)^2}{4 \ln 2}},
\end{split}
\end{equation}

where $I_k$ denotes the $k^{\rm th}$ modified Bessel function of the first kind.

\subsection{Polarimetric m-ring Model}
\label{sec:pol_mring}
The m-ring model can be easily generalized to include linear and circular polarization. Each polarization product has an image determined by an associated set of Fourier coefficients; we use $\left\{ \beta_{\mathcal{I},k} \right\}$ for total intensity, $\left\{ \beta_{\mathcal{P},k} \right\}$ for the linear polarization,\footnote{We generally work with the complex linear polarization field, $\mathcal{P} \equiv \mathcal{Q} + i \mathcal{U}$, rather than the individual Stokes parameters.} and $\left\{ \beta_{\mathcal{V},k} \right\}$ for circular polarization. Because the linear polarization is complex, there is no conjugation symmetry in the associated $\beta_{\mathcal{P},k}$. The circular polarization image is real, so $\beta_{\mathcal{V},-k} = \beta_{\mathcal{V},k}^\ast$. The image-integrated linear and circular polarization fractions are given by $m_{\mathrm{net}} \equiv\beta_{\mathcal{P},0} \in \mathbb{C}$ (a complex number) and $\mathcal{V}_{\mathrm{net}} \equiv\beta_{\mathcal{V},0} \in \mathbb{R}$ (a real number), respectively. 

The only difficulty in specifying parameter ranges for the m-ring model is that it is non-trivial to enforce image positivity ($\mathcal{I}(\vec{x})>0$) and a physical polarization limit ($\mathcal{I}(\vec{x})^2 \geq \left| \mathcal{P}(\vec{x}) \right|^2 + \mathcal{V}(\vec{x})^2$). To approximate these conditions, we typically require $\left|\beta_{\mathcal{I},k}\right| < 0.5$ (for $k \neq 0$), $\left|\beta_{\mathcal{P},k}\right| < 1$, and $\left|\beta_{\mathcal{V},k}\right| < 1$. Specifying the precise physical parameter domain is not problematic in practice because the total polarization is typically much smaller than the intensity: $\mathcal{I}(\vec{x})^2 \gg \left| \mathcal{P}(\vec{x}) \right|^2 + \mathcal{V}(\vec{x})^2$.

\begin{figure*}
    \centering
    \includegraphics[height=0.28\textwidth]{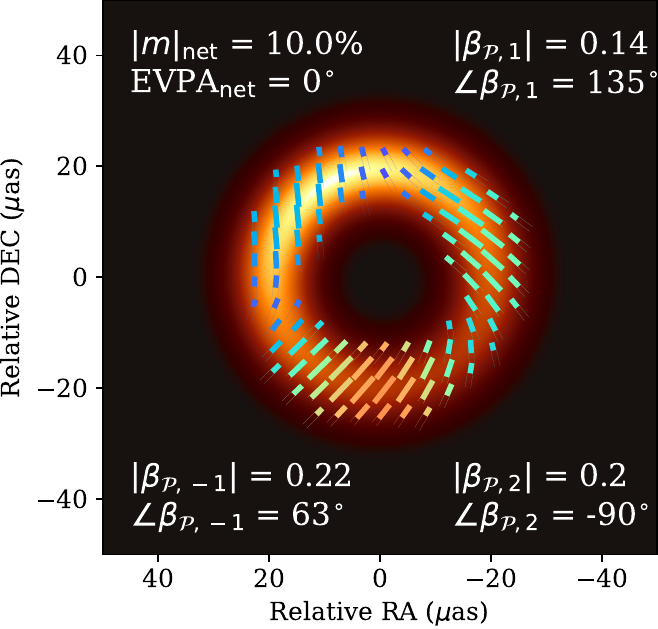}
    \includegraphics[height=0.28\textwidth]{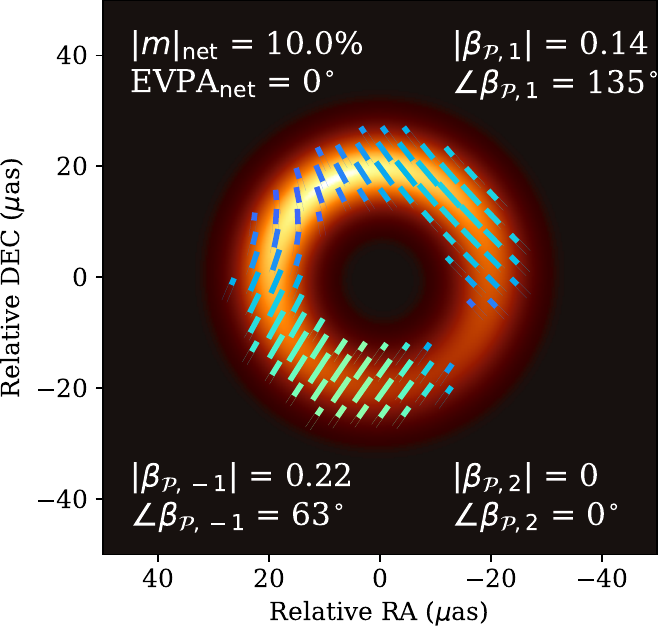}
    \includegraphics[height=0.28\textwidth]{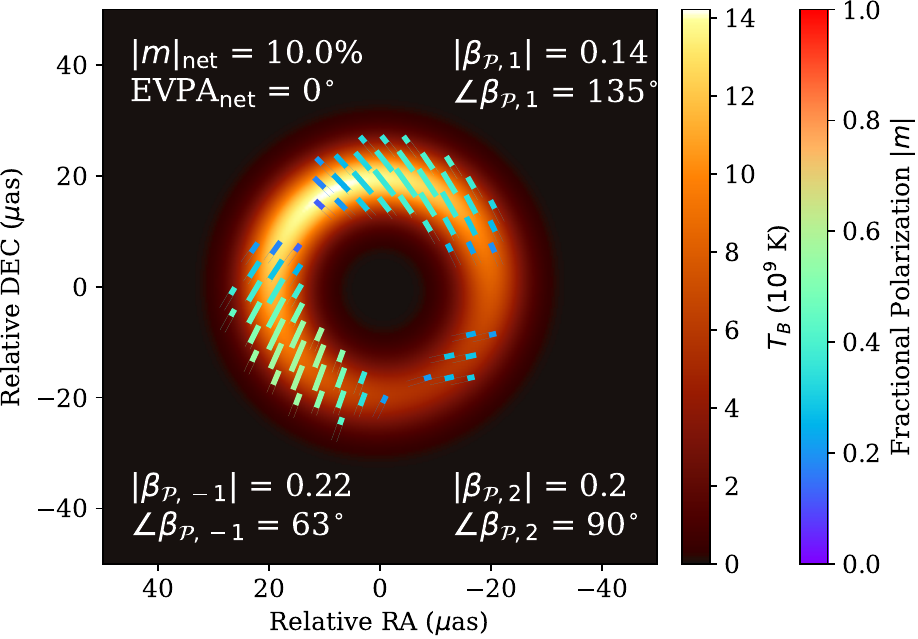}\\
    \includegraphics[height=0.28\textwidth]{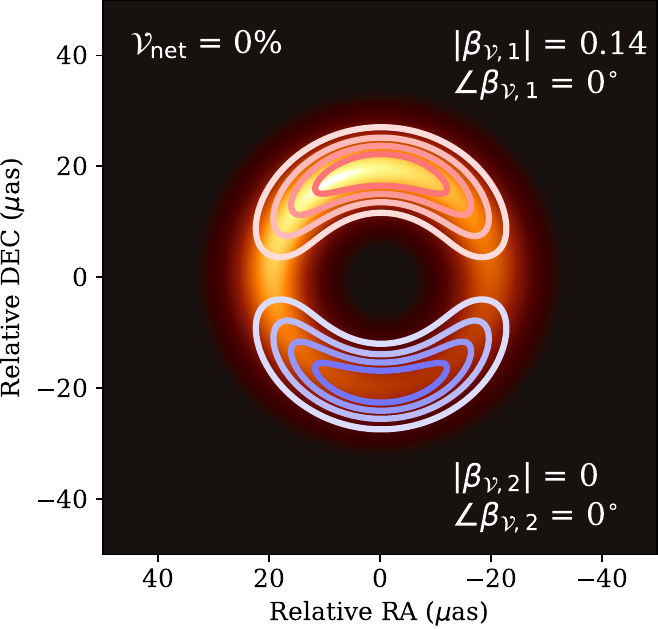}
    \includegraphics[height=0.28\textwidth]{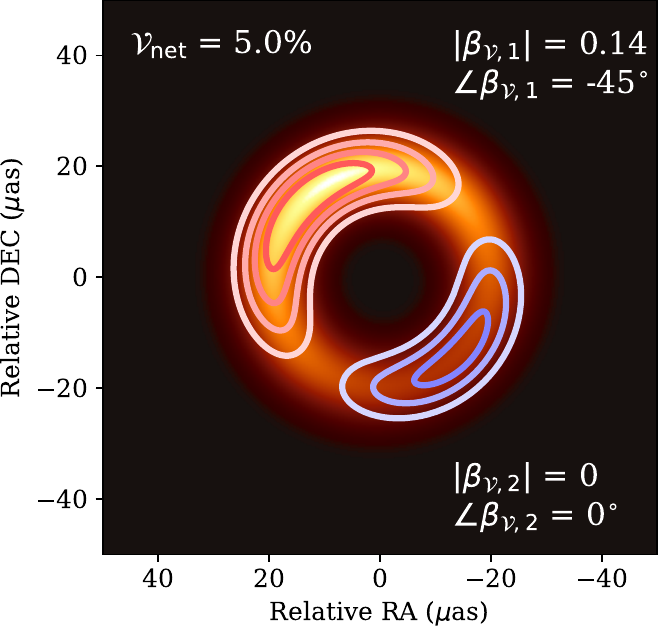}
    \includegraphics[height=0.28\textwidth]{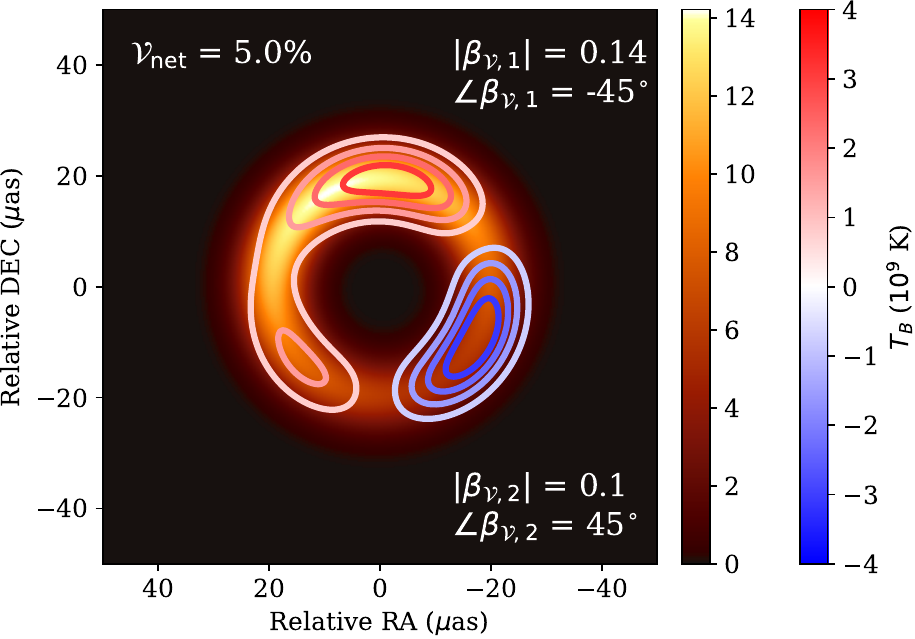}
    \caption{Examples of three m-ring models in Stokes $\mathcal{I}$ and $\mathcal{P}$ (upper panels), and Stokes $\mathcal{I}$ and $\mathcal{V}$ (lower panels). Throughout the panels, the Stokes $\mathcal{I}$ structure (heat map) is kept constant with $F=0.5$ Jy, $d=40$ $\mu$as, $\alpha=10$ $\mu$as, and $\beta_{\mathcal{I},1}=0.2-0.1i$. The top center panel shows a linear polarization structure with $m_{\rm net} \equiv \beta_{\mathcal{P},0}=0.1$, $\beta_{\mathcal{P},-1}=0.1+0.2i$, and $\beta_{\mathcal{P},1}=-0.1+0.1i$. In the top left and right panel, nonzero $\beta_{\mathcal{P},2}$ components have been added with opposite sign. The bottom left panel shows a dipolar circular polarization structure (contours) oriented towards the North ($\beta_{\mathcal{V},1}=0.14$). The net circular polarization is zero, so that the North and South half of the ring are identical with opposite sign in Stokes $\beta_{\mathcal{V}}$. In the bottom center panel, we have rotated the circular polarization structure by -45$^{\circ}$ and introduced a nonzero net circular polarization ($\mathcal{V}_{\rm net} \equiv \beta_{\mathcal{V},0}=0.05$), so that the symmetry is broken. Finally, in the bottom right panel we have added a nonzero $\beta_{\mathcal{V},2}$ component, increasing the complexity of the azimuthal structure in Stokes~$\mathcal{V}$. The model shown in the center panels is used for our geometric tests (\autoref{sec:synthdata_geomodels}, \autoref{fig:geometrictestsstokesv}).}
    \label{fig:mring_examples}
\end{figure*}

\autoref{fig:mring_examples} shows examples of polarized m-rings with different parameters.

\section{Polarimetric m-ring fitting procedure}
\label{sec:fitting}
\subsection{Data products}
\label{sec:data_products}

In VLBI observations, each antenna feed records the complex electric field $E$. For EHT antennas, these feeds are left and right circularly polarized\footnote{An exception is ALMA, which has linear feeds. The mixed-basis correlations are converted to a circular basis with {\tt PolConvert} \citep{MartiVidal2016}.}, so an antenna $j$ records the electric fields $E_{Lj}$ and $E_{Rj}$, respectively. For each pair of antennas in the array, these signals are then cross-correlated to form the cross-correlation matrix $\boldsymbol{\rho}_{jk}$, which in the absence of any other observational effects can be written in terms of the Stokes visibility components $\tilde{\mathcal{I}}$, $\tilde{\mathcal{Q}}$, $\tilde{\mathcal{U}}$, and $\tilde{\mathcal{V}}$ as

\begin{equation}
\begin{split}
\boldsymbol{\rho}_{jk} & =
\begin{pmatrix}
\langle E_{Rj}E_{Rk}^*\rangle & \langle E_{Rj}E_{Lk}^*\rangle \\
\langle E_{Lj}E_{Rk}^*\rangle & \langle E_{Lj}E_{Lk}^*\rangle
\end{pmatrix} 
\equiv
\begin{pmatrix}
R_jR_k^* & R_jL_k^* \\
L_jR_k^* & L_jL_k^* 
\end{pmatrix}\\
& =
\begin{pmatrix}
\tilde{\mathcal{I}}_{jk} + \tilde{\mathcal{V}}_{jk} & \tilde{\mathcal{Q}}_{jk} + i\tilde{\mathcal{U}}_{jk} \\
\tilde{\mathcal{Q}}_{jk} - i\tilde{\mathcal{U}}_{jk} & \tilde{\mathcal{I}}_{jk} - \tilde{\mathcal{V}}_{jk}
\end{pmatrix}\\
& \equiv
\begin{pmatrix}
\tilde{\mathcal{I}}_{jk} + \tilde{\mathcal{V}}_{jk} & \tilde{\mathcal{P}}_{jk} \\
\tilde{\mathcal{P}}_{jk}^* & \tilde{\mathcal{I}}_{jk} - \tilde{\mathcal{V}}_{jk}
\end{pmatrix}.
\end{split}
\end{equation}

However, this relation does not hold for imperfect instruments. Observational effects affecting this relation can be categorized in the gain matrix $\mathbf{G}$, the leakage matrix $\mathbf{D}$, and the field rotation matrix $\mathbf{\Phi}$. For each antenna $j$, they are combined in the Jones matrix
\begin{equation}
    \mathbf{J}_j = \mathbf{G}_j\mathbf{D}_j\mathbf{\Phi}_j = 
    \begin{pmatrix}
G_{jR} & 0 \\
0 & G_{jL} 
\end{pmatrix}
\begin{pmatrix}
1 & D_{jR} \\
D_{jL} & 1 
\end{pmatrix}
\begin{pmatrix}
e^{-i\phi_j} & 0 \\
0 & e^{i\phi_j} 
\end{pmatrix}.
\end{equation}
The measured correlation matrix is then given by the Radio Interferometer Measurement Equation \citep[RIME;][]{Hamaker1996, Smirnov2011}
\begin{equation}
\boldsymbol{\rho}'_{jk} = \mathbf{J}_j\boldsymbol{\rho}_{jk}\mathbf{J}^\dagger_k.
\end{equation}

The Stokes $\mathcal{I}$ and $\mathcal{V}$ information is primarily contained in the parallel-hand visibilities \citep{Roberts1994}:
\begin{equation}
\label{eq:rrll}
\begin{split}
            R_jR_k^*  & = G_{jR}G_{kR}^*\left[\left(\tilde{\mathcal{I}}_{jk} + \tilde{\mathcal{V}}_{jk}\right)e^{i\left(\phi_k-\phi_j\right)}\right. \\
            & + D_{jR}D_{kR}^*\left(\tilde{\mathcal{I}}_{jk} - \tilde{\mathcal{V}}_{jk}\right)e^{i\left(\phi_j-\phi_k\right)} \\
            & + D_{jR}\tilde{\mathcal{P}}_{kj}^*e^{i\left(\phi_j+\phi_k\right)} \\
            & \left. + D_{kR}^*\tilde{\mathcal{P}}_{jk}e^{i\left(-\phi_j-\phi_k\right)}\right], \\
            L_jL_k^* & = G_{jL}G_{kL}^*\left[\left(\tilde{\mathcal{I}}_{jk} - \tilde{\mathcal{V}}_{jk}\right)e^{i\left(\phi_j-\phi_k\right)}\right. \\
            & + D_{jL}D_{kL}^*\left(\tilde{\mathcal{I}}_{jk} + \tilde{\mathcal{V}}_{jk}\right)e^{i\left(\phi_k-\phi_j\right)} \\
            & + D_{jL}\tilde{\mathcal{P}}_{jk}e^{i\left(-\phi_j-\phi_k\right)} \\
            & \left. + D_{kL}^*\tilde{\mathcal{P}}_{kj}^*e^{i\left(\phi_j+\phi_k\right)}\right].
\end{split}
\end{equation}

From these parallel-hand visibilities, we can construct two data products that are especially suitable for fitting the Stokes $\mathcal{V}$ structure. First, we can fit to the parallel-hand visibility ratios, $R_jR_k^*$/$L_jL_k^*$. 
Assuming the leakage terms have been well corrected and the fractional linear polarization is small, dropping the terms proportional to $D^2$ and $D\tilde{\mathcal{P}}$, $R_jR_k^*$/$L_jL_k^*$ (hereafter referred to as RR/LL) to first order depends on the fractional circular polarization in the visibility domain:
\begin{equation}
\label{eq:rrllratio}
\begin{split}
   \frac{R_jR_k^*}{L_jL_k^*} & \approx \frac{G_{jR}G_{kR}^*}{G_{jL}G_{kL}^*}e^{2i\left(\phi_k-\phi_j\right)} \frac{\tilde{\mathcal{I}}_{jk} + \tilde{\mathcal{V}}_{jk}}{\tilde{\mathcal{I}}_{jk} - \tilde{\mathcal{V}}_{jk}} \\
   & =  \frac{G_{jR}G_{kR}^*}{G_{jL}G_{kL}^*}e^{2i\left(\phi_k-\phi_j\right)} \left(1 + 2\frac{\tilde{\mathcal{V}}_{jk}}{\tilde{\mathcal{I}}_{jk}} + \mathcal{O}\left(\frac{\tilde{\mathcal{V}}_{jk}^2}{\tilde{\mathcal{I}}_{jk}^2} \right)\right).
\end{split}
\end{equation}

This data product has the advantage of cancelling rapid atmospheric phase variations, since the atmosphere is not significantly birefringent at millimeter wavelengths (i.e., its refractive index is independent of polarization). However, the visibility ratios depend on the R/L gain ratios. These are often stable over many hours and can be corrected, although some EHT sites have shown rapidly variable R/L gain ratios \citepalias[see also \autoref{sec:m87cpolcasa}]{M87PaperIX}. Since the gain ratios are antenna based and multiplicative while the circular polarization signal is baseline-based and additive, complex circular polarization structure may be extracted from the RR/LL data product even if the R/L gain ratio calibration is imperfect \citep[see also][]{Homan1999}.

\begin{figure*}
    \centering
    \includegraphics[width=0.32\textwidth]{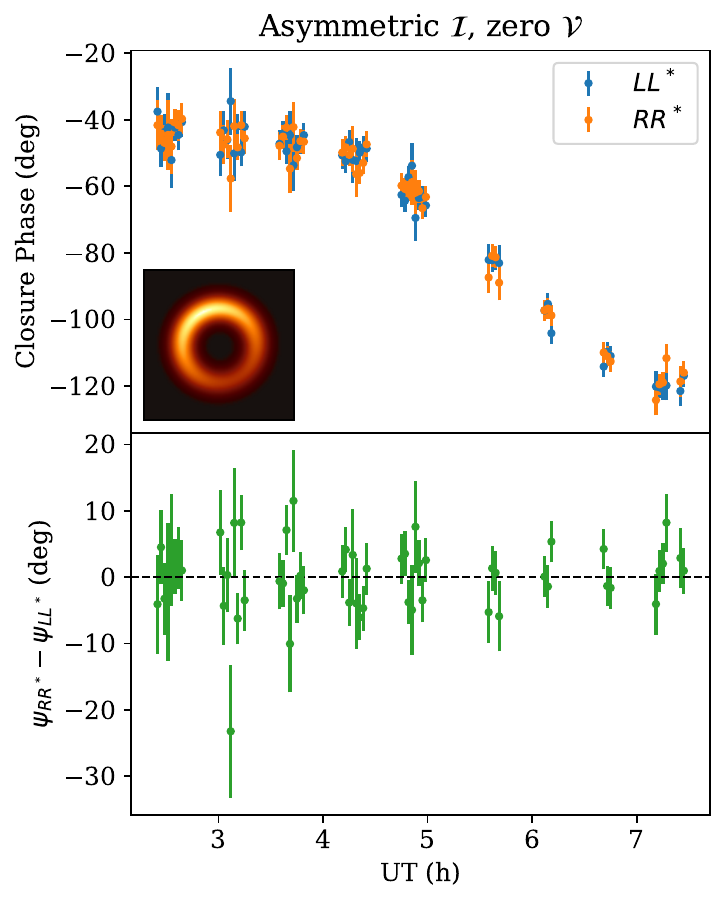}
    \includegraphics[width=0.32\textwidth]{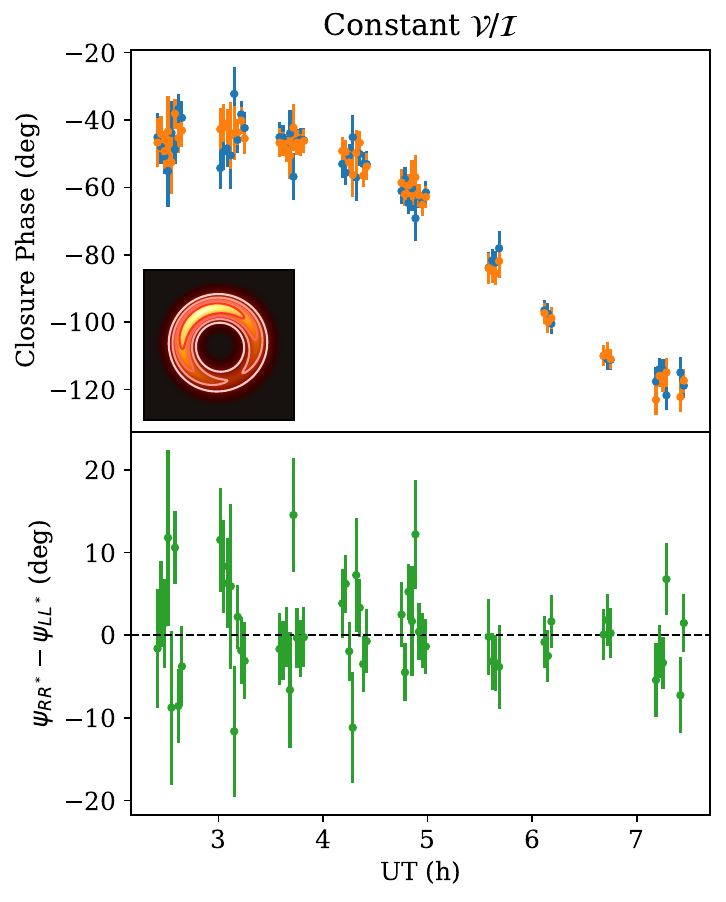}
    \includegraphics[width=0.32\textwidth]{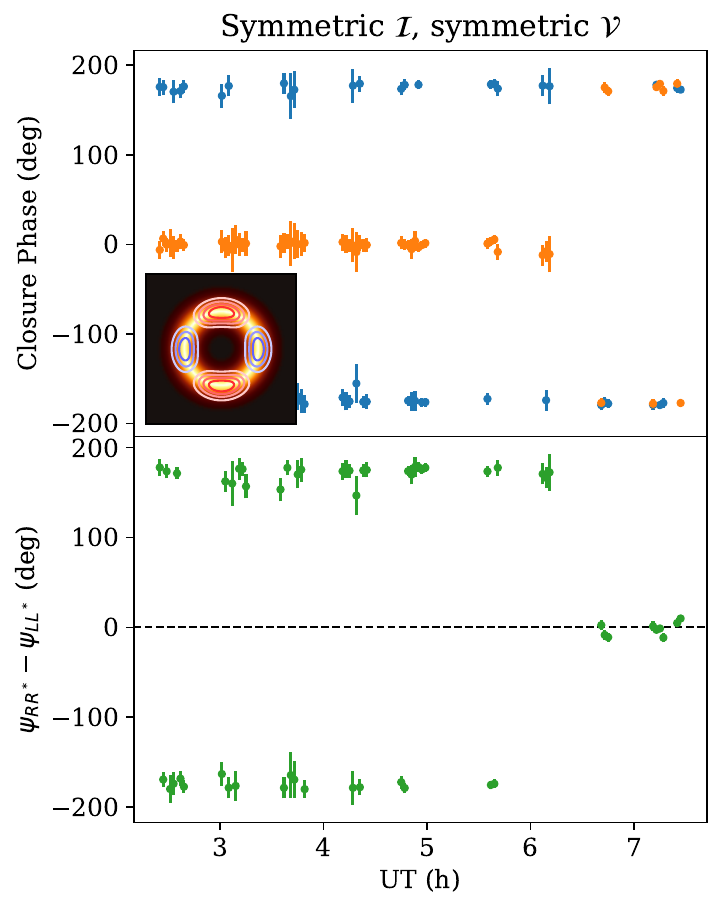}\\
    \includegraphics[width=0.32\textwidth]{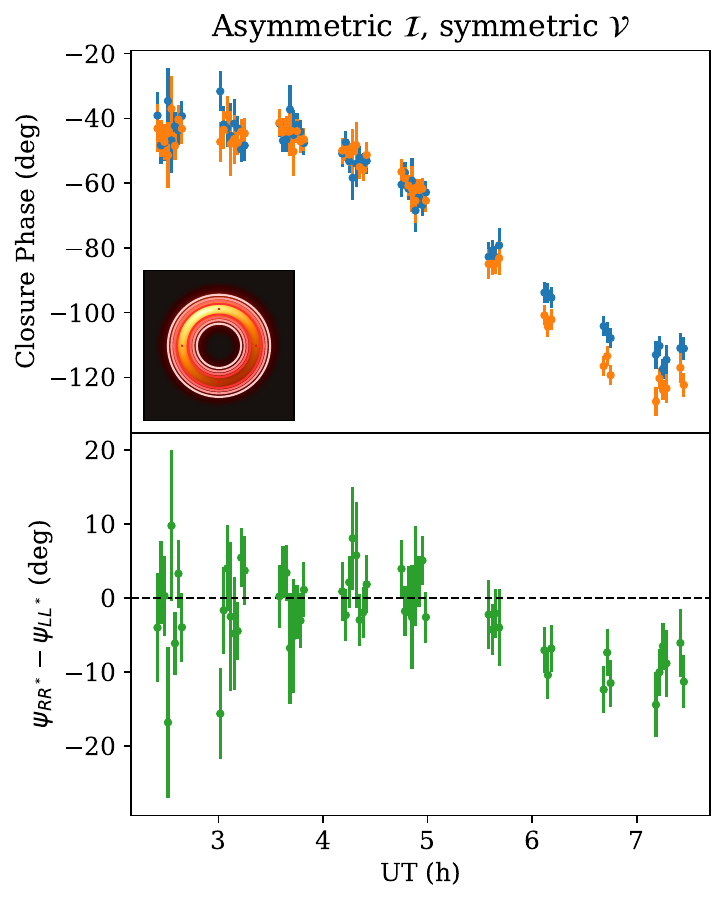}
    \includegraphics[width=0.32\textwidth]{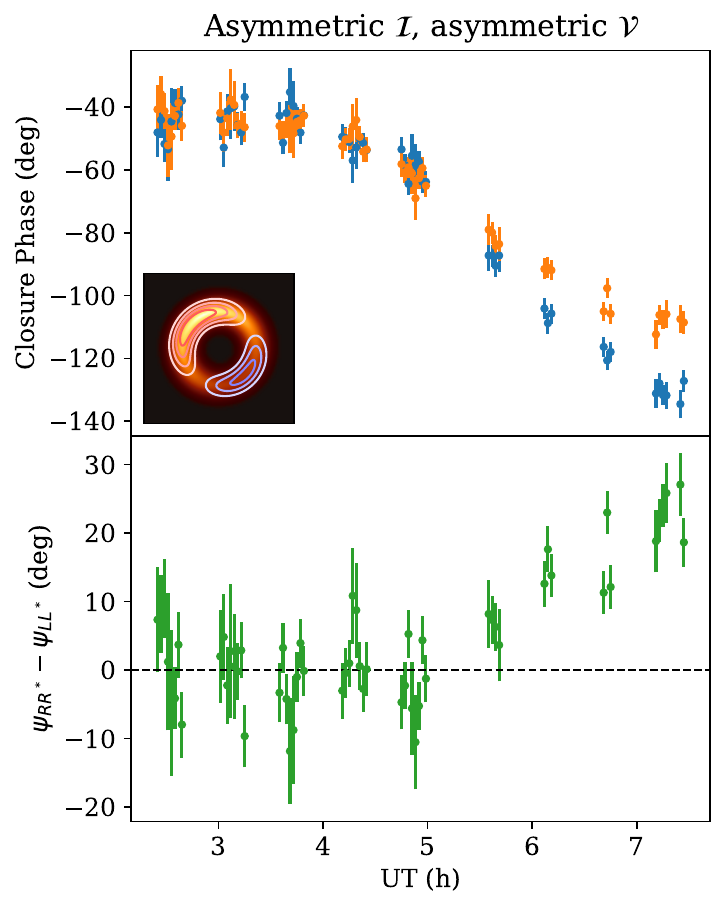}
    \includegraphics[width=0.32\textwidth]{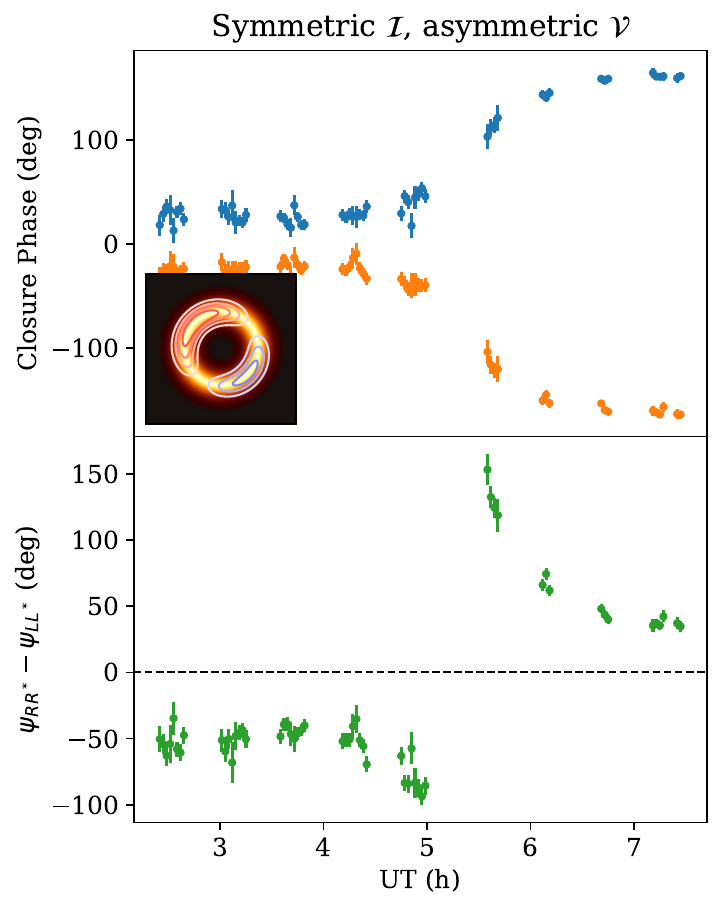}
    \caption{$LL^*$ (blue) and $RR^*$ (orange) closure phases and their differences (green) on the ALMA-SMT-LMT triangle for m-ring models with different combinations of point-symmetric and non-point-symmetric total intensity and circular polarization structures, simulated with 2017 April 11 EHT coverage and thermal noise. The $\beta_{\mathcal{I}}$ parameters are the same as those in \autoref{fig:mring_examples}, except for the rightmost panels, where $\beta_{\mathcal{I},1}=0$. In models with asymmetric Stokes $\mathcal{V}$ structure, the $\beta_{\mathcal{V}}$ parameters are the same as those in the bottom center panel of \autoref{fig:mring_examples}, and for the models with symmetric Stokes $\mathcal{V}$ structure, $\beta_{\mathcal{V},0}=0.1$ and $\beta_{\mathcal{V},1}=0$, with $\beta_{\mathcal{V},2}=0.2$ in the top right panel. The bottom center model is identical to the bottom center model in \autoref{fig:mring_examples}. A constant fractional circular polarization structure results in identical closure phases in both parallel hands (top left and middle panels). If the fractional circular polarization structure is not constant, the parallel-hand closure phase differences are non-zero (bottom and top right panels). In the top right panel, the fractional circular polarization is point-symmetric but not constant, and the closure phase differences are 0 or 180 degrees.}
    \label{fig:cphase}
\end{figure*}

Alternatively, we can fit to the parallel-hand closure phases and closure amplitudes. Closure phase is the sum of visibility phases on a triangle of baselines, and closure amplitudes are ratios of visibility products on a quadrangle of baselines \citep[e.g.,][]{TMS2017}. These data products are independent of multiplicative station-based calibration errors, including the gains as they cancel in the sums and products, respectively. The closure products are not independent of other station-based calibration errors, such as polarimetric leakage and bandpass errors. For EHT data, estimated residual leakages are only $\sim$1\% \citep{PaperVII}, and thus have a negligible effect unless polarization fractions are very high. Noting that $R_jR_k^*\approx \tilde{\mathcal{I}}_{jk}(1+\tilde{\mathcal{V}}_{jk}/\tilde{\mathcal{I}}_{jk})$ and $L_jL_k^*\approx \tilde{\mathcal{I}}_{jk}(1-\tilde{\mathcal{V}}_{jk}/\tilde{\mathcal{I}}_{jk})$, it becomes apparent that a non-constant fractional circular polarization leads to phase and amplitude differences between the parallel hands, which can be robustly detected by investigating the gain-invariant closure products. However, closure quantities contain less information than the baseline-based visibility ratios when prior knowledge on the gains is available \citep[e.g.][]{Blackburn2020}. In total intensity, closure phases deviating from 0 or 180 degrees indicate the presence of non-point-symmetric structure \citep[e.g.][]{Monnier2007}. In circular polarization, non-zero differences between $LL^*$ and $RR^*$ closure phases on a given triangle and time indicate the presence of non-constant fractional circular polarization structure (see \autoref{fig:cphase} for examples).

Thus, comparing the results of fits to these two types of data products brackets the range of uncertainty in the Stokes~$\mathcal{V}$ structure, with smaller uncertainties expected for the visibility ratios (tied to our confidence in the a-priori R/L gain calibration) and larger uncertainties for the closure products (with fewer calibration assumptions required).

In addition to this information in the parallel-hand visibilities, the cross-hand visibilities contain information about the linear polarization structure of the source \citep[see, e.g.,][]{Roberts1994}:
\begin{equation}
\label{eq:rl}
\begin{split}
            R_jL_k^*  & = G_{jR}G_{kL}^*\left[\tilde{\mathcal{P}}_{jk}e^{i\left(-\phi_k-\phi_j\right)}\right. \\
            & + D_{jR}D_{kL}^*\tilde{\mathcal{P}}_{kj}^*e^{i\left(\phi_j+\phi_k\right)} \\
            & + D_{jR}\left(\tilde{\mathcal{I}}_{jk} - \tilde{\mathcal{V}}_{jk}\right)e^{i\left(-\phi_k+\phi_j\right)} \\
            & \left. + D_{kL}^*\left(\tilde{\mathcal{I}}_{jk} + \tilde{\mathcal{V}}_{jk}\right)e^{i\left(-\phi_k+\phi_j\right)}\right], \\  
            L_jR_k^*  & = G_{jL}G_{kR}^*\left[\tilde{\mathcal{P}}_{kj}e^{i\left(\phi_k+\phi_j\right)}\right. \\
            & + D_{jL}D_{kR}^*\tilde{\mathcal{P}}_{jk}^*e^{i\left(-\phi_j-\phi_k\right)} \\
            & + D_{jL}\left(\tilde{\mathcal{I}}_{jk} + \tilde{\mathcal{V}}_{jk}\right)e^{i\left(-\phi_k+\phi_j\right)} \\
            & \left. + D_{kR}^*\left(\tilde{\mathcal{I}}_{jk} + \tilde{\mathcal{V}}_{jk}\right)e^{i\left(\phi_k-\phi_j\right)}\right].           
\end{split}
\end{equation}

Since the leakage terms here enter in products $D\tilde{\mathcal{I}}$ rather than $D\tilde{\mathcal{P}}$ (\autoref{eq:rrll}), it is of greater importance to calibrate them for a faithful linear polarization source reconstruction. Neglecting $D^2$, $D\tilde{\mathcal{P}}$, and $\mathcal{V}$ terms, we can use Equations~\ref{eq:rrll} and \ref{eq:rl} to construct the cross-hand to parallel-hand visibility ratios \citep{Roberts1994, Johnson2015}

\begin{equation}
\begin{split}
 \frac{L_jR_k^*}{L_jL_k^*} &\approx \frac{G_{kR}^*}{G_{kL}^*} \left[\breve{m}_{kj}^*e^{2i\phi_k} + D_{jL}e^{-2i\left(\phi_j-\phi_k\right)} + D_{kR}^*\right],\\
\frac{L_jR_k^*}{R_jR_k^*} &\approx \frac{G_{jL}}{G_{kR}} \left[\breve{m}_{kj}^*e^{2i\phi_j} + D_{jL} + D_{kR}^*e^{2i\left(\phi_j-\phi_k\right)}\right],\\
\frac{R_jL_k^*}{L_jL_k^*} &\approx \frac{G_{jR}}{G_{jL}} \left[\breve{m}_{jk}e^{2i\phi_j} + D_{jR} + D_{kL}^*e^{-2i\left(\phi_j-\phi_k\right)}\right],\\
\frac{R_jL_k^*}{R_jR_k^*} &\approx \frac{G_{kL}^*}{G_{kR}^*} \left[\breve{m}_{jk}e^{2i\phi_k} + D_{jR}e^{2i\left(\phi_j-\phi_k\right)} + D_{kL}^*\right].
\end{split}
\end{equation}

Here, $\breve{m}_{jk}=\tilde{\mathcal{P}}_{jk}/\tilde{\mathcal{I}}_{jk}$ is the polarimetric ratio in the visibility domain. This quantity, while not the Fourier transform of its image-domain analog $m=\mathcal{P}/\mathcal{I}$, has proven useful in capturing linear polarization source structure \citep{Johnson2015, Gold2017}. Phase fluctuations cancel in the correlation ratios, and the correlation ratios only depend on left-right gain ratios rather than left and right gains separately. Instrumental polarization leakage remains a source of systematic error, and it can be calibrated by observing the source and a set of calibrators across a wide range of field rotation angle.

Another calibration-invariant data product containing polarization information is the closure trace, which is formed on a quadrangle of baselines \citep{Broderick2020}. Closure traces are a natural extension of closure amplitudes and closure phases into full Stokes. They are insensitive to both station gains and polarimetric leakage, and hence allow for maximum calibration freedom. They may also provide an unambiguous detection of non-trivial polarization image structure, although linear and circular polarization cannot be distinguished in such a test. Since we have a decent handle on the EHT 2017 D-terms and gains, we utilize that information in our fits by fitting to $\breve{m}$ rather than closure traces. We use closure traces for consistency checks in \citetalias{M87PaperIX}.

\subsection{Fitting implementation}
Our polarimetric model fitting methods have been implemented in {\tt eht-imaging} \citep{Chael2016,Chael2018,Chael2019}. This Python library contains many utilities for the analysis of VLBI data, including functions for common operations (e.g., data flagging, averaging, network calibration) on VLBI datasets, plotting utilities, static and dynamic imaging tools, synthetic data generation tools, an interstellar scattering module, and most recently geometric modeling \citep[e.g.,][]{SgrAEHTCIV}. Within the modeling module, several geometric models have been implemented apart from the m-ring model, including point sources, Gaussians, disks, crescents, and rings. These models can be flexibly combined to form a multi-component source model. For our m-ring fits, we set flat priors (uniform between minimum and maximum values) for all source parameters. For posterior exploration, we use the {\tt dynesty} sampler \citep{Speagle2020}.

\subsection{Data pre-processing and treatment of zero-baseline parameters}
Before performing any fitting, we pre-process our datasets in several ways. As typically done for analysis of EHT datasets \citep[e.g.][]{PaperIV, SgrAEHTCII, SgrAEHTCIII, SgrAEHTCIV}, we add a small fraction \citep[estimated at around 2\%; ][]{PaperIII} of the visibility amplitudes to the thermal error budget in quadrature, effectively acting as a regularization and imposing a maximal signal-to-noise ratio in order to represent residual systematic errors. These systematic uncertainties are added to each data point, thus they impact both visibilities and interferometric closure products. For our fits to RR/LL visibility ratios, this fractional noise absorbs systematic uncertainties in the R/L gain ratios. For our fits to closure data products, the fractional noise covers non-closing errors. Apart from polarimetric leakage, the main source of these additional non-closing uncertainties is related to gain-calibration of wide frequency bands \citep[e.g., ][]{Natarajan2022}. We add the 2\% noise to GRMHD synthetic datasets generated without non-closing errors as well in order to maintain consistency in the data pre-processing, but the effect on the closure product noise budget is generally small. Apart from adding fractional noise, we also pre-process our data by performing scan-averages. This operation increases signal-to-noise ratios and reduces the number of data points that need to be fit, making the process more efficient. Finally, we may rescale zero-baseline fluxes depending on the dataset. During the generation of the synthetic GRMHD datasets \citepalias[][\autoref{sec:grmhd}]{M87PaperIX}, a large-scale component was added to the visibilities to mimic large-scale structure seen in the EHT \m87 data. For our fits to these datasets, we added a large-scale circular Gaussian (FWHM 2 mas) model component with the same total flux and polarization parameters as this added component, and kept these parameters fixed while fitting the compact structure with the polarized m-ring model. For the real EHT \m87 data, we used the datasets where the large-scale structure was taken out by rescaling the zero baselines \citepalias{M87PaperIX}.

Circular polarization fitting to the parallel-hand closure products does not constrain the integrated fractional circular polarization $\mathcal{V}_{\mathrm{net}}/\mathcal{I_{\mathrm{tot}}}$. The zero-baseline LL visibilities measure $\mathcal{I_{\mathrm{tot}}}+\mathcal{V_{\mathrm{net}}}$ and the RR visibilities are sensitive to $\mathcal{I_{\mathrm{tot}}}-\mathcal{V_{\mathrm{net}}}$ (\autoref{eq:rrll}), but the closure products containing zero-baselines cannot distinguish between these. During the fitting, we therefore fix $\mathcal{V}_{\mathrm{net}}$ to the ground-truth value for synthetic datasets, and to the measured $\mathcal{V}_{\mathrm{net}}$ from zero-baseline observations \citep[ALMA-only,][]{Goddi2021} for real \m87 data. For consistency, we also fix $\mathcal{V}_{\mathrm{net}}$ for our fits to right-left visibility ratios. We investigate the effect of varying $\mathcal{V}_{\mathrm{net}}$ on our \m87 fits in \autoref{sec:m87cpol}.

\section{Tests on synthetic data from geometric models}
\label{sec:synthdata_geomodels}
To further outline, motivate, and test our polarimetric fitting procedure, we start by fitting polarized m-rings to synthetic EHT data generated from the same model. Thus, the model specification is perfect for these tests. In particular, we use these tests on geometric models to establish a preferred fitting procedure that is free of biases, at least in these idealized cases. As shown below, biases may be introduced by not fitting the Stokes $\mathcal{I}$ and $\mathcal{V}$ structure simultaneously, by fitting the Stokes $\mathcal{V}$ structure without taking the linear polarization structure and leakage effects into account, and by fitting to the RR/LL data product in the presence of non-unity R/L gain ratios. Once the fitting procedures have been established and tested, we move on to the more realistic case of GRMHD models in \autoref{sec:grmhd}.

\subsection{Model description and synthetic data generation}
For the geometric model tests, we used a circularly polarized m-ring model with $F=0.5$ Jy, $d=40$ $\mu$as, $\alpha=10$ $\mu$as, $\beta_{\mathcal{I},1}=0.2-0.1i$, $\beta_{\mathcal{V},0}=0.05$, and $\beta_{\mathcal{V},1}=0.1-0.1i$. This model is shown in the middle panel of \autoref{fig:mring_examples}. The net circular polarization fraction of 5\% is substantially higher than observed for most VLBI sources, as is the resolved circular polarization fraction of up to $\sim$40\% (\autoref{sec:grmhd} shows more realistic cases). In order to test the effect of linear polarization structure and leakage, we also generated a model with added linear polarization by setting $\beta_{\mathcal{P},0}=0.1$, $\beta_{\mathcal{P},-1}=0.1+0.2i$, and $\beta_{\mathcal{P},1}=-0.1+0.1i$. 

Synthetic data were generated with {\tt eht-imaging} \citep{Chael2016, Chael2018}, using the $uv$-coverage and thermal noise from the synthetic datasets generated for the imaging and modeling method tests in \citetalias{M87PaperIX} (see also \autoref{sec:grmhddata}), corresponding to the low-band EHT \m87 dataset from 11~April 2017. Since these tests focus on the effect of fitting procedure choices and the presence of linear polarization and leakage, we did not introduce any systematic gain offsets for these tests. Such effects are introduced in our GRMHD model fits (\autoref{sec:grmhd}). For the tests with linear polarization, we set the left leakage terms $D_{jL} = 0.04+0.04i$, and the right leakage terms $D_{jR} = 0.03+0.03i$ for all stations.

\subsection{Fitting procedures and results}
\label{sec:results_geometric_models}

\begin{figure*}
    \centering
    \includegraphics[width=0.47\textwidth]{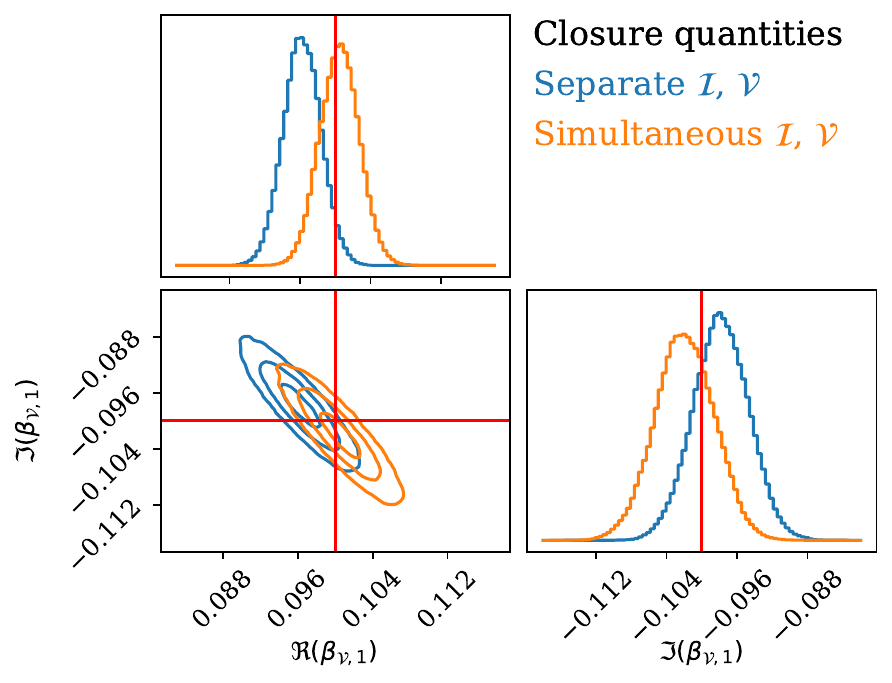}
    \includegraphics[width=0.47\textwidth]{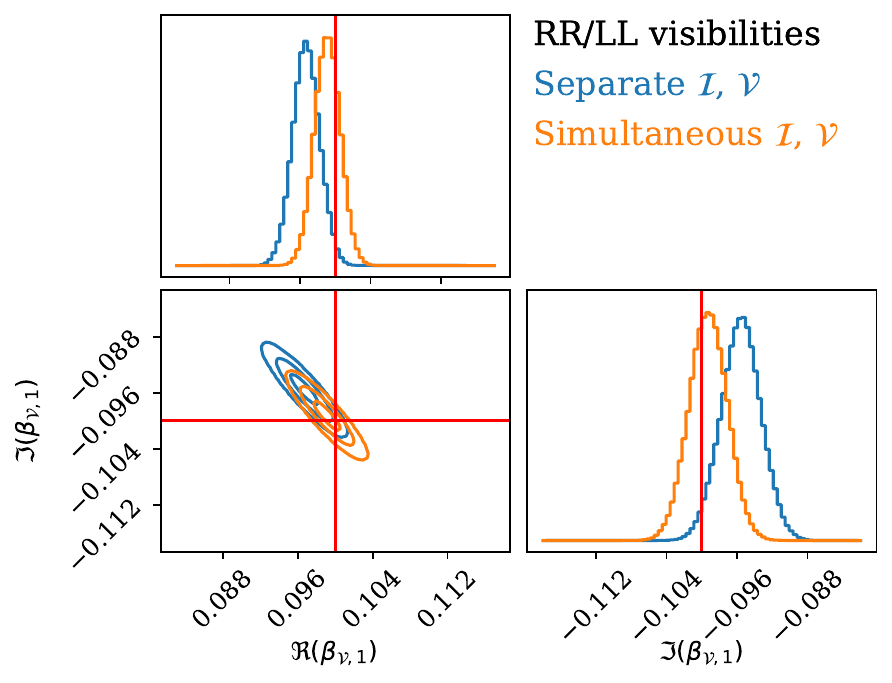} \\
    \includegraphics[width=0.47\textwidth]{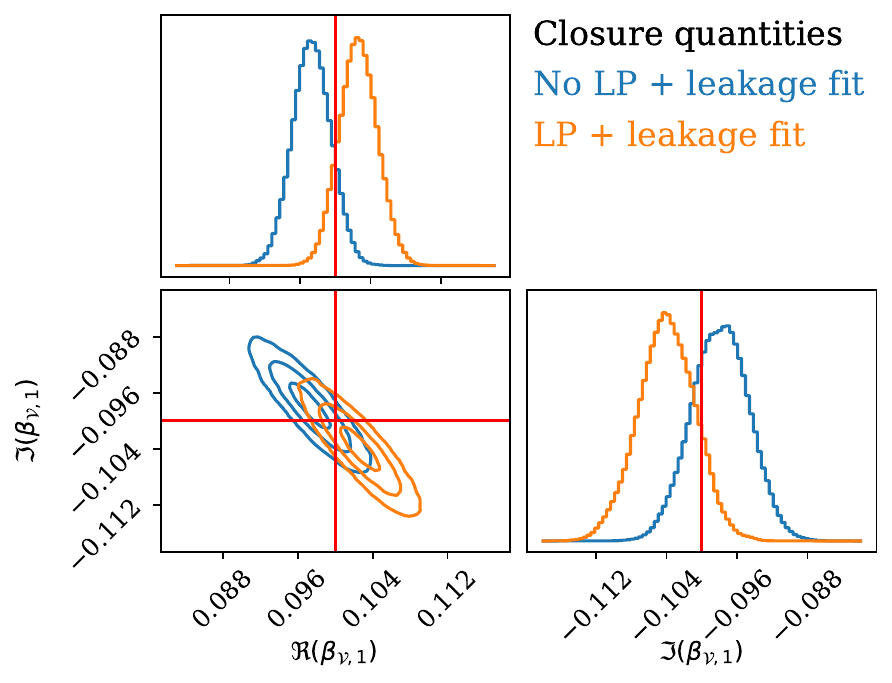}
    \includegraphics[width=0.47\textwidth]{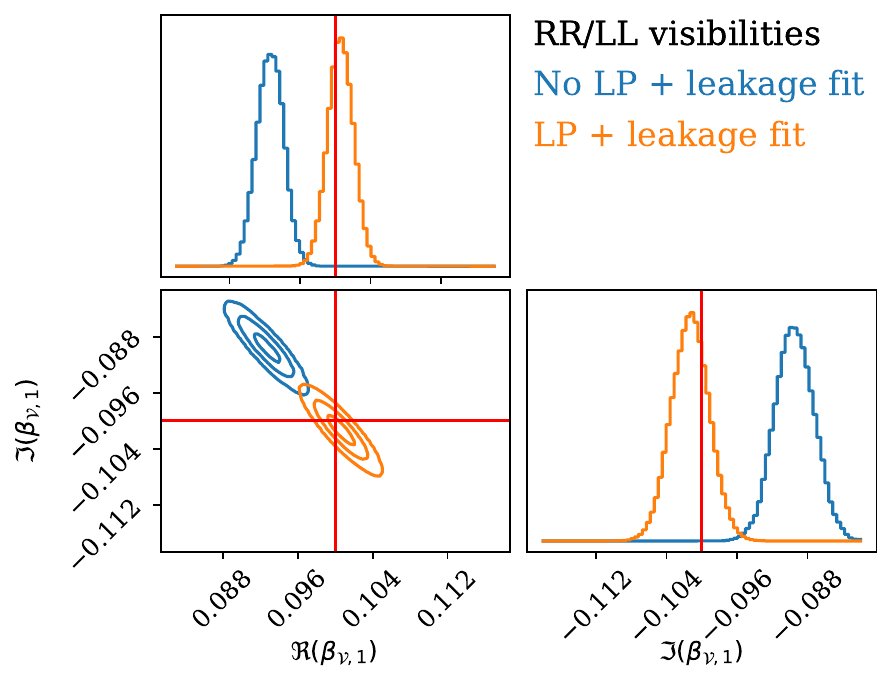}\\
    \includegraphics[width=0.47\textwidth]{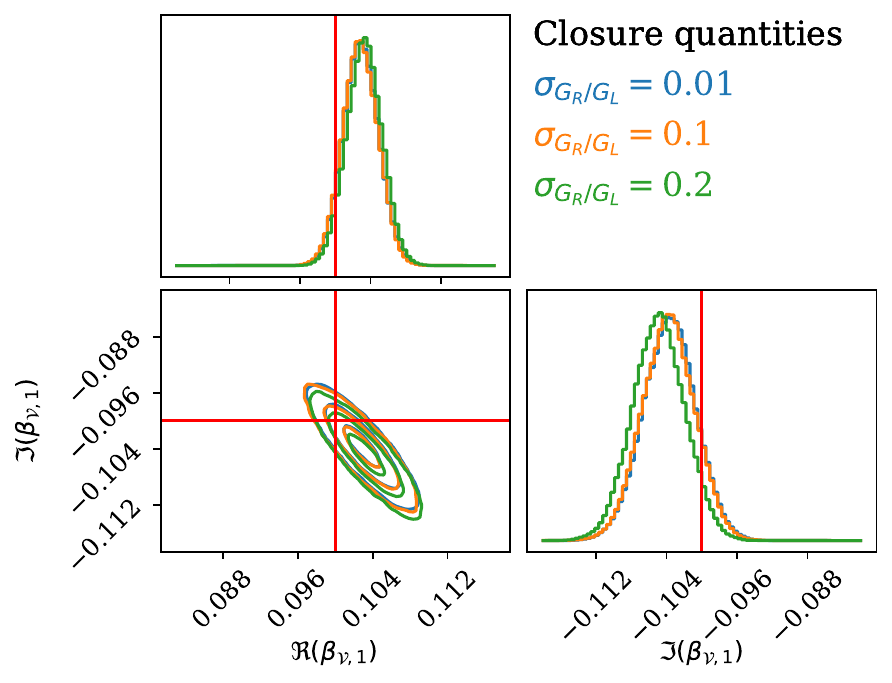}
    \includegraphics[width=0.47\textwidth]{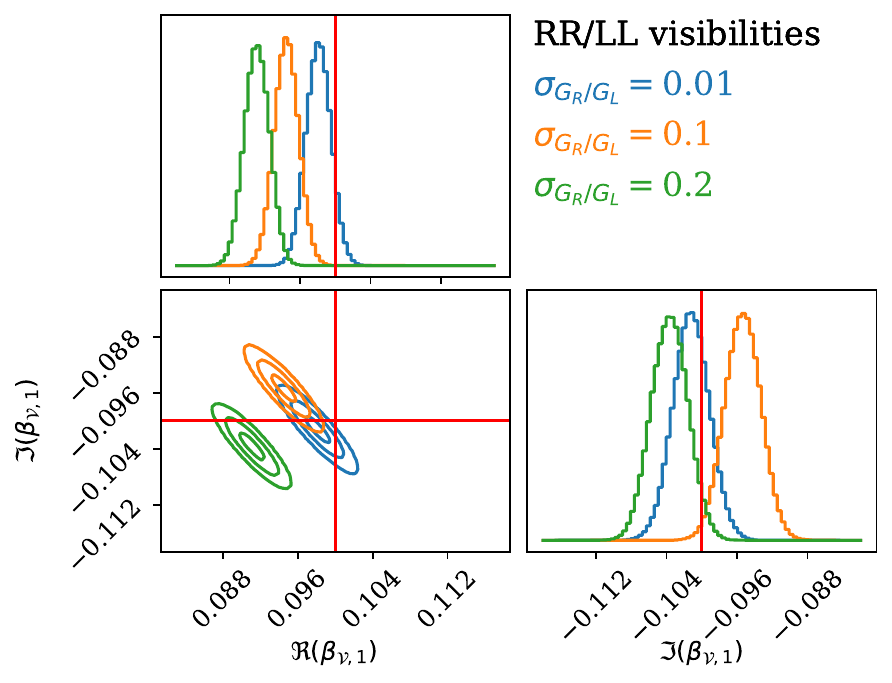}
    \caption{$\beta_{\mathcal{V},1}$ posteriors from fitting a circularly polarized m-ring to synthetic EHT 2017 data generated from two different m-ring models. Ground truth values are indicated with red vertical and horizontal lines and correspond to the middle panel of \autoref{fig:mring_examples}. Fits using closure quantities are shown on the left, and fits with RR/LL visibility ratios are shown on the right. The posteriors on the top row were computed from fits to data generated from a model with zero linear polarization and without any leakage corruptions added to the data. Each top row panel compares separate (consecutive) Stokes $\mathcal{I}$ and $\mathcal{V}$ fits (blue) versus simultaneous Stokes $\mathcal{I}$ and $\mathcal{V}$ fits (orange). The posteriors in the middle row were computed from simultaneous Stokes $\mathcal{I}$ and $\mathcal{V}$ fits to data generated from a model with nonzero linear polarization and with leakage corruptions added to the data. Each middle row panel compares fits ignoring linear polarization (blue) to fits including linear polarization and leakage fits (orange). The bottom row shows comparisons between applying different R/L gain ratio offsets to the synthetic data before fitting. The contours show 1$\sigma$, 2$\sigma$, and 3$\sigma$ levels.
    In general, the fits using only closure quantities have slightly weaker constraints, and both data products show biases when fitting Stokes $\mathcal{V}$ separate from $\mathcal{I}$; RR/LL visibility ratios show biases when ignoring the presence of linear polarization and leakage effects, and when R/L gain offsets are present in the data. 
    }
    \label{fig:geometrictestsstokesv}
\end{figure*}

\autoref{fig:geometrictestsstokesv} shows $\beta_{\mathcal{V},1}$ posteriors from fits to our geometric models using different fitting procedures and data products. In general, fits to closure quantities result in wider posteriors than fits to RR/LL visibility ratios, which is expected given that there is more information in the latter. 

The top row in \autoref{fig:geometrictestsstokesv} shows the importance of fitting for the Stokes $\mathcal{I}$ and $\mathcal{V}$ structure simultaneously. Polarimetric reconstructions are often made by first reconstructing the Stokes $\mathcal{I}$ structure, and keeping that frozen while reconstructing the polarimetric structure \citep[e.g.,][]{PaperVII}. Taking a similar approach here results in the blue curves in the top row panels of \autoref{fig:geometrictestsstokesv}. Here, the Stokes $\mathcal{I}$ m-ring parameters were fixed to the posterior maximum (MAP) of the Stokes $\mathcal{I}$ fit before fitting the Stokes $\mathcal{V}$ structure. This strategy results in a small (a few percent) but statistically significant bias. Small errors in the Stokes $\mathcal{I}$ parameters propagate to a biased estimate of the Stokes $\mathcal{V}$ parameters, as the parallel-hand visibilities, which are the data products used for the fits, contain contributions from both (\autoref{eq:rrll}). Fitting for the Stokes $\mathcal{I}$ and $\mathcal{V}$ structure simultaneously (orange curves) is computationally more expensive, but removes the biases.

The middle row in \autoref{fig:geometrictestsstokesv} shows another potential source of biases in the Stokes~$\mathcal{V}$ posteriors. For these fits, the ground truth source model included linear polarization structure, and polarization leakage was introduced in the synthetic data generation (\autoref{sec:synthdata_geomodels}). The blue curves show posteriors resulting from fitting the Stokes $\mathcal{I}$ and $\mathcal{V}$ structure simultaneously, but ignoring the linear polarization structure and leakage. While the fit to closure quantities is acceptable, a significant bias is present for the fit to the RR/LL visibility ratios. Even though the linear polarization structure and leakage enters the parallel-hand visibilities only as a second-order effect (\autoref{eq:rrll}), they may still cause the circular polarization fits to be biased and therefore should be included in the Stokes~$\mathcal{V}$ fitting process. The orange curves were obtained by first fitting the Stokes $\mathcal{I}$ structure, then fitting the linear polarization structure and leakage parameters, and subsequently fixing the linear polarization structure and leakage parameters to the MAP while fitting for the Stokes $\mathcal{I}$ and $\mathcal{V}$ structure simultaneously. This strategy removes the biases introduced when ignoring the linear polarization structure and leakage effects.

Finally, the bottom row in \autoref{fig:geometrictestsstokesv} shows that when introducing R/L gain amplitude offsets (here set to be constant in time), the calibration-invariant closure-only posteriors are not affected, while the R/L gain-sensitive RR/LL posteriors (\autoref{eq:rrll}) show increasing biases with increasing R/L gain offsets. In practice, it is thus important that the R/L gain ratios are calibrated as well as possible when fitting to RR/LL visibility ratios from real datasets. Since closure products are not affected by these gain offsets, checking for consistency with closure-only fits is recommended (see also \autoref{sec:m87}).

\section{Application to synthetic data from GRMHD models}
\label{sec:grmhd}
In this section, we apply our polarimetric m-ring fitting procedures to synthetic EHT data from a set of three GRMHD models, investigating in particular how well the basic (asymmetric) linear and circular structure can be recovered for different GRMHD parameters, and how the geometric fits behave as a function of the Stokes~$\mathcal{V}$ m-ring order $m_{\mathcal{V}}$. The synthetic datasets were generated for the circular polarization imaging and modeling tests described in \citetalias{M87PaperIX}. Here, we discuss the m-ring modeling results in greater depth.

\subsection{Model description and synthetic data generation}
\label{sec:grmhddata}
The three GRMHD models 1 (MAD, $a_*$ = -0.5), 2 (MAD, $a_*$ = 0.5), and 3 (SANE, $a_*$ = 0) have different levels of resolved circular polarization. The image-averaged, total-intensity weighted circular polarization fraction, $\langle|\mathcal{V/I}|\rangle$, is approximately 0.5\%, 2\%, and 4\%, respectively, for the three models. All models pass linear and total circular polarization constraints from \citet{PaperVII} and \citet{Goddi2021}. 

Synthetic data were generated using {\tt eht-imaging} for low-band EHT \m87 $uv$-coverage on 11 April 2017. Thermal noise and systematic gain and leakage terms were added, and non-unity left-right gain ratios $G_R/G_L$ were introduced for all sites except ALMA. The $G_R/G_L$ amplitudes were sampled from a Gaussian distribution with unity mean and a 20\% standard deviation, with a 2-hour correlation timescale. For the $G_R/G_L$ phases, a standard deviation of 10$^{\circ}$ (40$^{\circ}$ for the SMA station) and a 24-hour correlation timescale were used. These numbers were motivated by a priori limits estimated for the 2017 EHT data \citepalias{M87PaperIX}. More details on the GRMHD models and data generation are reported in \citetalias{M87PaperIX}. 

To quantitatively compare our fits to the ground-truth model, we compute the ground-truth $\beta_{\mathcal{P}, k}$ (and analogously the $\beta_{\mathcal{V}, k}$) in the image domain \citep[see, e.g.,][]{Palumbo2020} as
\begin{equation}
    \beta_{\mathcal{P},k} = \frac{1}{\mathcal{I}_{\mathrm{tot}}}\int_{\rho_{\mathrm{min}}}^{\rho_{\mathrm{max}}}\int_0^{2\pi}\mathcal{P}(\rho,\phi)e^{-ik\phi}\rho d\phi d\rho,
\end{equation}
where we set the inner radius $\rho_{\mathrm{min}}$ to zero and the outer radius $\rho_{\mathrm{max}}$ to a large value (outside the field of view) in order to capture the full model image.

\subsection{Linear polarization results}
\begin{figure*}
    \centering
    \includegraphics[height=0.3\textwidth]{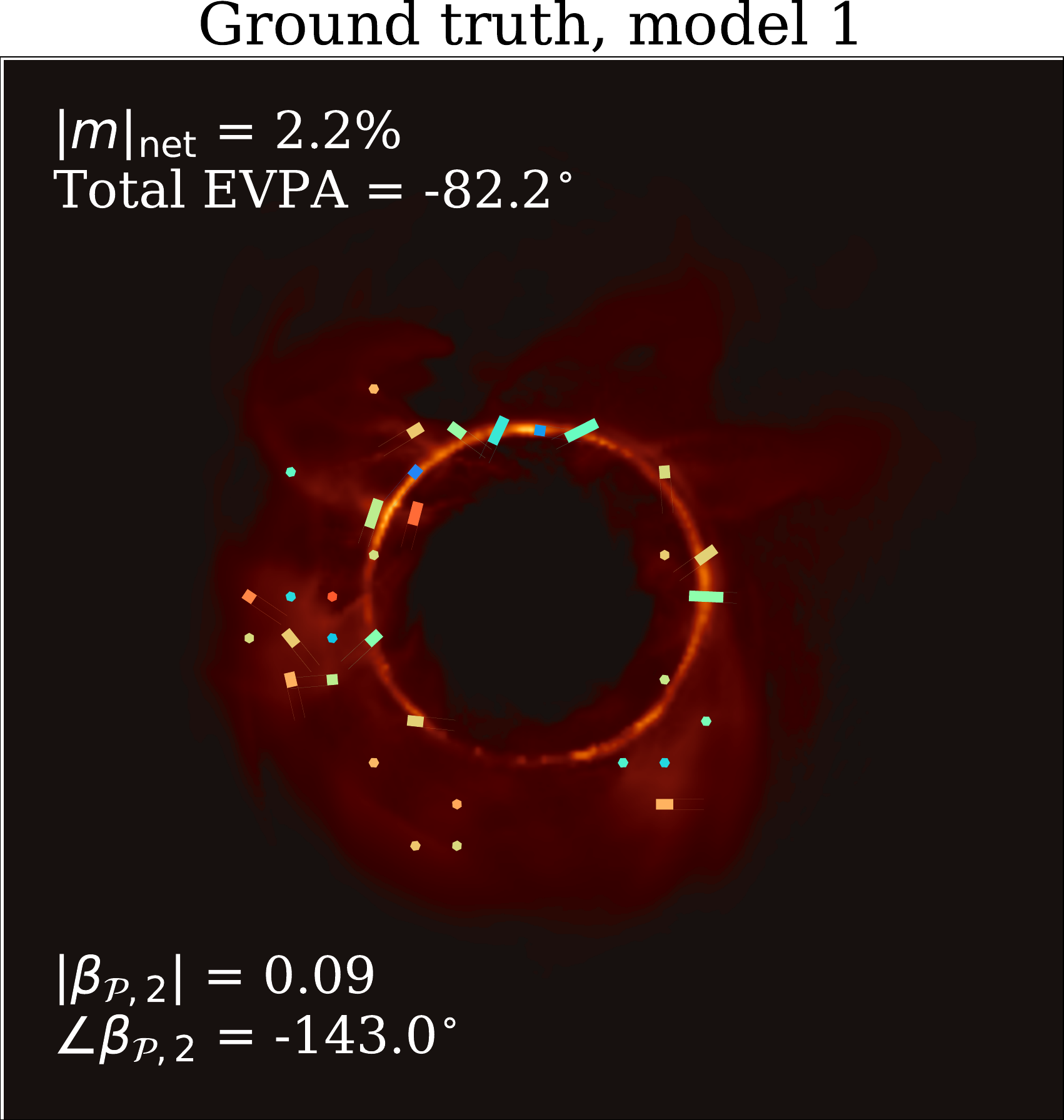}
    \includegraphics[height=0.3\textwidth]{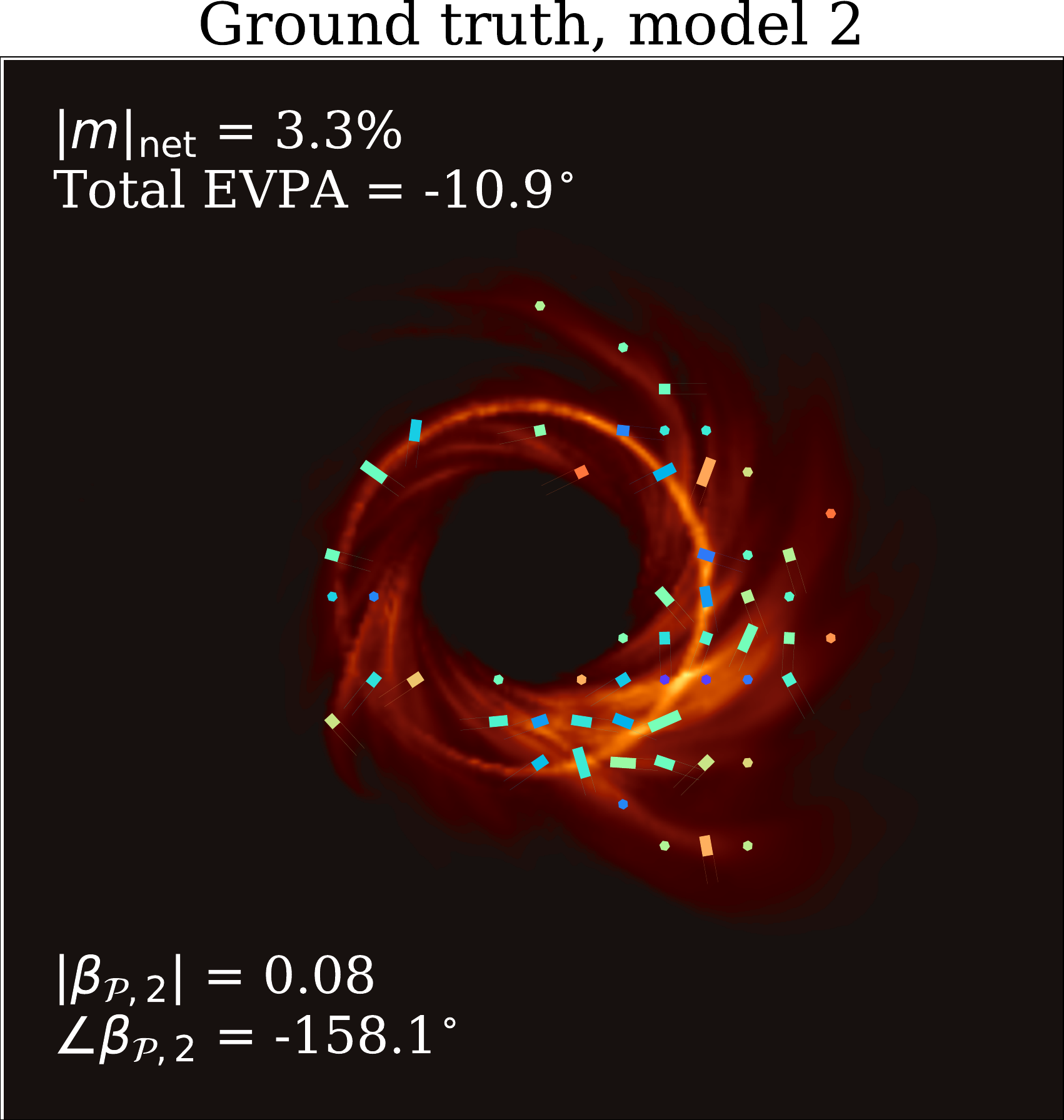}
    \includegraphics[height=0.3\textwidth]{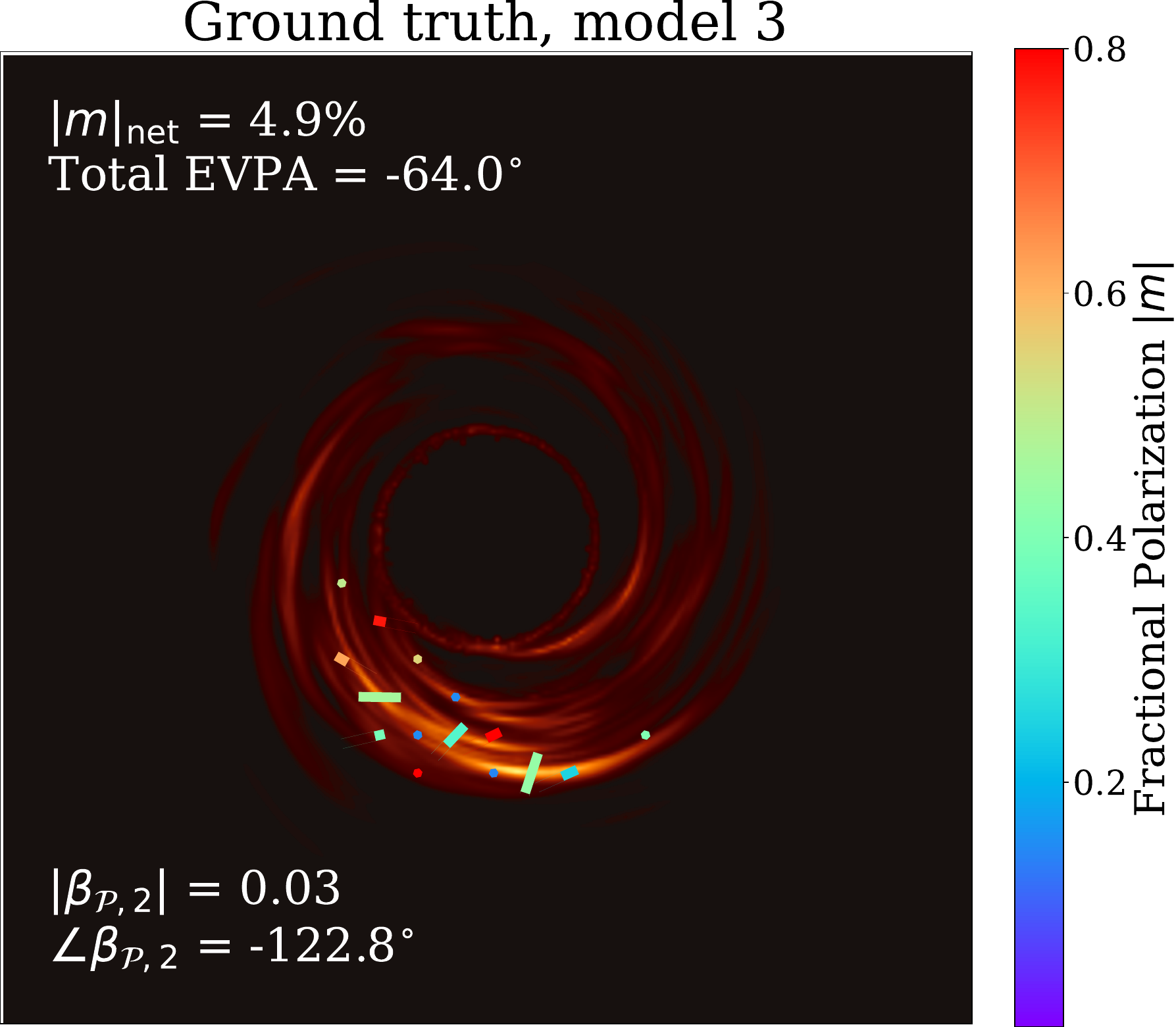} \\ \vspace{3mm}
    \includegraphics[height=0.3\textwidth]{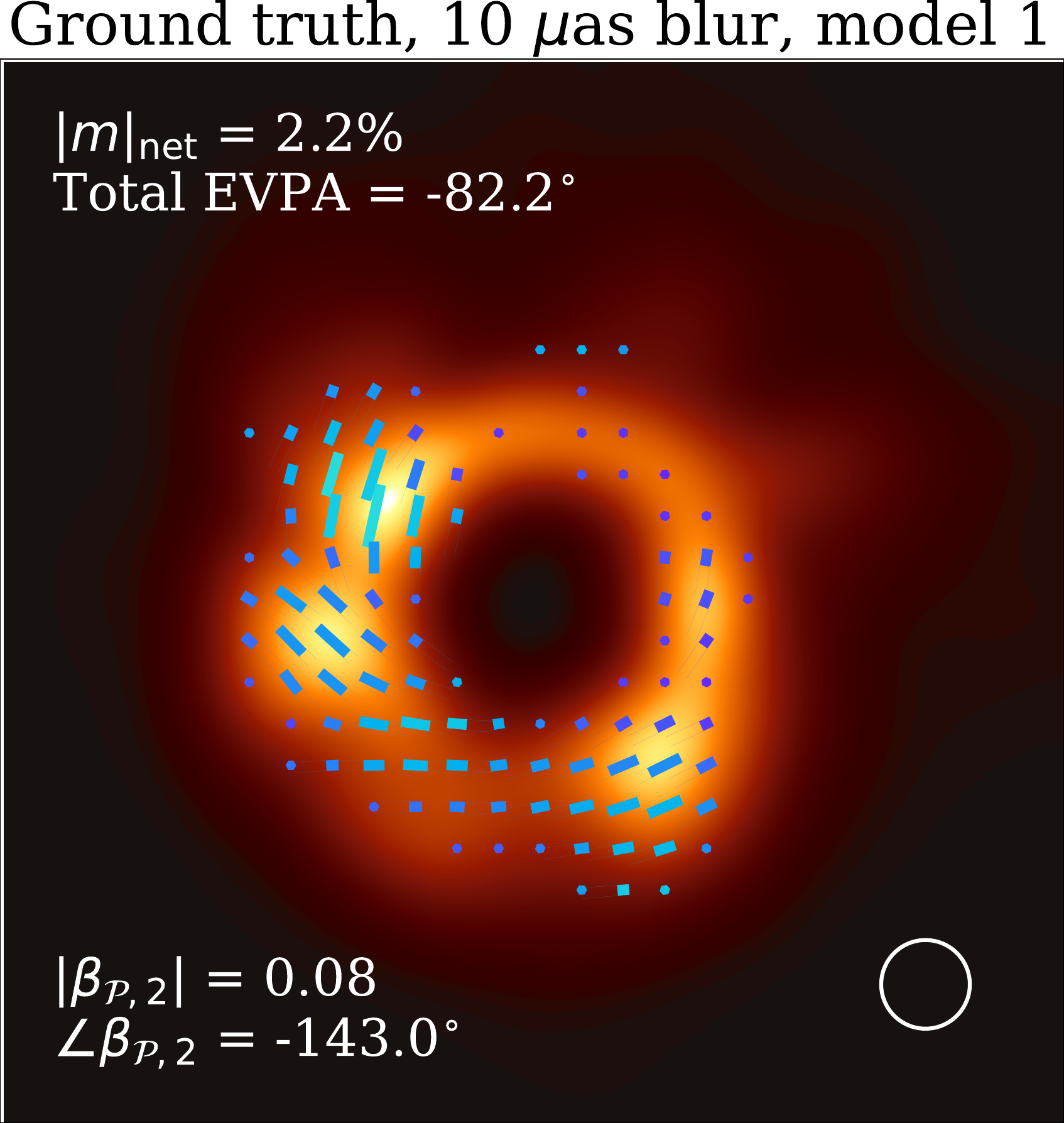}
    \includegraphics[height=0.3\textwidth]{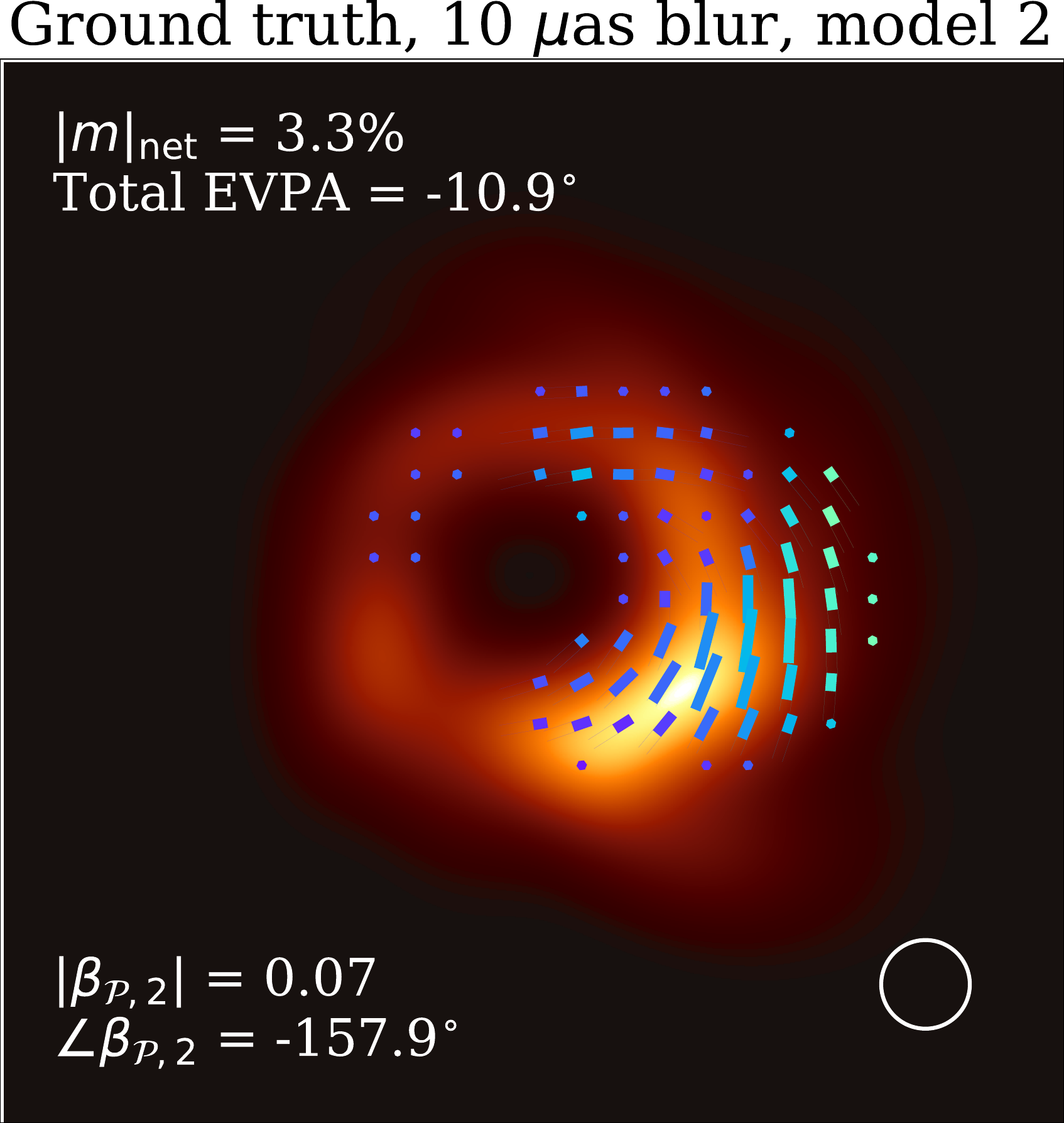}
    \includegraphics[height=0.3\textwidth]{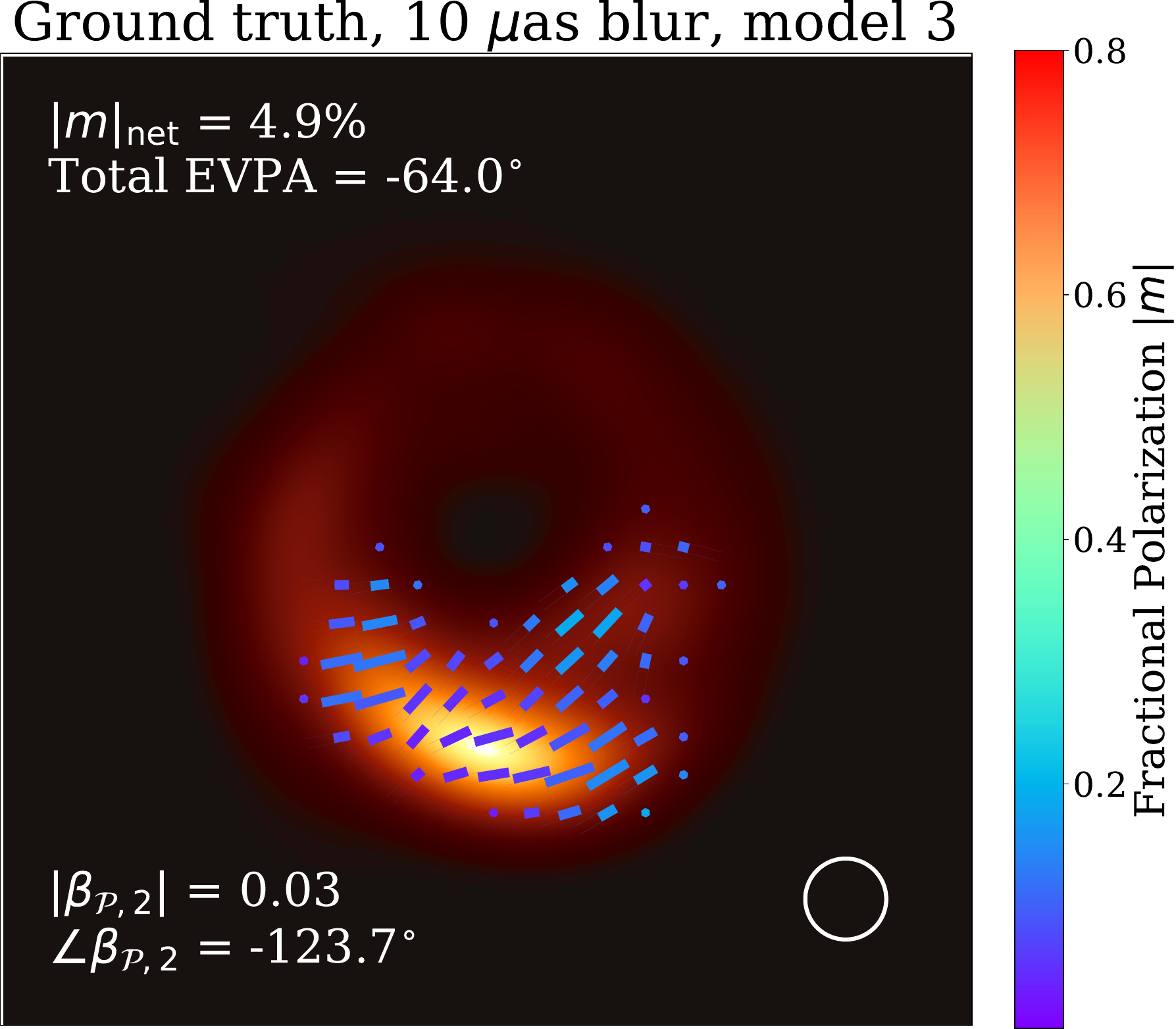} \\ \vspace{3mm}
    \includegraphics[height=0.3\textwidth]{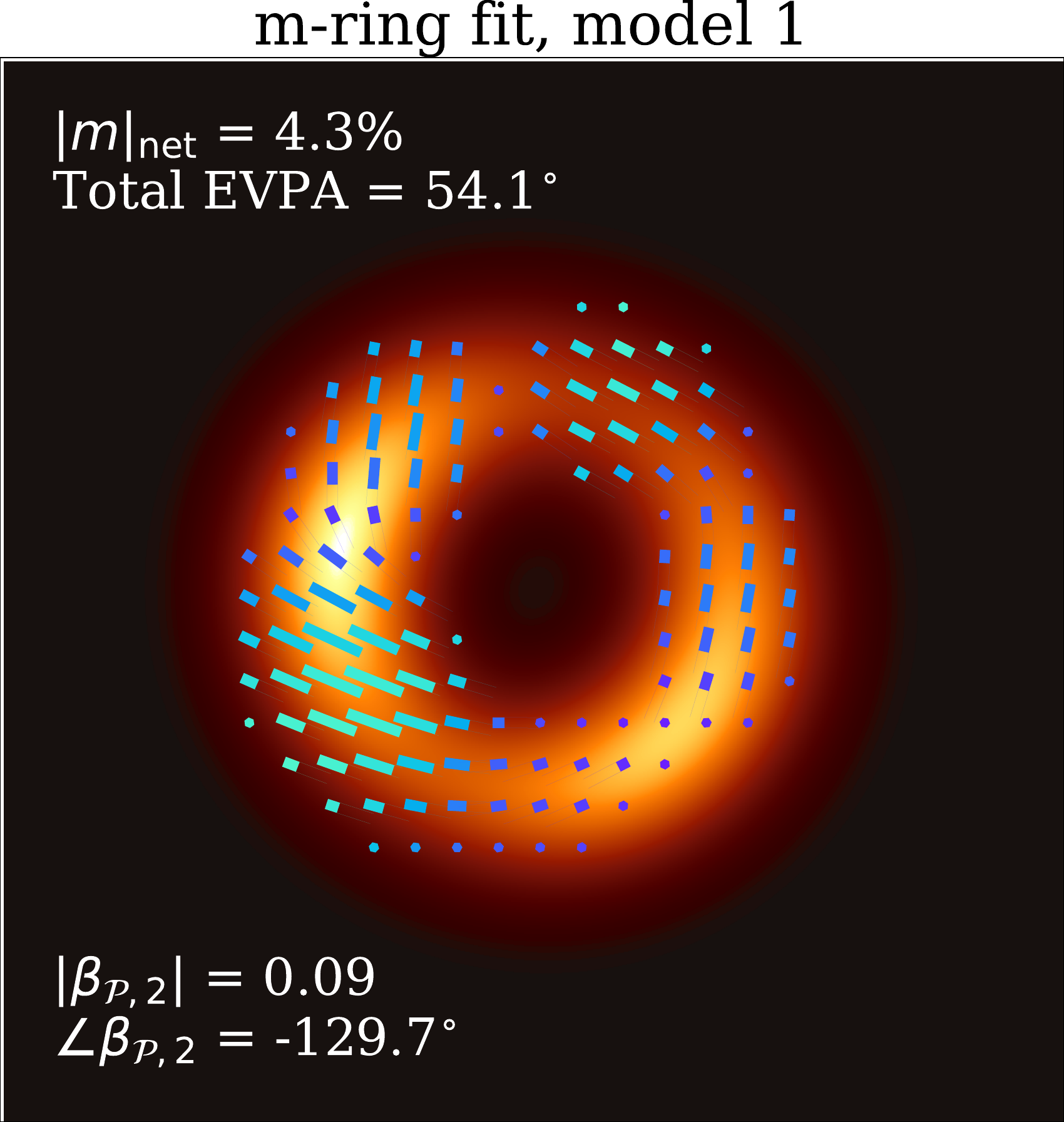}
    \includegraphics[height=0.3\textwidth]{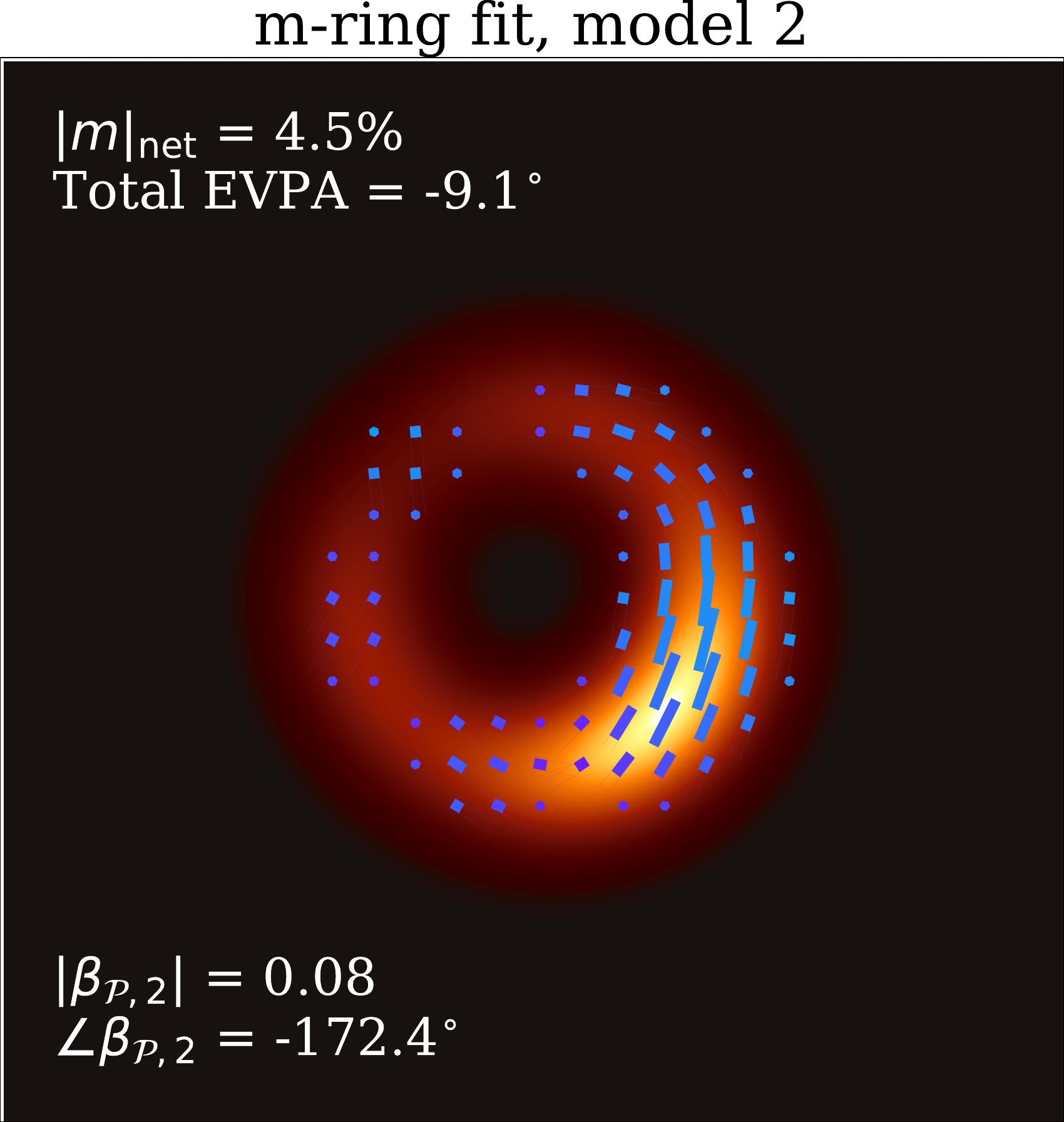}
    \includegraphics[height=0.3\textwidth]{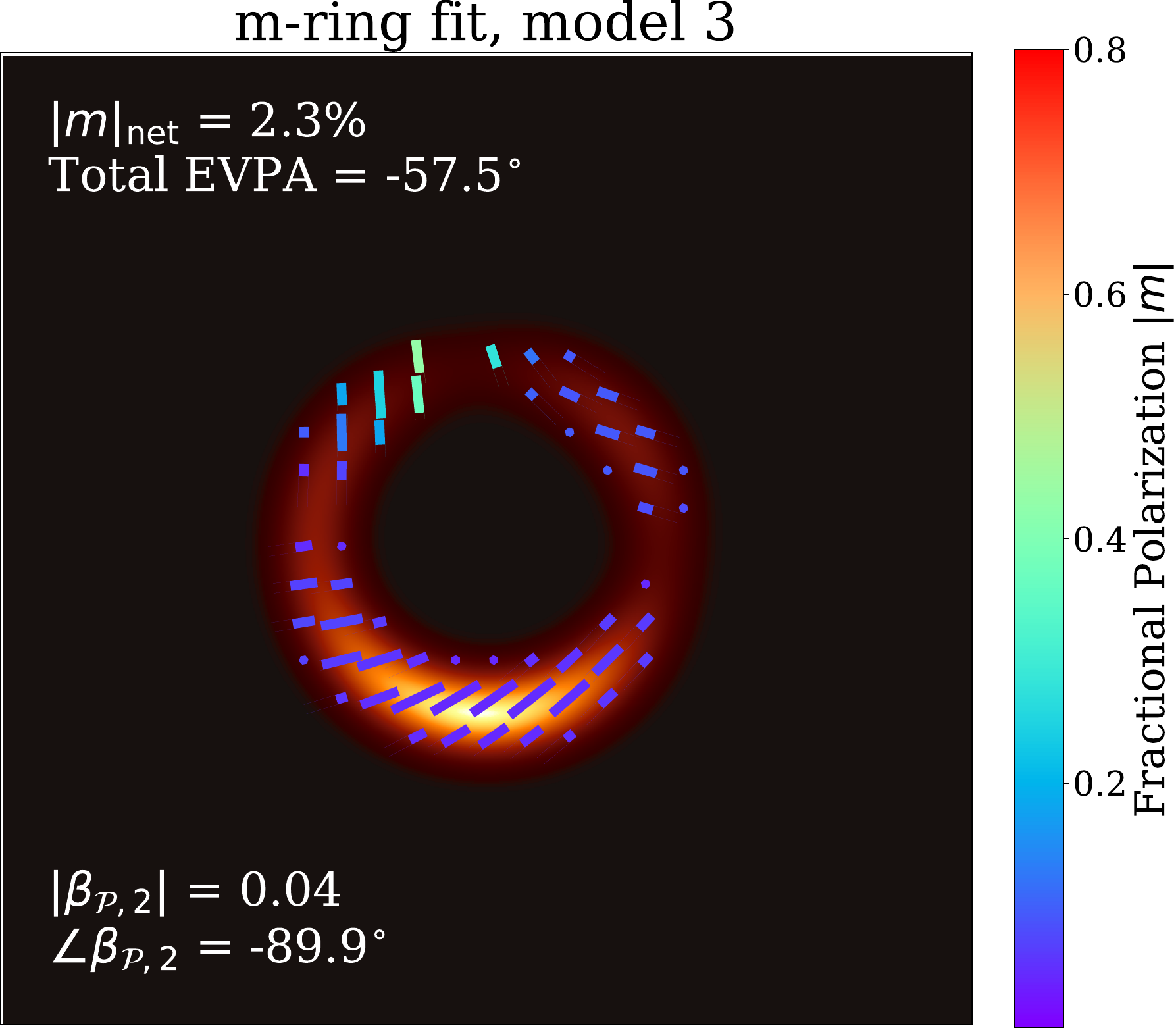}
    \caption{Total intensity and linear polarization reconstructions of our three GRMHD models (left to right), from synthetic data with EHT 2017 coverage. The unblurred and blurred (FWHM 10 $\mu$as) ground truth images are shown in the top and middle rows, respectively. The bottom row shows the posterior maxima of m-ring fits with $m_{\mathcal{I}}=3$ and $m_{\mathcal{P}}=3$. In each panel, the Stokes $\mathcal{I}$ structure is indicated by the heat map, and the scale is normalized to the brightest pixel in each panel. The tick length indicates the polarized intensity, the tick orientation indicates the EVPA, and the tick color indicates the fractional linear polarization. The m-ring fit posteriors for several key parameters are compared to the ranges found by \citet{PaperI, PaperIV, PaperVII} in \autoref{tab:fitcomp}.}
    \label{fig:grmhd_linpol}
\end{figure*}

\autoref{fig:grmhd_linpol} shows total intensity and linear polarization m-ring fits ($m_{\mathcal{I}}=3$, $m_{\mathcal{P}}=3$) to synthetic data from the three GRMHD models. While a comparison by eye shows many low-order features being recovered by the modeling, some systematic offsets clearly remain, which we attribute to model misspecification. The total polarization fraction is recovered least accurately for model 3, which was challenging to fit with the m-ring model due to the extended emission outside the photon ring and the high degree of asymmetry, concentrating most total intensity and polarized emission in the South. The net EVPA is recovered within a few degrees for models 2 and 3 but not for model 1, which has small net polarization fraction and the most complex EVPA structure. $\angle\beta_{\mathcal{P},2}$ is recovered to within $12-34$ degrees, and $|\beta_{\mathcal{P},2}|$ to within 0.01. 

Overall, model 2 is fit best, which can be attributed to a simple twisty polarization pattern centered on the photon ring, which lends itself especially well to m-ring modeling. Such a polarization pattern is seen in many MAD GRMHD models. Models 1 and 3 both show most emission outside the photon ring, due to model 1 being a model with retrograde spin and model 3 being a SANE model with zero spin \citep[SANE and zero-spin models have been ruled out for \m87,][]{M87PaperV, PaperVIII}. As is made clear from these results, the ability of m-ring modeling to constrain the polarization structure depends on the similarity of the model to the ground truth, which is indeed the most important caveat for any geometric modeling result. As shown in \autoref{sec:m87}, it appears that \m87 as observed by the EHT is well-suited for linear polarization m-ring modeling, given the excellent agreement between imaging and modeling results.

\subsection{Circular polarization results}
\begin{figure*}
    \centering
    \includegraphics[height=0.200\textwidth]{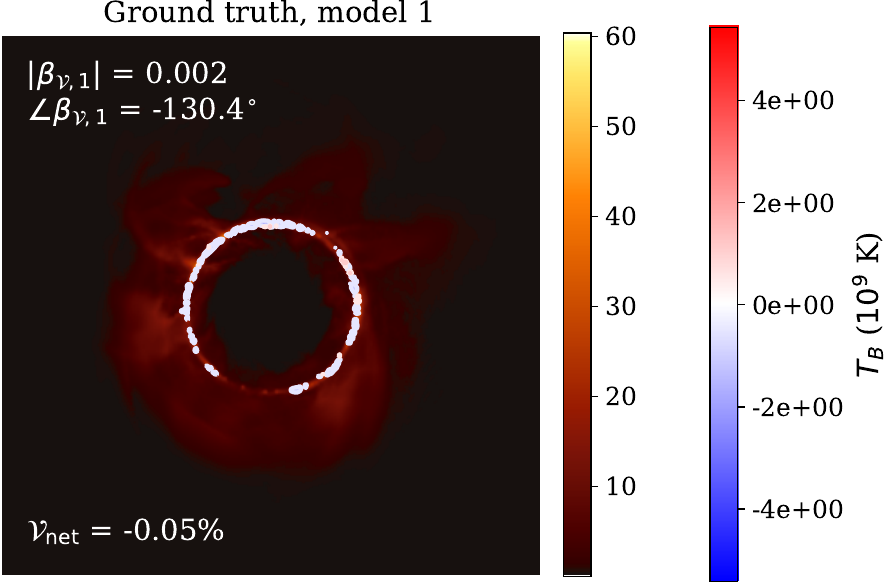} 
    \includegraphics[height=0.200\textwidth]{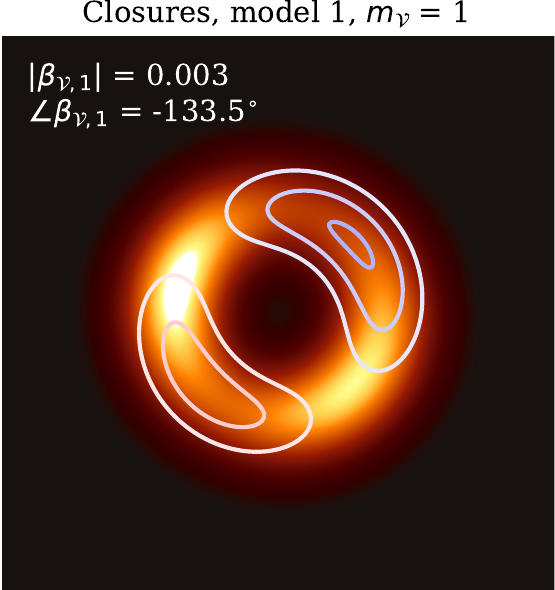}
    \includegraphics[height=0.200\textwidth]{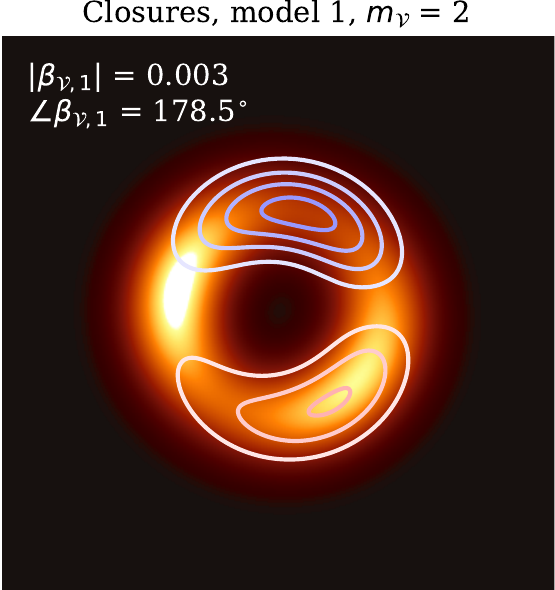}
    \includegraphics[height=0.200\textwidth]{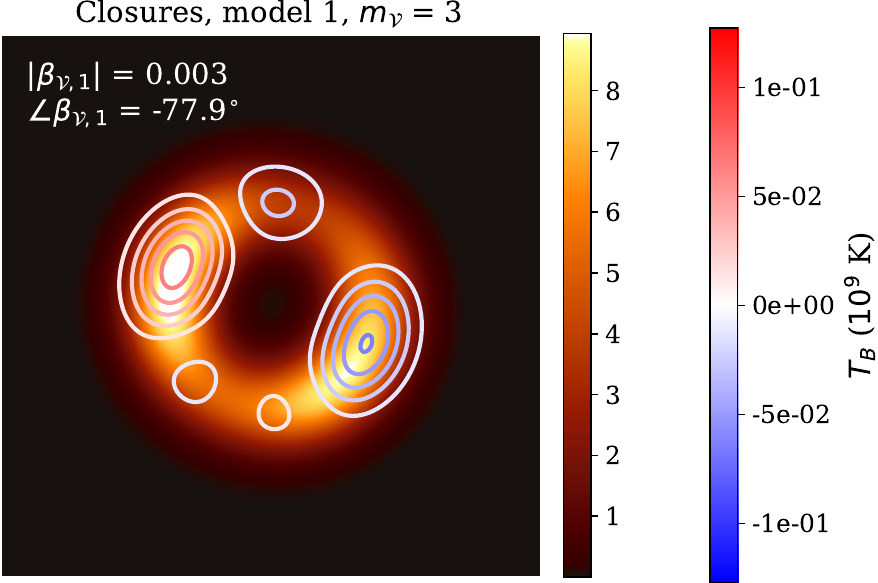} \\
    \includegraphics[height=0.200\textwidth]{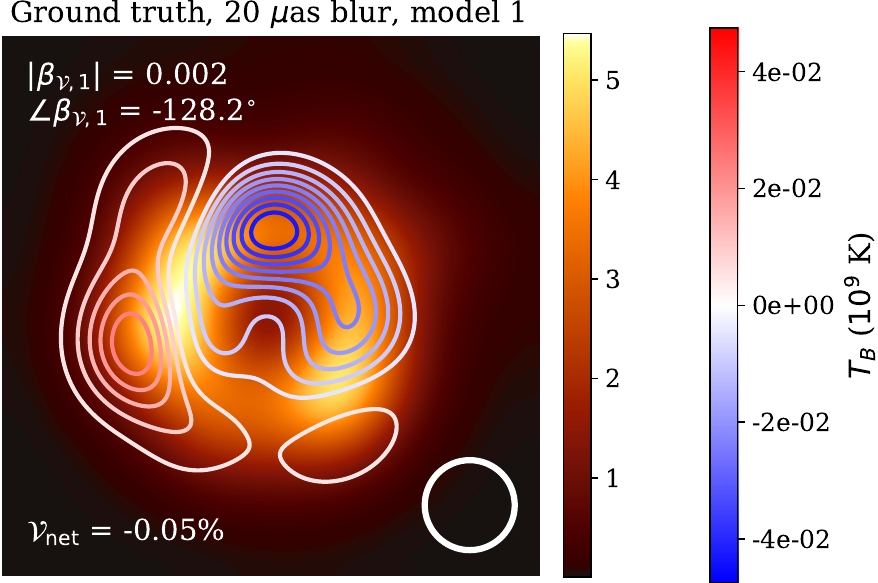} 
    \includegraphics[height=0.200\textwidth]{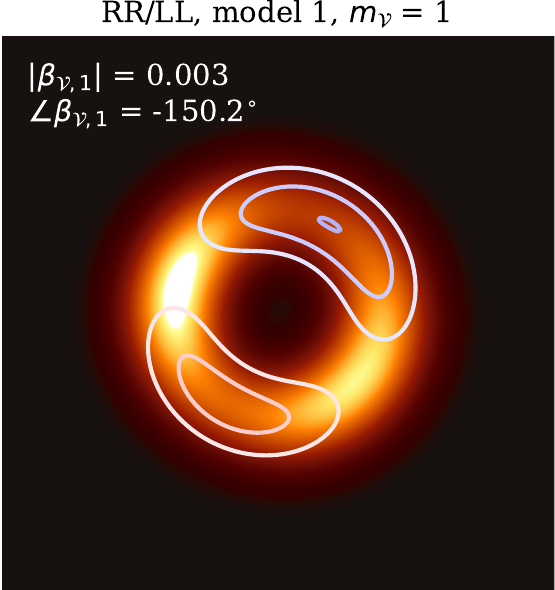}
    \includegraphics[height=0.200\textwidth]{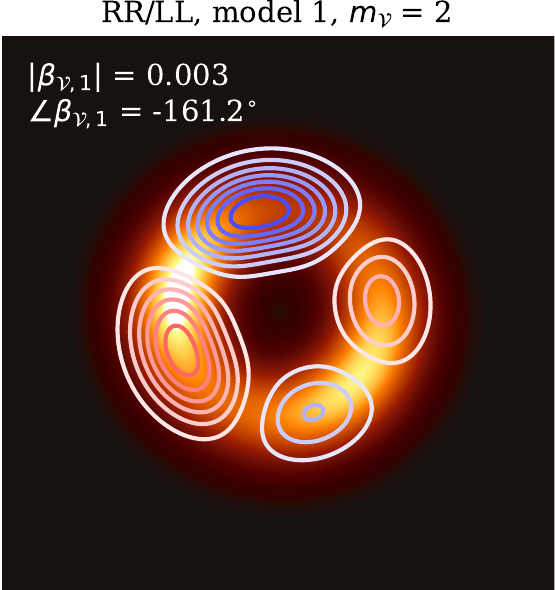}
    \includegraphics[height=0.200\textwidth]{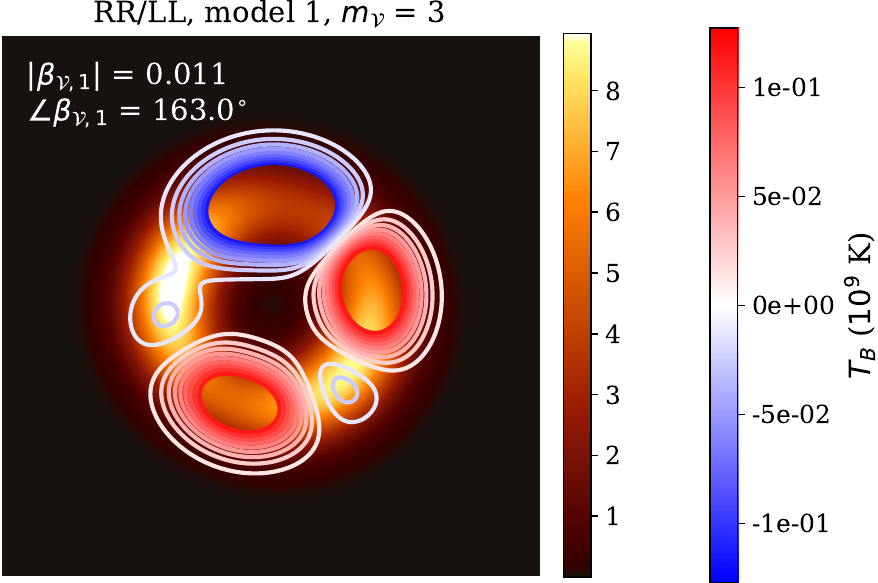} \\
    \vspace{6mm}
    \includegraphics[height=0.200\textwidth]{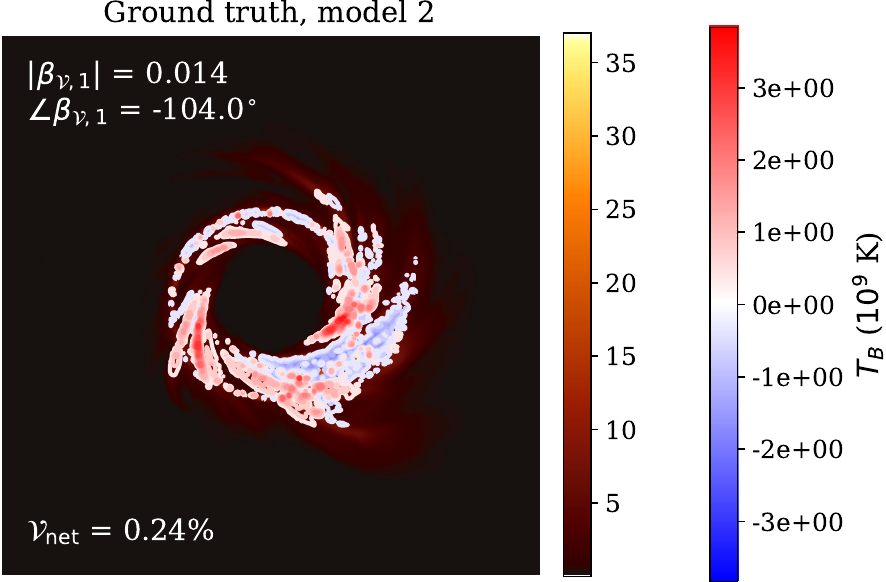} 
    \includegraphics[height=0.200\textwidth]{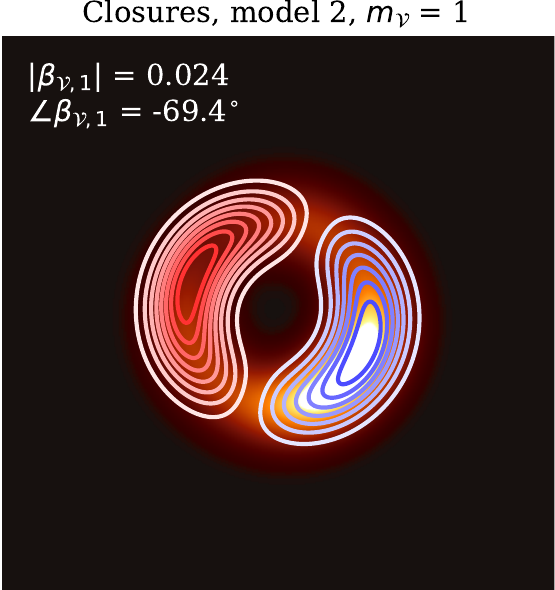}
    \includegraphics[height=0.200\textwidth]{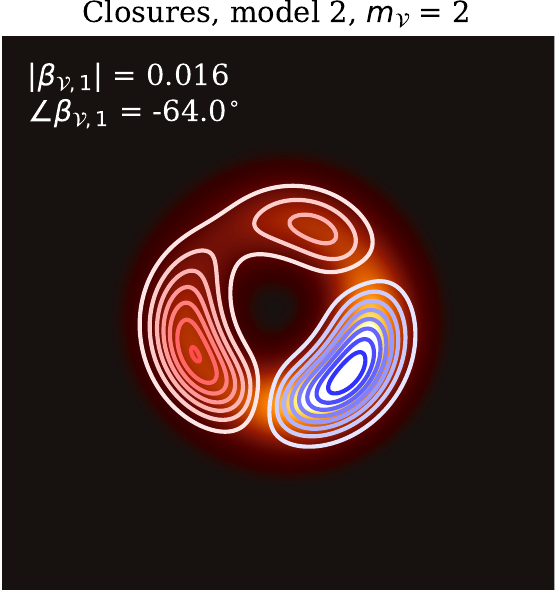}
    \includegraphics[height=0.200\textwidth]{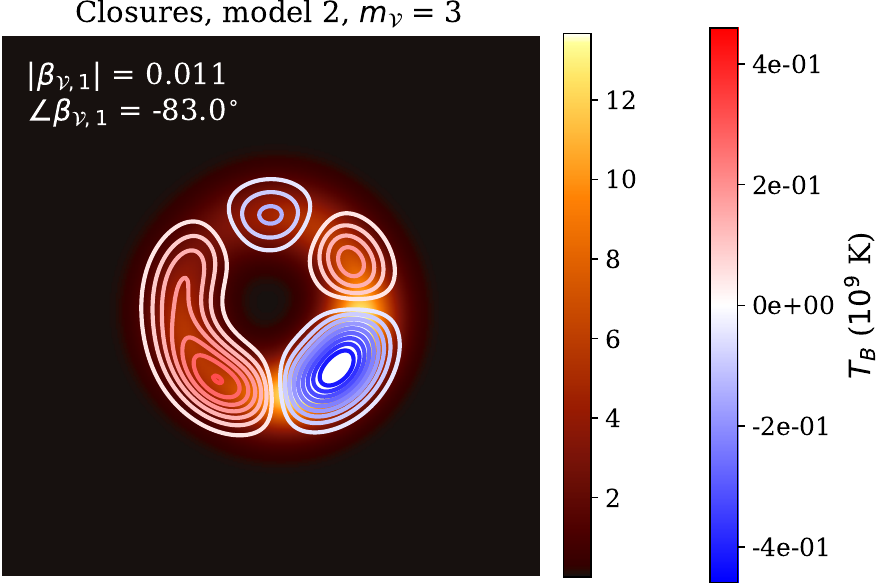} \\
    \includegraphics[height=0.200\textwidth]{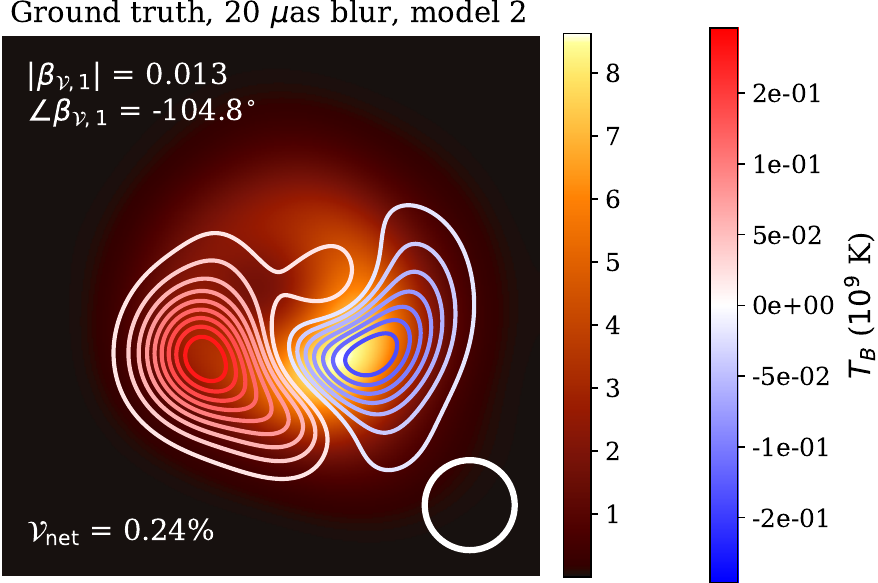} 
    \includegraphics[height=0.200\textwidth]{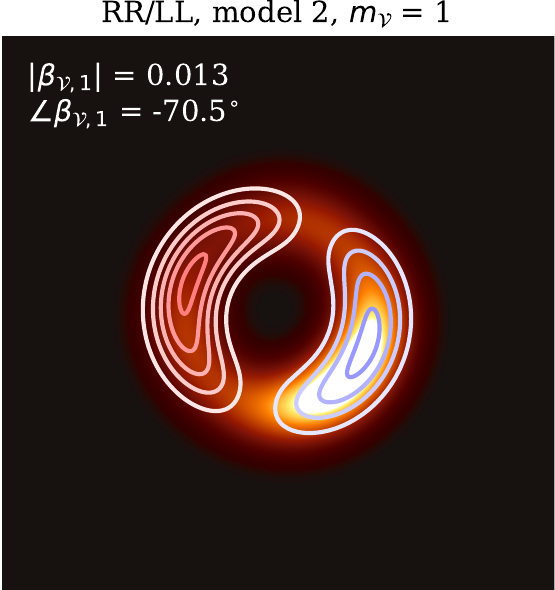}
    \includegraphics[height=0.200\textwidth]{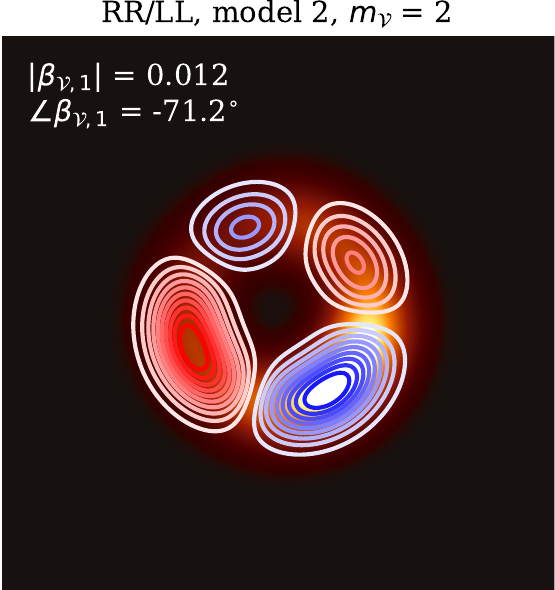}
    \includegraphics[height=0.200\textwidth]{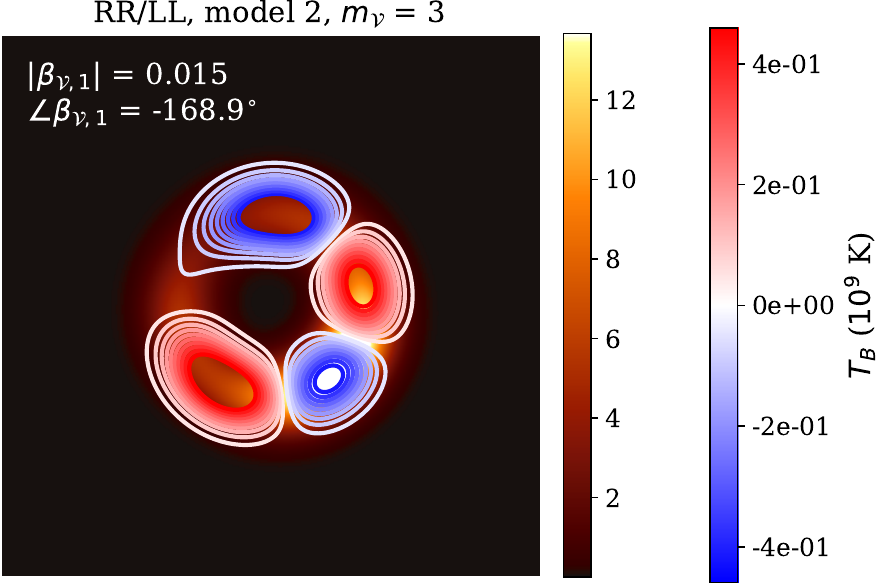} \\
    \vspace{6mm}
    \includegraphics[height=0.200\textwidth]{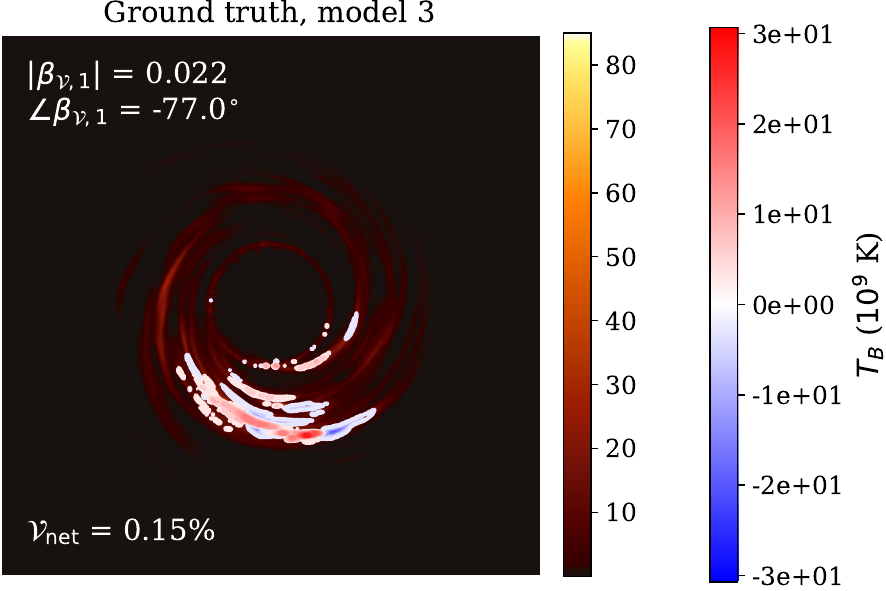} 
    \includegraphics[height=0.200\textwidth]{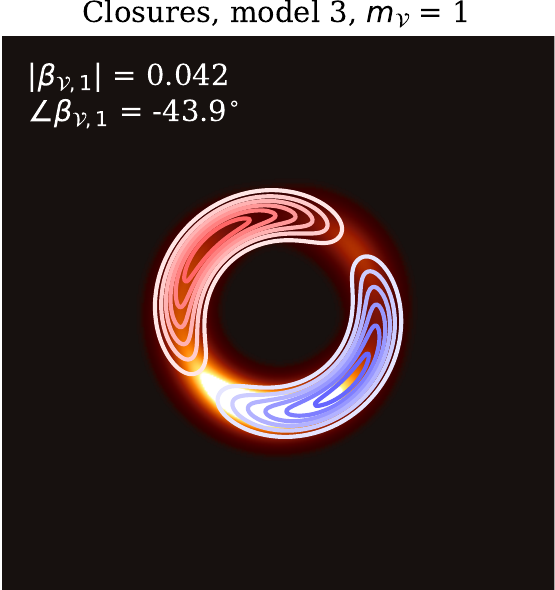}
    \includegraphics[height=0.200\textwidth]{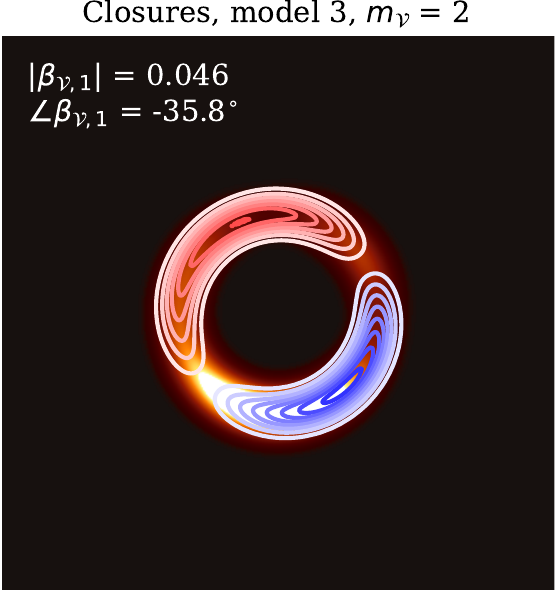}
    \includegraphics[height=0.200\textwidth]{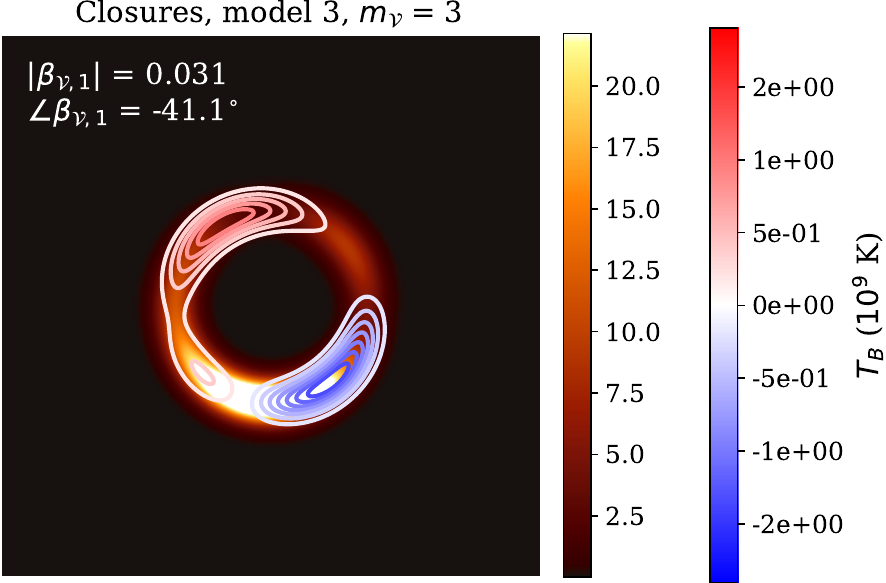} \\
    \includegraphics[height=0.200\textwidth]{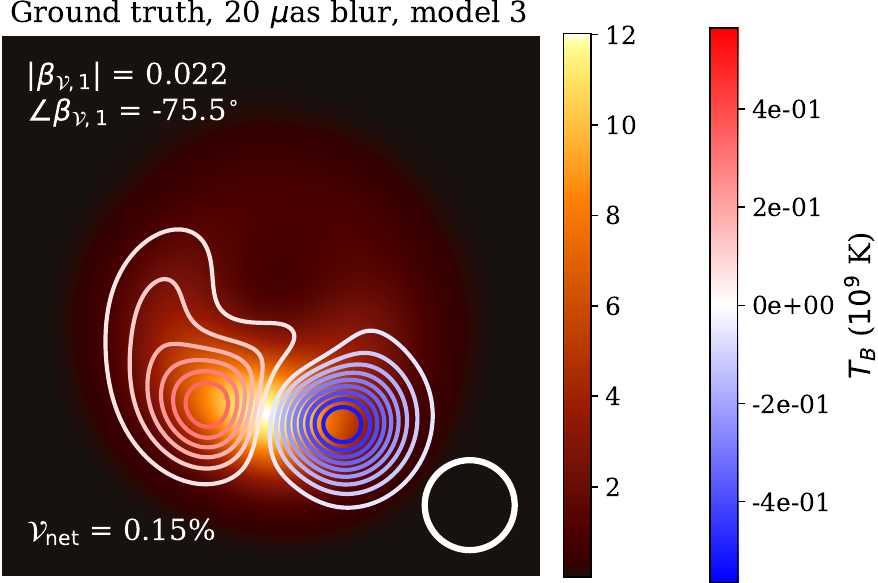} 
    \includegraphics[height=0.200\textwidth]{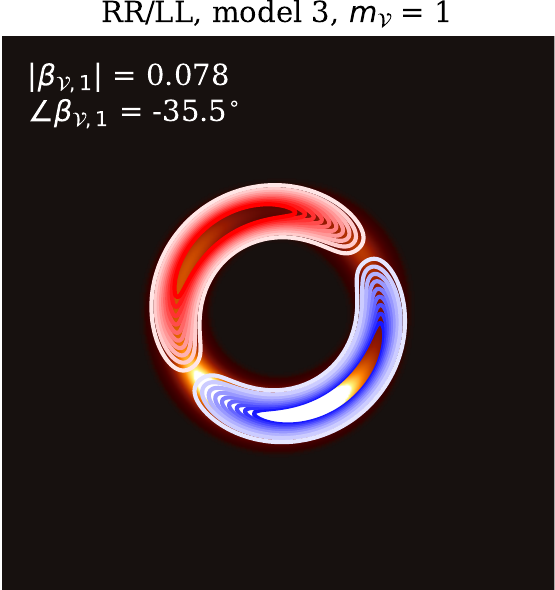}
    \includegraphics[height=0.200\textwidth]{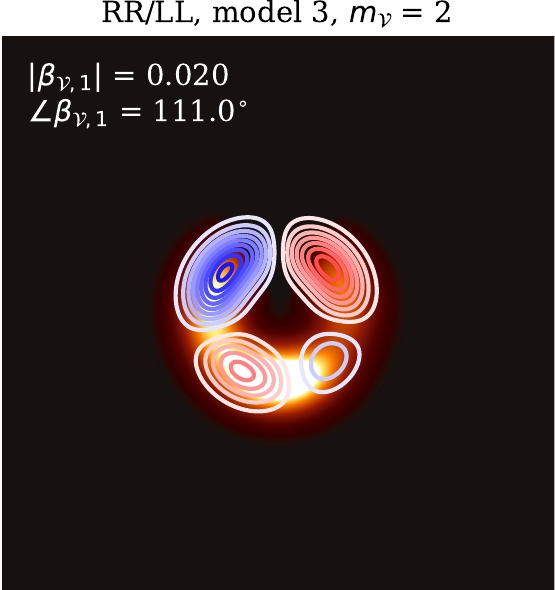}
    \includegraphics[height=0.200\textwidth]{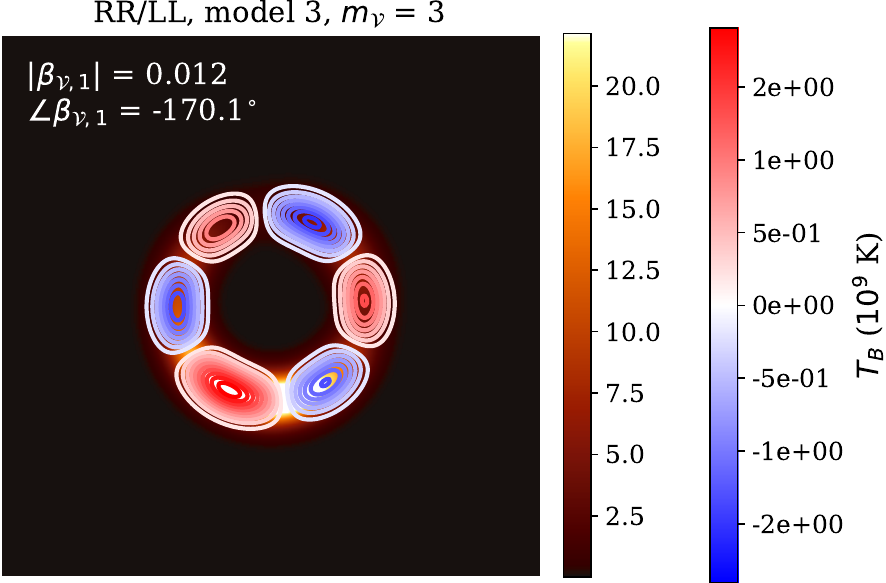} \\

    \caption{GRMHD ground truth Stokes $\mathcal{V}$ images and m-ring fits from EHT April 11 2017 coverage (low band). Each model covers a set of two rows. The left column shows the ground truth without (upper panels) and with (lower panels) Gaussian blurring (FWHM 20 $\mu$as). The right three columns show the Stokes $\mathcal{V}$ m-ring fits for $m_{\mathcal{V}}=1, 2$, and 3, fitting to closure quantities (upper panels for each model) and RR/LL visibility ratios (lower panels for each model).}
    \label{fig:grmhd_stokesv_maps}
\end{figure*}

\begin{figure*}
    \centering
    \includegraphics[height=0.24\textwidth]{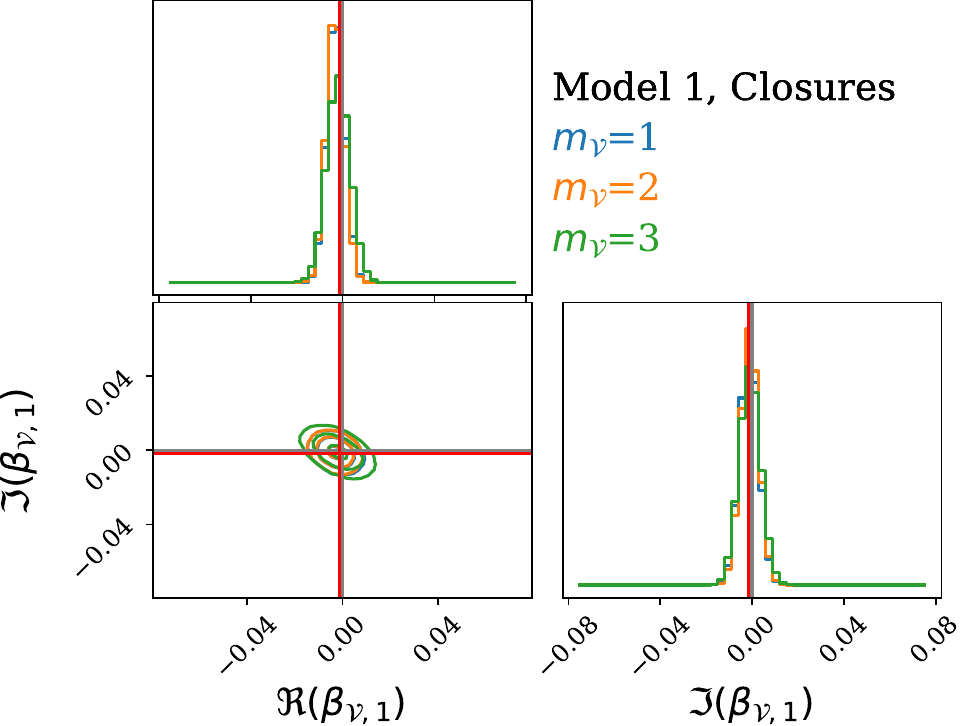}
    \includegraphics[height=0.24\textwidth]{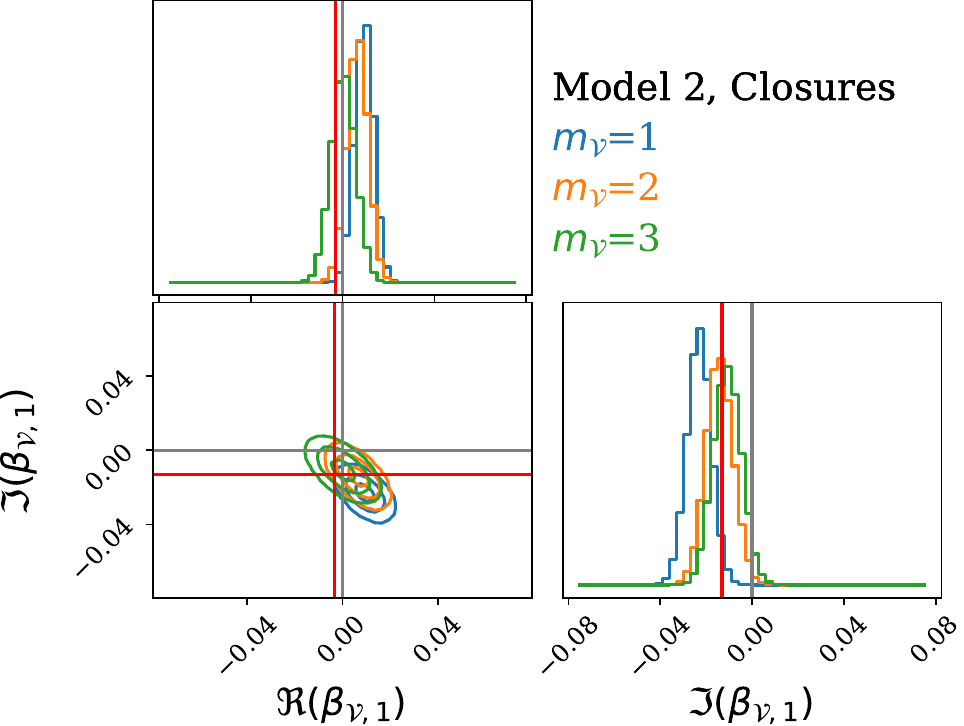}
    \includegraphics[height=0.24\textwidth]{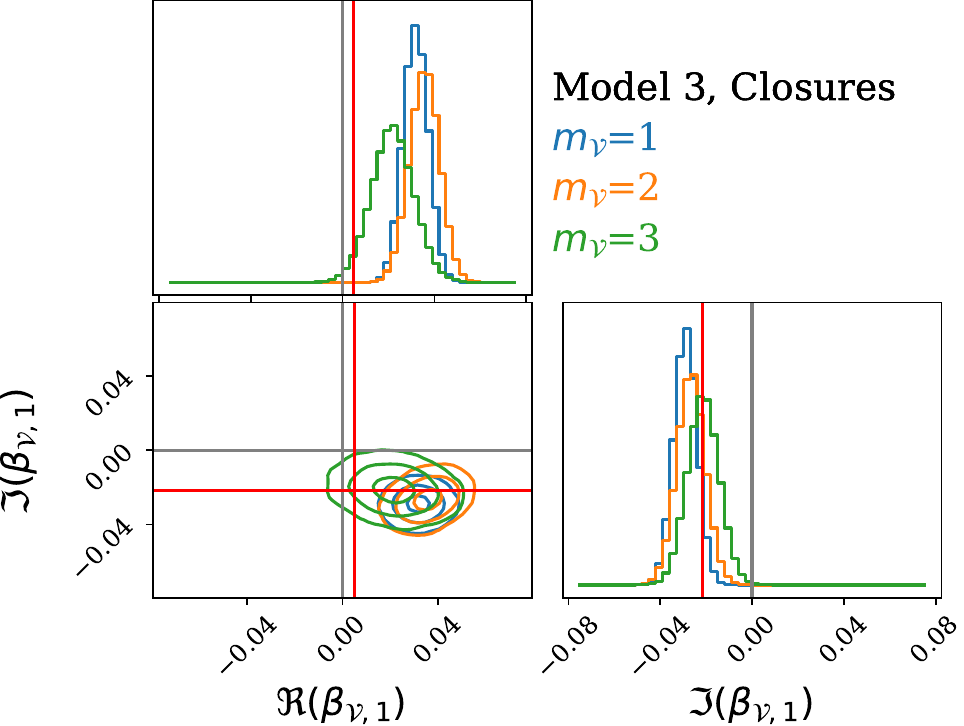}\\
    \includegraphics[height=0.24\textwidth]{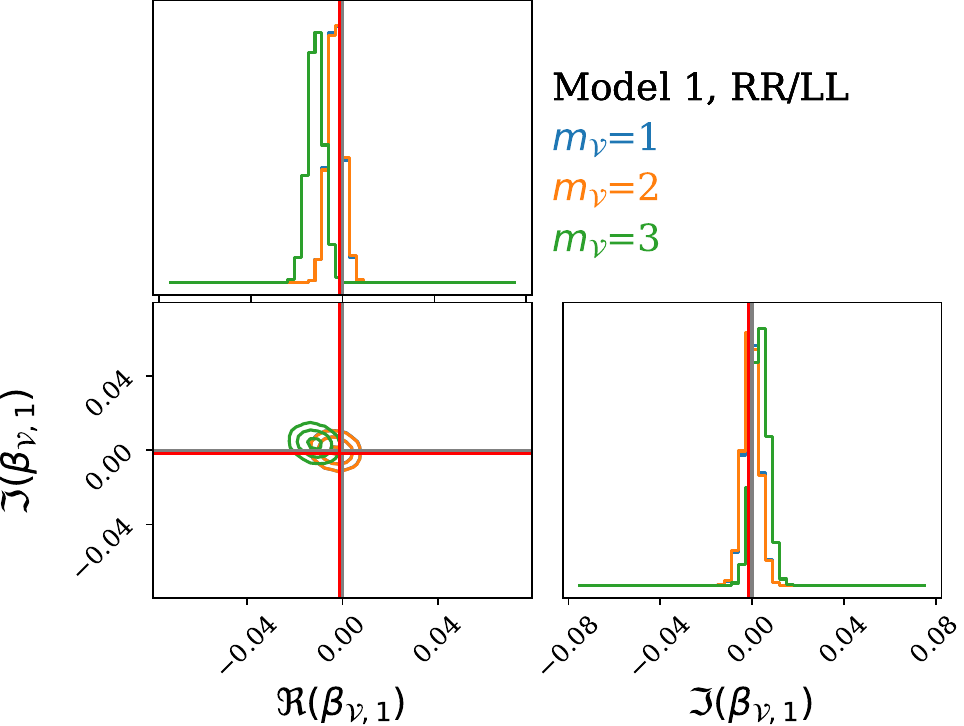}
    \includegraphics[height=0.24\textwidth]{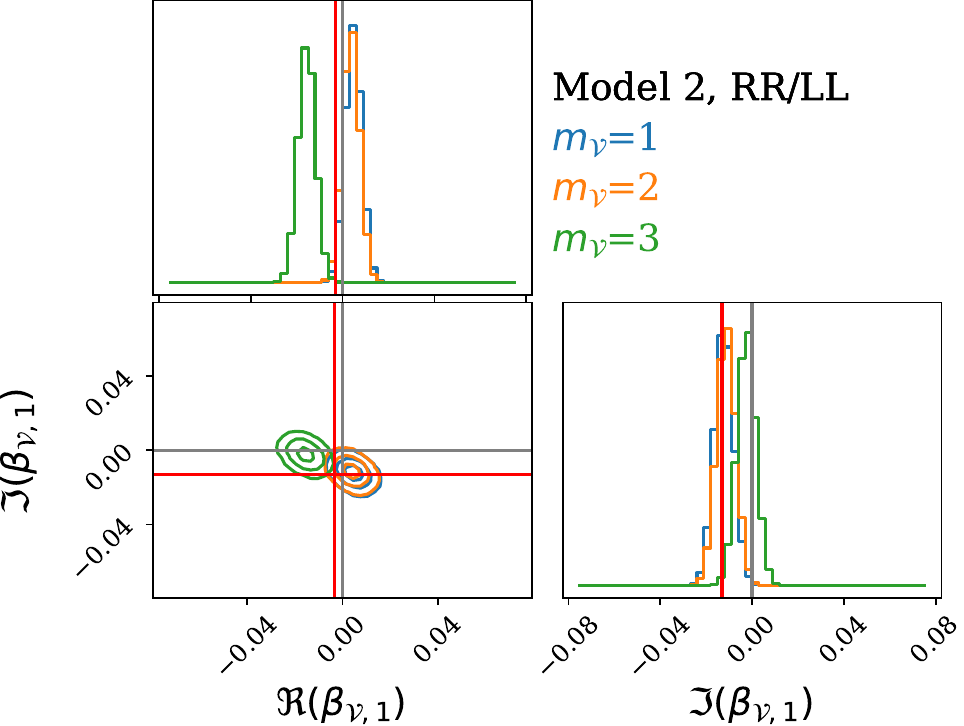}
    \includegraphics[height=0.24\textwidth]{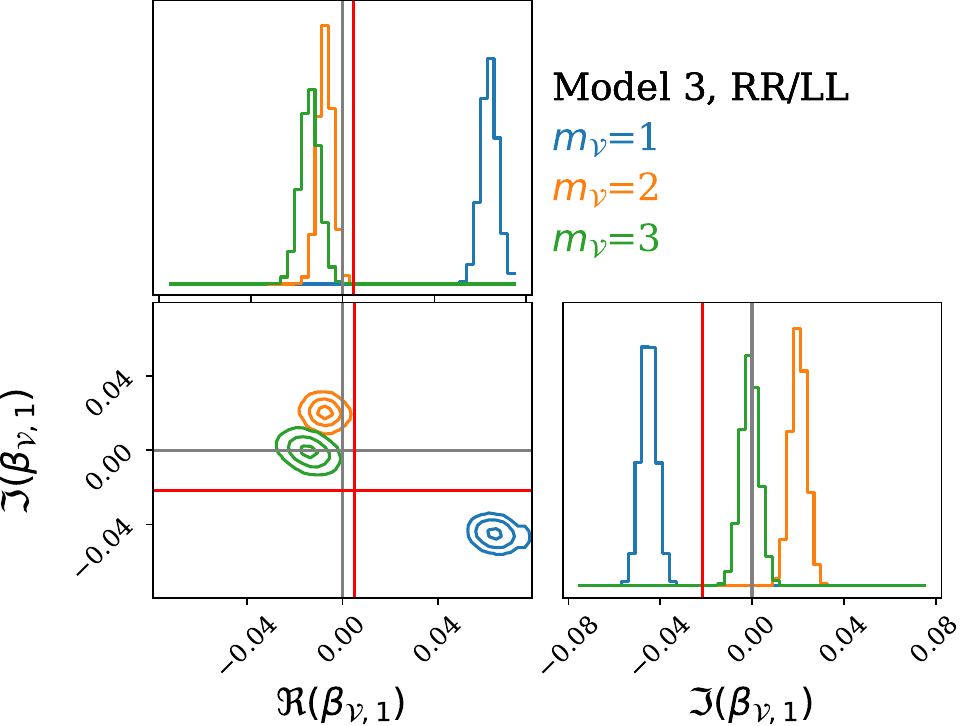}
    \caption{$\beta_{\mathcal{V},1}$ posteriors for the Stokes $\mathcal{V}$ m-ring fits to closure quantities (top row) and RR/LL visibility ratios (bottom row) from synthetic observations (EHT April 11 2017, low band) of our GRMHD models 1, 2, and 3 (left to right). The contours show 1$\sigma$, 2$\sigma$, and 3$\sigma$ levels. The red lines indicate the ground truth values.}
    \label{fig:grmhd_betav_posteriors}
\end{figure*}

\autoref{fig:grmhd_stokesv_maps} shows the posterior maxima of the Stokes $\mathcal{V}$ m-ring fits to synthetic data from the three GRMHD models. These were produced as outlined in \autoref{sec:results_geometric_models}, with the assumption of a perfect leakage calibration. The m-ring model is suitable for recovering the basic, low-order properties of the complex Stokes $\mathcal{V}$ structure in the GRMHD models, although like for linear polarization the performance varies depending on the ground-truth model. For model 1, the first-order orientation of the Stokes $\mathcal{V}$ structure, $\angle\beta_{\mathcal{V},1}$, is reproduced to within a few degrees for $m_{\mathcal{V}}=1$ (see also \autoref{fig:grmhd_betav_posteriors}, which shows the complex $\beta_{\mathcal{V},1}$ posteriors compared to the ground-truth values), and to within $30-50$ degrees for $m_{\mathcal{V}}=2$. The $\angle\beta_{\mathcal{V},1}$ of the $m_{\mathcal{V}}=3$ fit to the visibility ratios deviates most significantly, which is likely due to the multi-lobe structure of the $m_{\mathcal{V}}=3$ model in combination with a low total circular polarization fraction, so that a small deviation in one of the lobes can result in a large deviation of the first-order orientation $\angle\beta_{\mathcal{V},1}$. 

The Stokes $\mathcal{V}$ structure of model 2 is more complex, but $\angle\beta_{\mathcal{V},1}$ is nevertheless recovered to within less than 40 degrees. The higher m-order fits are more informative here, recovering the alternating regions of positive and negative Stokes $\mathcal{V}$ along the ring.

The Stokes $\mathcal{V}$ structure of model 3 is less ring-like than for the other models, with most of the Stokes $\mathcal{V}$ emission concentrated in the South. The $\beta_{\mathcal{V},1}$ posteriors (\autoref{fig:grmhd_betav_posteriors}) are further away from the ground truth than for the other models, despite the fact that the image-averaged circular polarization fraction $\langle|\mathcal{V/I}|\rangle$ is highest for this model. The low m-order fits nevertheless limit the first-order orientation offset to within $30-40$ degrees. Interestingly, the higher m-order fits to the visibility ratios favor a much smaller ring size in order to capture the compact Stokes~$\mathcal{V}$ emission in the South.

To summarize, the m-ring model is able to recover low-order Stokes $\mathcal{V}$ structures for these three quite distinct GRMHD models with varying accuracy. Since the Stokes $\mathcal{V}$ structure is often complex, there are systematics due to model misspecification (the low-order m-ring model does not fully describe the physical characteristics of the Stokes $\mathcal{V}$ emission), but the first-order asymmetry orientation is recovered to within a few tens of degrees for $m_{\mathcal{V}}=1,2$. Higher-order m-ring fits perform well if the ground-truth Stokes $\mathcal{V}$ structure is well described by alternating lobes of positive and negative Stokes $\mathcal{V}$ (model 2), but these fits are often less consistent in other cases and hence should be applied with caution in practice. We do not see a clear trend of increasing Stokes $\mathcal{V}$ reproducibility with increasing $\langle|\mathcal{V/I}|\rangle$ for these datasets. As discussed in \citetalias{M87PaperIX}, imaging methods showed similar and often greater difficulty in recovering the circular polarization structure from these datasets, and m-ring modeling often outperforms imaging in the recovery of quantities like the image-averaged circular polarization fraction or the first-order orientation, especially when the net circular polarization fraction is low. m-ring modeling is therefore a useful tool for studying low-order circular polarization structures in mm-VLBI observations of black holes.

\section{Application to EHT data of \m87}
\label{sec:m87}
In this Section, we apply our polarimetric m-ring modeling framework to EHT 2017 observations of \m87. We use data calibrated through the \texttt{EHT-HOPS} pipeline \citep{Blackburn2019, PaperIII, M87PaperIX}, unless otherwise specified (\autoref{sec:m87cpol}). We use both low-band (LO, with a central frequency of 227.1 GHz) and high-band (HI, with a central frequency of 229.1 GHz) data for our analyses. Both bands have a bandwidth of 2 GHz. All datasets were leakage-calibrated using the estimated $D$-terms from \citet{PaperVII, Issaoun2022}, which have estimated residual leakages of only $\sim$1\%. We show our recovery of the previously imaged linear polarization structure \citep{PaperVII} in \autoref{sec:m87linpol}. In \autoref{sec:m87cpol}, we then show our circular polarization fits to \m87 data, providing an extension and more detailed exploration of the fits presented in \citetalias{M87PaperIX}.

\subsection{Linear polarization}
\label{sec:m87linpol}
\begin{figure*}
    \centering
    \includegraphics[height=0.22\textwidth]{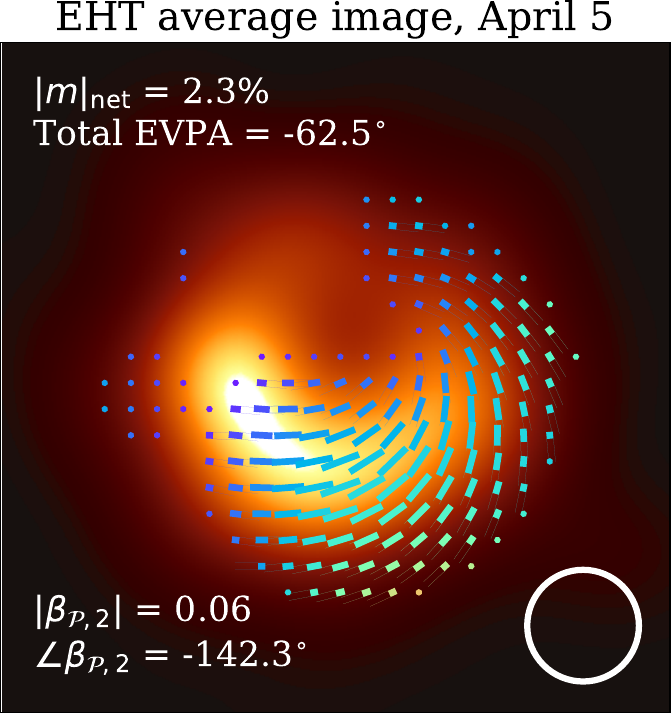}
    \includegraphics[height=0.22\textwidth]{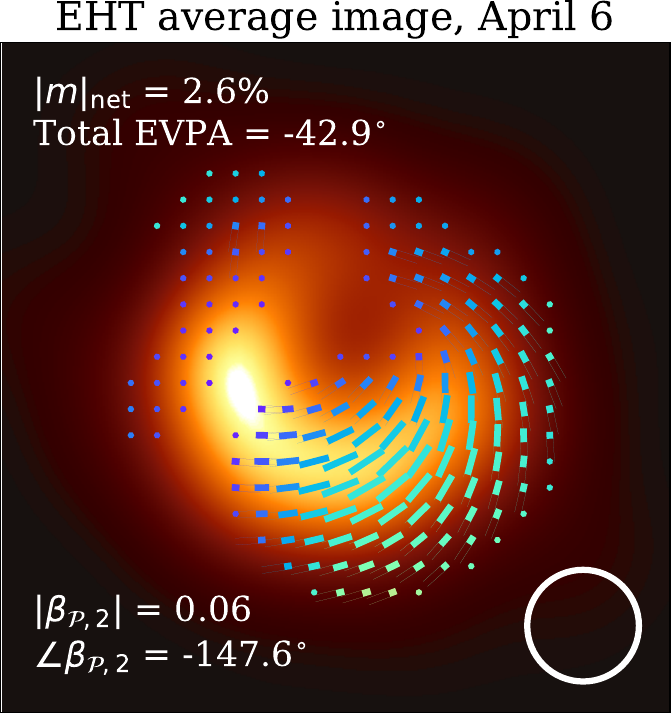}
    \includegraphics[height=0.22\textwidth]{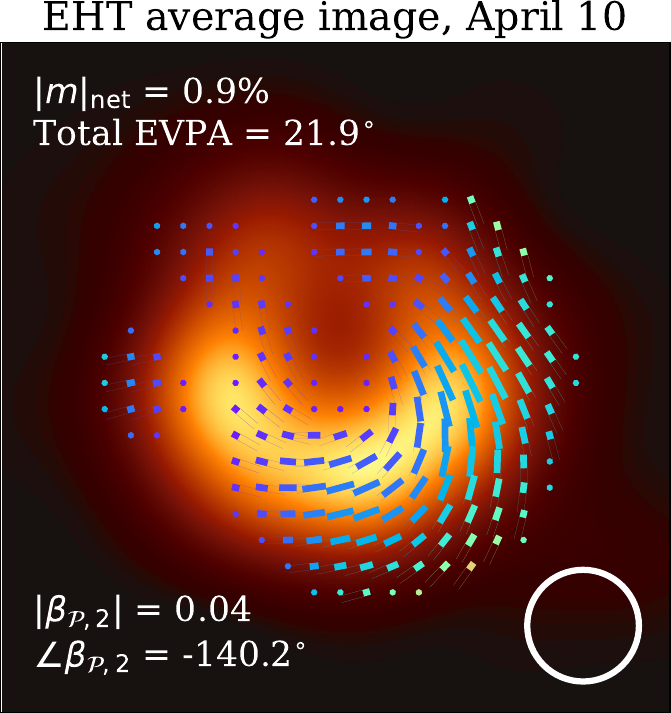}
    \includegraphics[height=0.22\textwidth]{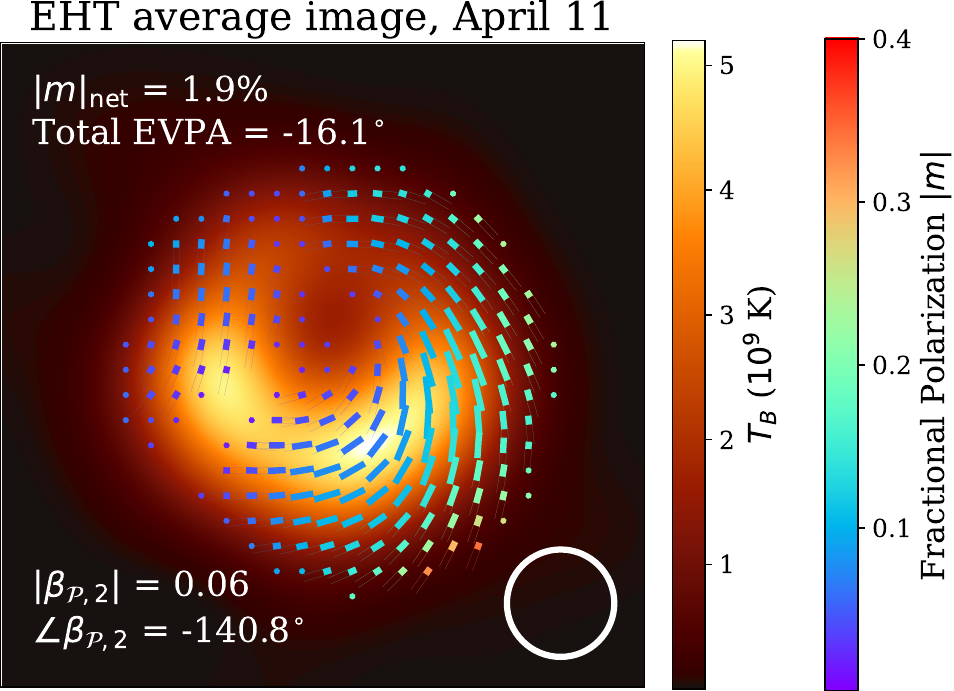} \\
    \includegraphics[height=0.22\textwidth]{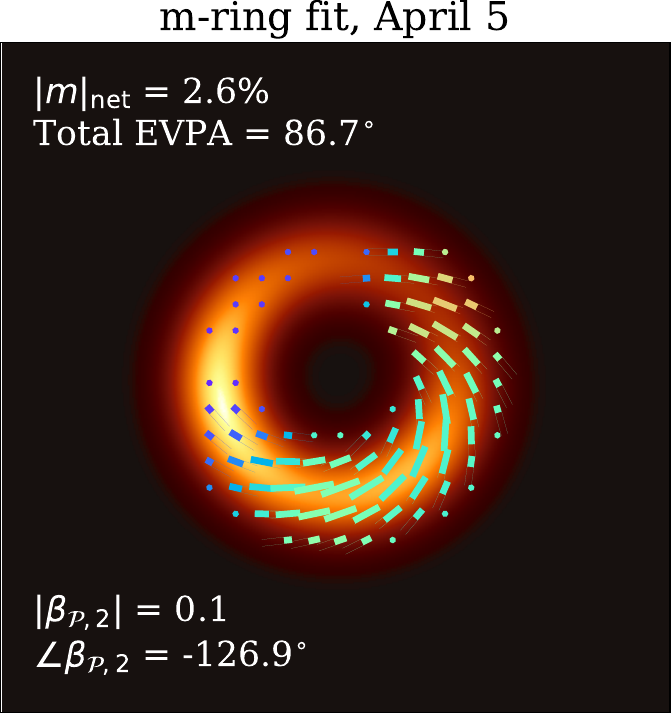}
    \includegraphics[height=0.22\textwidth]{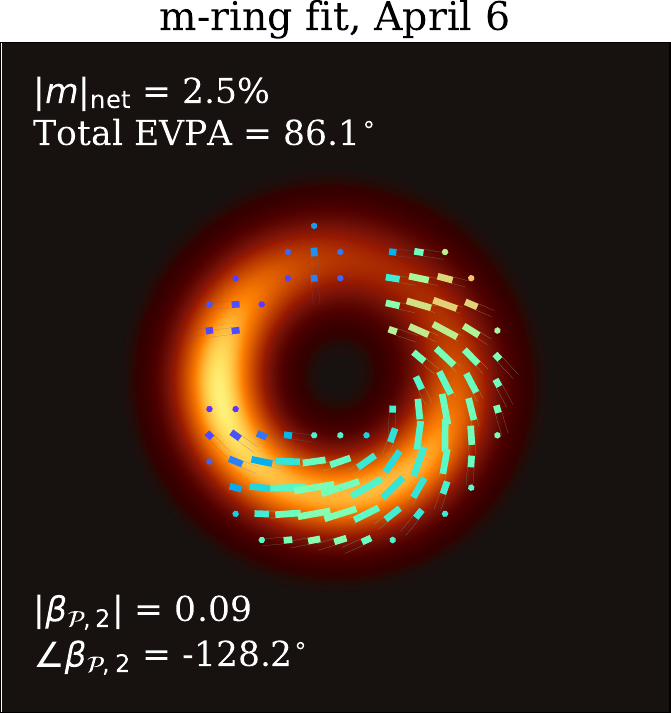}
    \includegraphics[height=0.22\textwidth]{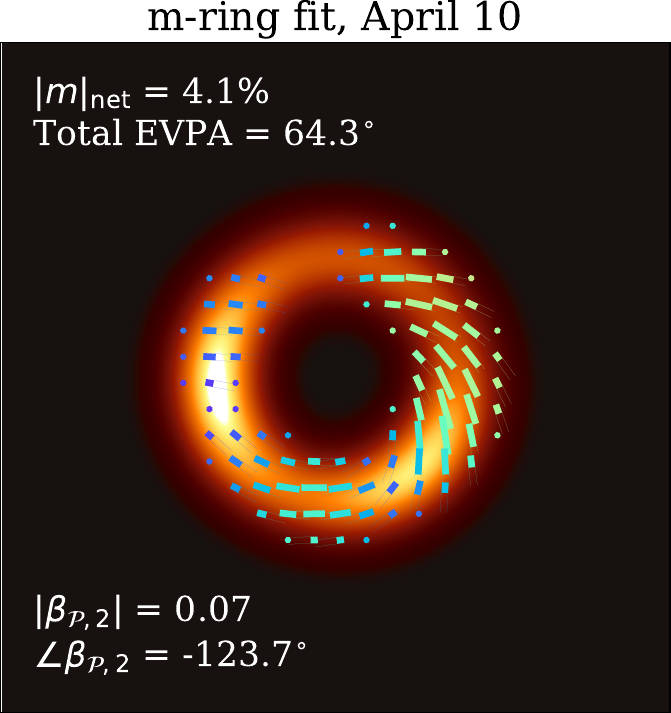} 
    \includegraphics[height=0.22\textwidth]{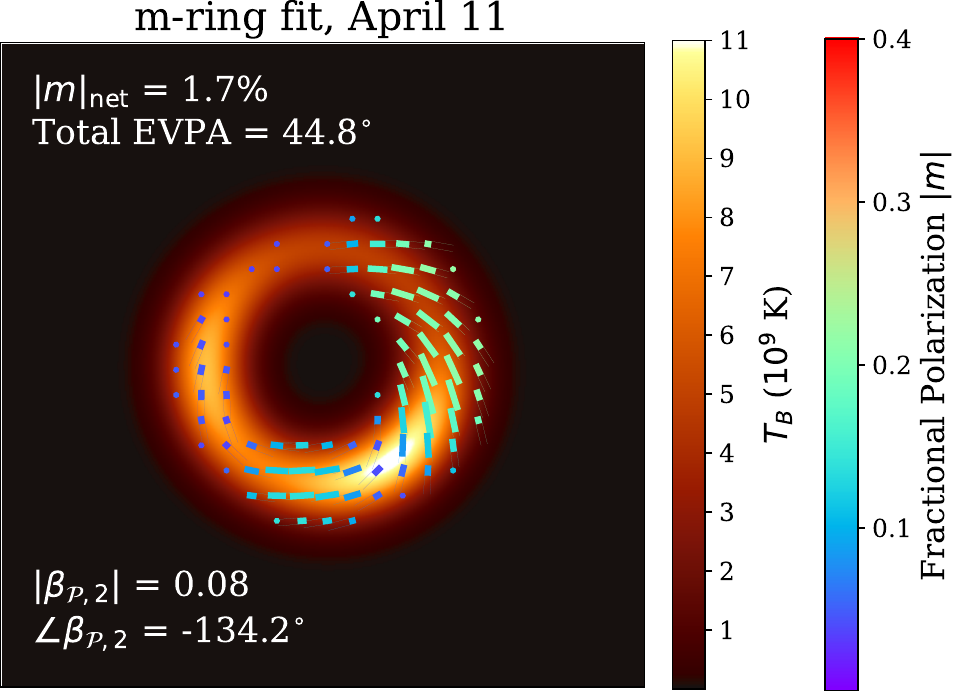} \\
    \includegraphics[height=0.22\textwidth]{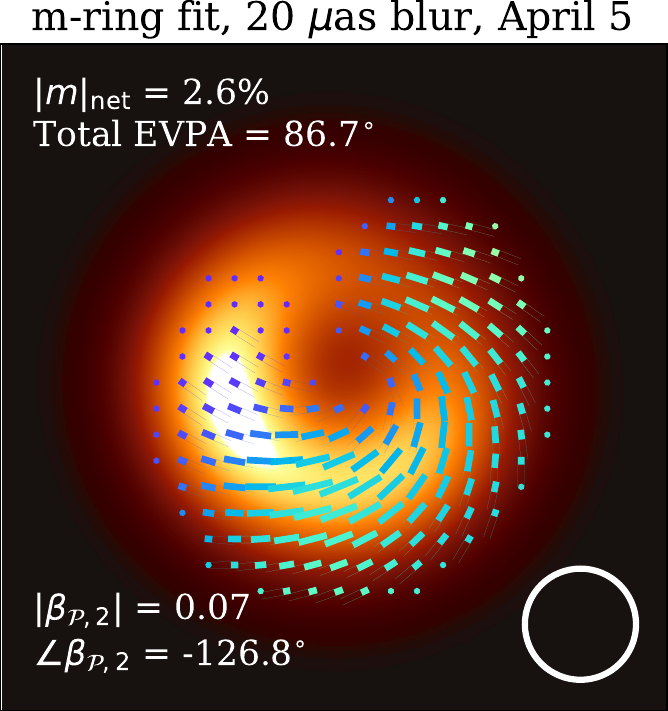}
    \includegraphics[height=0.22\textwidth]{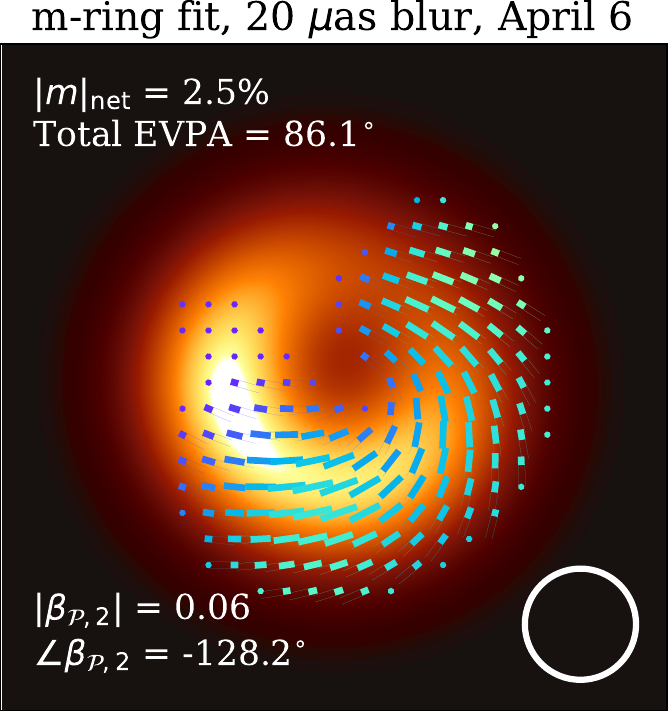}
    \includegraphics[height=0.22\textwidth]{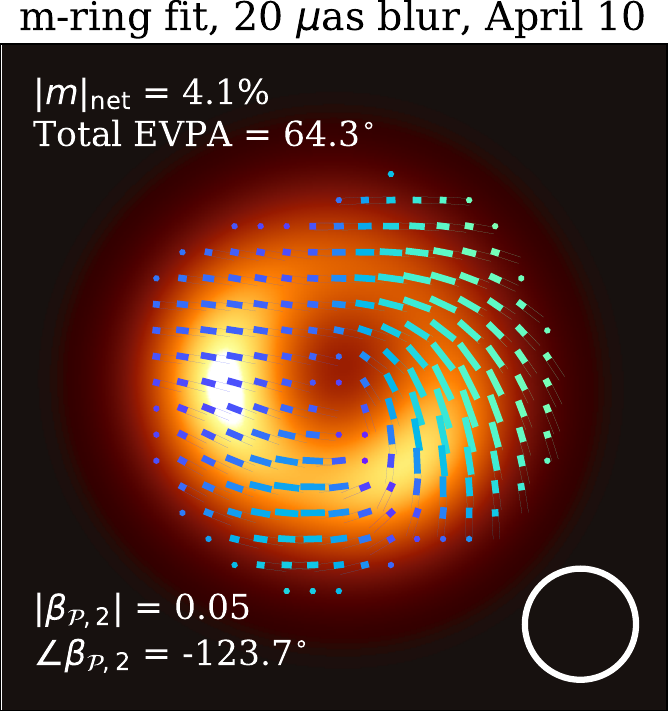} 
    \includegraphics[height=0.22\textwidth]{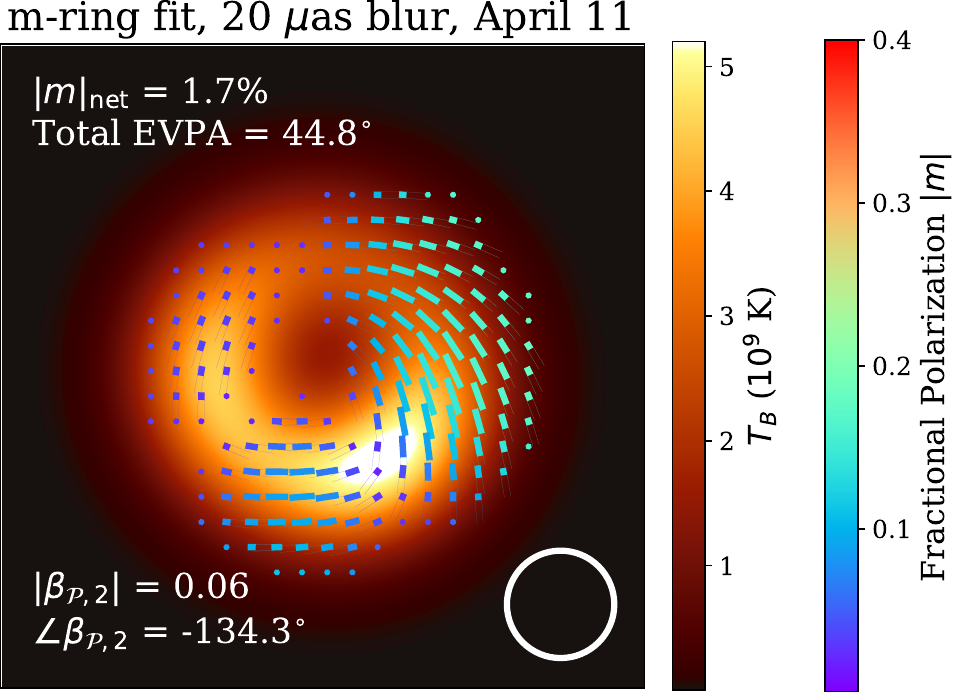}
    \caption{EHT 2017 method-averaged total intensity and linear polarization images of \m87 \citep[top row;][]{PaperVII} on April 5, 6, 10, and 11 (left to right), compared to our total intensity and linear polarization m-ring fits (posterior maxima; $m_{\mathcal{I}}=3$, $m_{\mathcal{P}}=3$) to the same data, without (middle row) and with (bottom row) blurring with a Gaussian kernel with a FWHM of 20 $\mu$as.}
    \label{fig:m87_linpol}
\end{figure*}

\autoref{fig:m87_linpol} shows our total intensity and linear polarization fits ($m_{\mathcal{I}}=3$, $m_{\mathcal{P}}=3$) to EHT 2017 \m87 data on four days (April 5, 6, 10, and 11), with and without blurring the images with a Gaussian kernel with a FWHM of 20 $\mu$as. This blurring kernel is representative for the blurring kernels used for the images in \citet{PaperIV, PaperVII}. \autoref{tab:fitcomp} compares our fitted parameters to previous EHT results \citep{PaperI, PaperIV, PaperVI, PaperVII}. The structure recovered with our m-ring fits is remarkably consistent across days and in excellent agreement with previous EHT imaging and modeling results, especially when our fits are blurred. 

For consistency with \citet{PaperI}, we report the shifted diameter
\begin{equation}
    d'=d-\frac{1}{4\ln{2}}\frac{\alpha^2}{d}
\end{equation}
in \autoref{tab:fitcomp}. This shifted diameter accounts for the change in peak brightness radius for a thick ring with FWHM $\alpha$ as compared to an infinitesimally thin ring \citep{PaperIV}. The diameter change is of order 2 $\mu$as for the values reported in \autoref{tab:fitcomp}. 

For a blurring kernel with FWHM $W$, the change in ring thickness can be approximated as \citep{PaperIV}
\begin{equation}
\label{eq:moranbias}
    \alpha_{\mathrm{blur}} \approx \sqrt{\alpha^2+W^2}. 
\end{equation}
For the blurred m-ring posterior ranges reported in \autoref{tab:fitcomp} (5th column), our $\alpha$-posteriors were transformed following this approximation, with $W=20$ $\mu$as. These values were then used to compute the blurred $d'$ posterior ranges following \autoref{eq:moranbias}. The $\beta$-values for the blurred fits were computed by generating 1000 image samples from the posteriors, blurring them with a 20 $\mu$as beam, and computing the values from the blurred image samples. Most fitted quantities are insensitive or only weakly sensitive to image blurring. An exception is $|\beta_{\mathcal{P},2}|$, which reduces significantly after blurring. 

As found in the previous EHT analyses, the peak brightness in total intensity moves from the southeast towards the southwest for the later days. The total linear polarization between about 1.5 and 4.3\% is consistent, within error bars, with that recovered by the polarimetric imaging methods in \citet{PaperVII}, which reported values between 1.0 and 3.7\%. 

Our fitted $\angle\beta_{\mathcal{P},2}$ between about -136 and -121$^{\circ}$ is also in agreement with the EHT imaging results, which reported values between -163$^{\circ}$ and -127$^{\circ}$ (except for a slight offset for the fit to high-band data on April 6). There is a significant discrepancy for $|\beta_{\mathcal{P},2}|$, but as shown in \autoref{fig:m87_linpol} and \autoref{tab:fitcomp} this quantity is sensitive to the applied blurring kernel, and the values are in full agreement when blurring our models with a 20 $\mu$as Gaussian kernel. Our posterior widths (\autoref{tab:fitcomp}) are generally much smaller than the ranges spanned by the EHT imaging methods, indicating that systematic offsets (model misspecifications) are likely dominant over the statistical uncertainties from fitting a specific model to the data. For all quantities, the statistical uncertainty is largest for April 10 data, which indeed has the smallest number of \m87 scans (amounting to less than 30\,minutes on-source).

\begin{table*}[]
\caption{Parameters describing the total intensity and polarization structure of \m87 as observed by the EHT in 2017 extracted through imaging and crescent fitting \citep{PaperI, PaperIV, PaperVII} and m-ring fitting (this work). The fitted model has $m_{\mathcal{I}}=3$, $m_{\mathcal{P}}=3$, and $m_{\mathcal{V}}=1$, with circular polarization fits to closure quantities. The m-ring fit values indicate the posterior means with 2$\sigma$ ranges. The $\langle|\mathcal{V/I}|\rangle$ values were computed from posterior image samples. The blurred m-ring parameters were obtained by propagation of the fit posteriors for $d'$, $\alpha$, and $|m|_{\mathrm{net}}$, and from blurred image samples for the $\beta$ parameters and $\langle|\mathcal{V/I}|\rangle$ (see text for further details).}
\begin{tabular}{lllllll}
\multirow{2}{*}{Parameter}           & \multirow{2}{*}{Day}    & EHTC  &  m-ring Fitting  &  m-ring Fitting &  m-ring Fitting  &  m-ring Fitting\\ 
                    &  & Imaging + Modeling & LO  &  HI  &   LO, 20 $\mu$as blur &   HI, 20 $\mu$as blur\\ \hline
\multirow{4}{*}{$d'$ ($\mu$as)} & Apr 5  & \multirow{4}{*}{$42\pm3$ \tablenotemark{a}}                  & $42.07\pm0.56$ &  $41.261\pm0.68$                        & $38.74\pm0.60$  & $38.27\pm0.73$\\
                   & Apr 6  &                   &   $42.25\pm0.43$   & $42.52\pm0.58$                      &  $38.93\pm0.54$ & $39.25\pm0.62$\\
                   & Apr 10 &                   &   $43.24\pm0.91$  &  $43.53\pm0.94$                      & $39.98\pm0.97$ & $40.32\pm0.94$\\
                   & Apr 11 &                   &   $43.53\pm0.47$  &  $43.01\pm0.33$                      &  $40.30\pm0.50$ & $39.72\pm0.39$\\ \hline
\multirow{4}{*}{$\alpha$ ($\mu$as)} & Apr 5 &  \multirow{4}{*}{$<20$ \tablenotemark{a}}                 &  $12.31\pm0.78$  &  $13.71\pm0.79$                       &  $23.49\pm0.41$ & $24.25\pm0.45$\\
                   & Apr 6 &                   &   $12.16\pm0.69$  &  $13.68\pm0.61$                      &  $23.41\pm0.36$ & $24.23\pm0.61$\\
                   & Apr 10 &                  &  $11.04\pm1.26$  &  $13.28\pm0.98$                        & $22.85\pm0.60$ & $24.01\pm0.54$\\
                   & Apr 11 &                   &  $11.97\pm0.67$  &  $12.48\pm0.60$                       & $23.22\pm0.34$ & $23.57\pm0.32$\\ \hline
\multirow{4}{*}{$|\beta_{\mathcal{I},1}|$} & Apr 5 &    \multirow{4}{*}{0.16 -- 0.32 \tablenotemark{b}}                 & $0.197\pm0.010$  &   $0.208\pm0.013$                        &  $0.171\pm0.008$ & $0.176\pm0.011$\\
                   & Apr 6 &                   &   $0.203\pm0.009$ &  $0.215\pm0.011$                       &  $0.178\pm0.007$ & $0.184\pm0.008$\\
                   & Apr 10 &                   &  $0.138\pm0.015$ &  $0.155\pm0.018$                        & $0.121\pm0.013$ & $0.134\pm0.014$\\
                   & Apr 11 &                   &  $0.166\pm0.012$ &  $0.172\pm0.012$                         & $0.138\pm0.010$ & $0.141\pm0.009$\\ \hline
\multirow{4}{*}{$\angle\beta_{\mathcal{I},1}$ (deg)} & Apr 5 &  \multirow{4}{*}{-200 -- -150 \tablenotemark{a}}              &  $-139.2\pm4.3$ & $-145.1\pm4.9$                         &  $-134.3\pm4.4$ & $-140.4\pm5.2$\\
                   & Apr 6 &                   & $-141.8\pm3.95$ &  $-144.2\pm4.1$                         &  $-137.0\pm4.0$ & $-139.8\pm4.3$\\
                   & Apr 10 &                   & $-137.4\pm11.9$ & $-139.3\pm10.6$                          & $-130.6\pm12.4$ & $-133.6\pm11.1$\\
                   & Apr 11 &                   & $-172.3\pm3.7$ &  $-174.6\pm3.4$                         & $-167.8\pm4.2$ & $-170.2\pm3.7$\\ \hline        
\multirow{4}{*}{$|m|_{\mathrm{net}}$ (\%)} & Apr 5 & \multirow{4}{*}{1 -- 3.7 \tablenotemark{c}}     & $2.51\pm0.61$ & $2.40\pm0.64$                           &  $2.51\pm0.61$ & $2.40\pm0.64$\\
                   & Apr 6 &                   &   $2.53\pm0.50$  & $2.51\pm0.50$                       &  $2.53\pm0.50$ & $2.51\pm0.50$\\
                   & Apr 10 &                   &  $4.25\pm0.88$  &  $3.38\pm0.87$                       & $4.25\pm0.88$ & $3.38\pm0.87$\\
                   & Apr 11 &                   &  $1.68\pm0.51$  &  $1.46\pm0.49$                       & $1.68\pm0.51$ & $1.46\pm0.49$\\ \hline
 \multirow{4}{*}{$|\beta_{\mathcal{P},2}|$} & Apr 5 & \multirow{4}{*}{0.04 -- 0.07 \tablenotemark{c}}  &   $0.110\pm0.006$ &  $0.110\pm0.007$                       &  $0.068\pm0.004$ & $0.065\pm0.004$\\
                   & Apr 6 &                   &  $0.103\pm0.006$  &   $0.106\pm0.006$                       &  $0.063\pm0.004$ & $0.064\pm0.004$\\
                   & Apr 10 &                   &  $0.084\pm0.011$ &   $0.080\pm0.012$                        &  $0.053\pm0.007$ & $0.050\pm0.008$\\
                   & Apr 11 &                   &  $0.090\pm0.007$  &  $0.090\pm0.007$                        & $0.058\pm0.005$ & $0.057\pm0.004$\\ \hline
\multirow{4}{*}{$\angle\beta_{\mathcal{P},2}$ (deg)} & Apr 5 & \multirow{4}{*}{-163 -- -127 \tablenotemark{c}} & $-126.2\pm3.3$ & $-128.1\pm3.7$                          &  $-125.8\pm3.3$ & $-128.0\pm3.7$\\
                   & Apr 6 &                   & $-129.4\pm3.3$ &   $-123.0\pm3.4$                        &  $-129.0\pm3.2$ & $-122.8\pm3.3$\\
                   & Apr 10 &                   &  $-125.5\pm7.7$ &  $-121.4\pm8.9$                        & $-126.0\pm7.9$ & $-122.2\pm9.0$\\
                   & Apr 11 &                   &  $-135.7\pm4.5$ &   $-133.3\pm4.6$                       & $-136.3\pm4.4$ & $-134.0\pm4.5$\\ \hline 
\multirow{4}{*}{$|\beta_{\mathcal{V},1}|$} & Apr 5 & \multirow{4}{*}{$\lesssim0.02$ \tablenotemark{d}} & $0.013\pm0.008$ & $0.010\pm0.009$                           &  $0.011\pm0.007$ &  $0.009\pm0.008$\\
                   & Apr 6 &                   & $0.013\pm0.010$ &  $0.012\pm0.007$                          &  $0.012\pm0.008$ & $0.010\pm0.006$\\
                   & Apr 10 &                   &  $0.024\pm0.018$ &  $0.007\pm0.014$                        & $0.021\pm0.016$ & $0.005\pm0.012$\\
                   & Apr 11 &                   &  $0.022\pm0.010$ &  $0.016\pm0.009$                        & $0.019\pm0.009$ & $0.014\pm0.008$\\ \hline     
\multirow{4}{*}{$\angle\beta_{\mathcal{V},1}$ (deg)} & Apr 5 & \multirow{4}{*}{$\sim$ -90 -- 180 \tablenotemark{d}} & $12.01\pm45.4$ &  $6.7\pm66.7$                         &  $13.1\pm45.5$ & $2.2\pm43.9$\\
                   & Apr 6 &                   & $-28.0\pm36.6$   &  $13.7\pm50.9$                       &  $-27.7\pm36.6$ & $14.9\pm50.5$\\
                   & Apr 10 &                   &  $-30.0\pm37.0$ &  $-33.2\pm138.2$                        & $-29.7\pm37.4$ & $-30.8\pm149.4$\\
                   & Apr 11 &                   &  $-27.8\pm17.4$ &  $-25.9\pm25.3$                        & $-27.7\pm18.0$ & $-26.2\pm24.4$\\ \hline  
\multirow{4}{*}{$\langle|\mathcal{V/I}|\rangle$ (\%)} & Apr 5 & \multirow{4}{*}{$<3.7$ \tablenotemark{d}} & $1.7\pm1.0$ & $1.5\pm1.1$                           &  $1.5\pm0.9$ & $1.3\pm1.0$\\
                   & Apr 6 &                   & $1.7\pm1.1$  &  $1.6\pm0.9$                        &  $1.5\pm1.0$ & $1.4\pm0.8$\\
                   & Apr 10 &                   &  $3.0\pm2.2$  &  $1.4\pm1.6$                       & $2.8\pm2.0$ & $1.3\pm1.5$\\
                   & Apr 11 &                   &  $2.8\pm1.2$  &  $2.1\pm1.1$                       & $2.5\pm1.1$ & $1.9\pm1.0$\\ \hline                 
\end{tabular}
\label{tab:fitcomp}
\tablenotetext{a}{\citet{PaperI}}
\tablenotetext{b}{\citet{PaperIV}}
\tablenotetext{c}{\citet{PaperVII}}
\tablenotetext{d}{\citetalias{M87PaperIX}}
\end{table*}

\subsection{Circular polarization}
\label{sec:m87cpol}
\subsubsection{Fit results}
\label{sec:m87cpolhops}
\begin{figure*}
    \centering
    \includegraphics[height=0.23\textwidth]{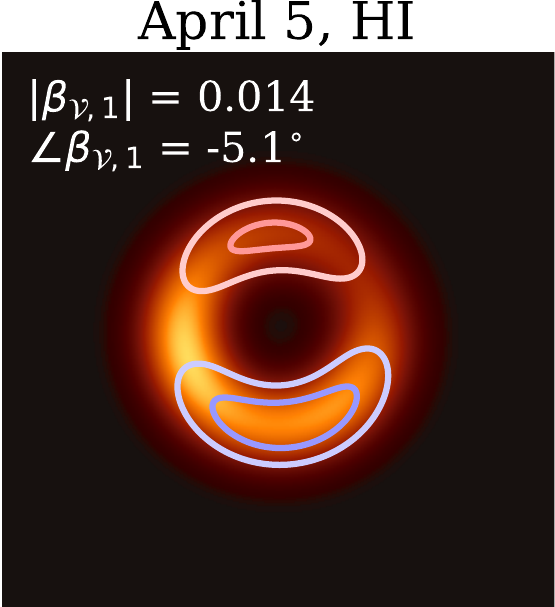}
    \includegraphics[height=0.23\textwidth]{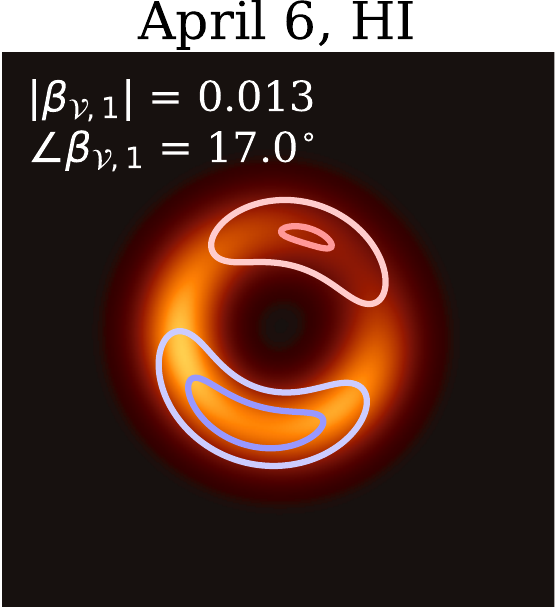}
    \includegraphics[height=0.23\textwidth]{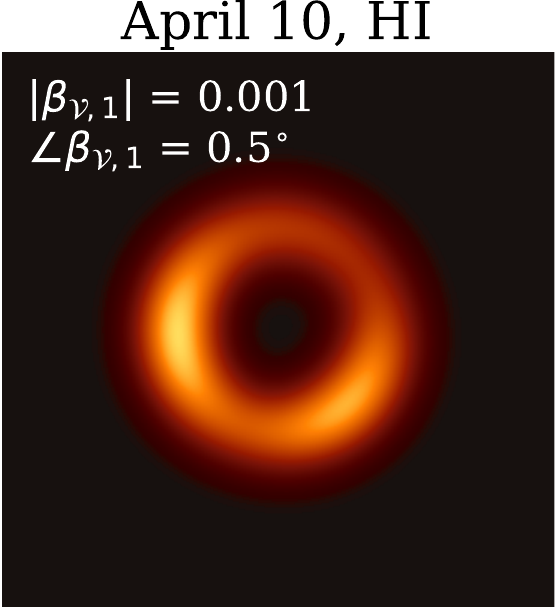}
    \includegraphics[height=0.23\textwidth]{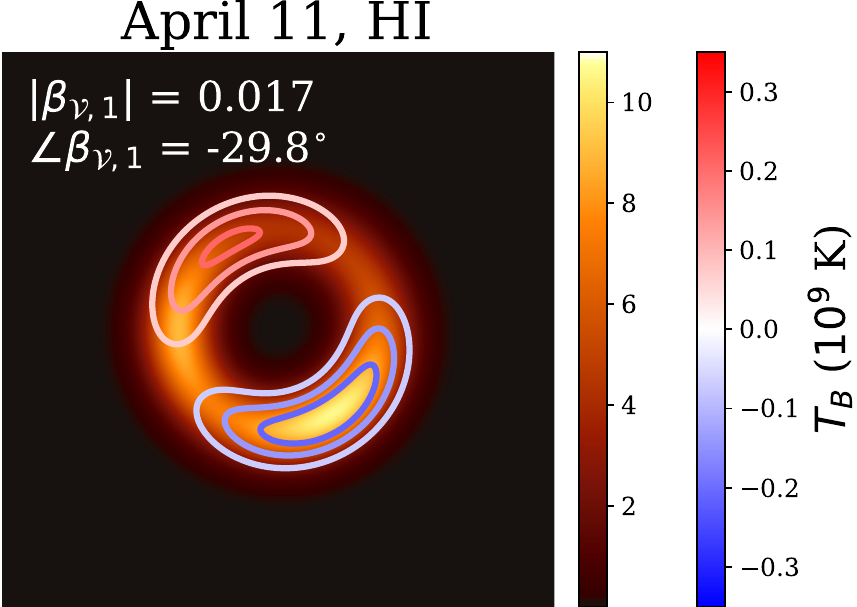}\\
    \includegraphics[height=0.23\textwidth]{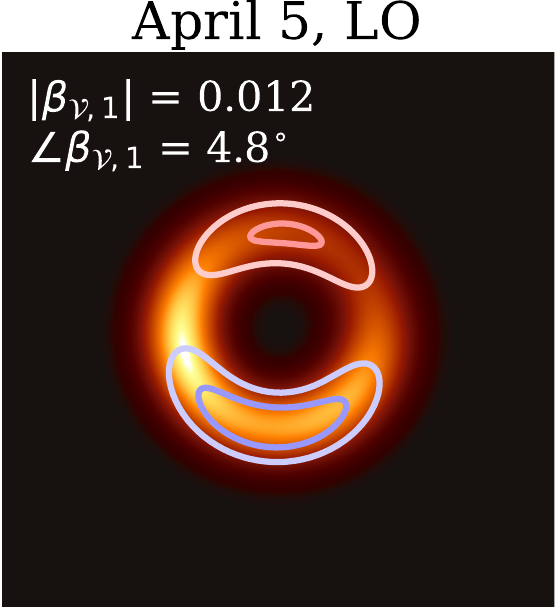}
    \includegraphics[height=0.23\textwidth]{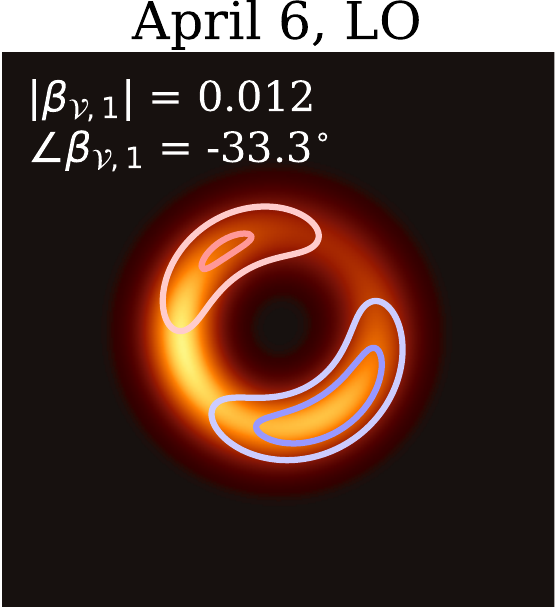}
    \includegraphics[height=0.23\textwidth]{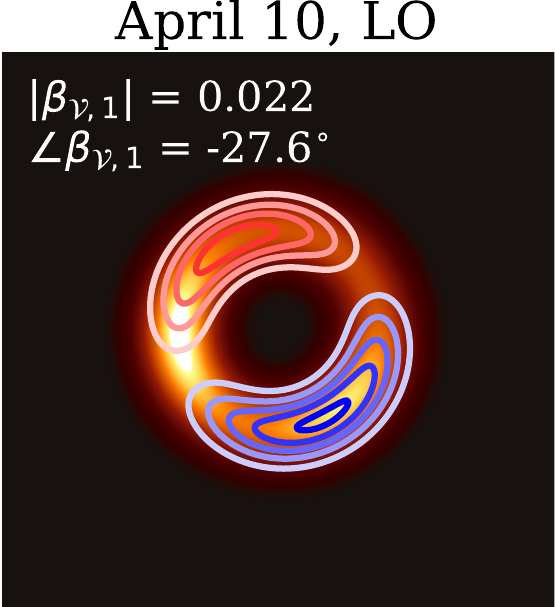}
    \includegraphics[height=0.23\textwidth]{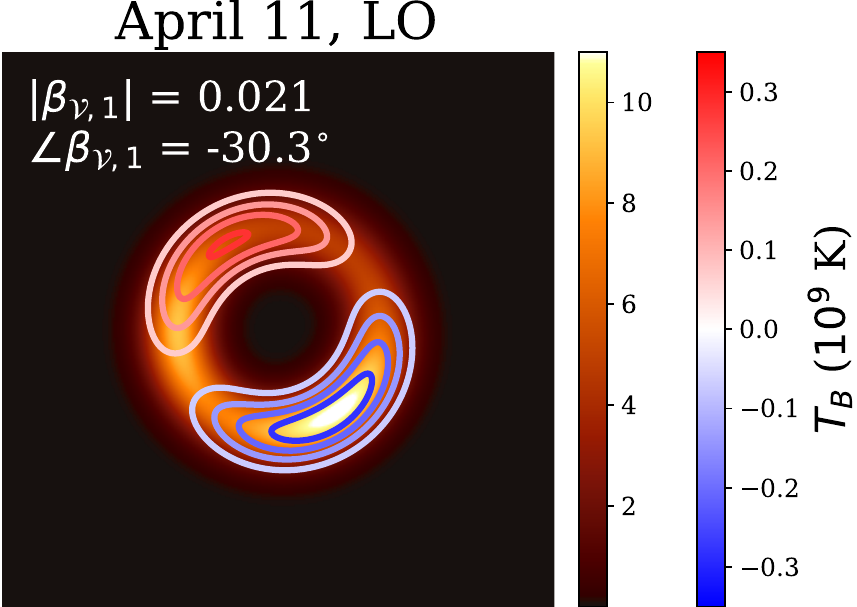} 
    \caption{Circular polarization m-ring fits ($m_{\mathcal{V}}=1$, $m_{\mathcal{I}}=3$; posterior maxima) to closure products of  low-band EHT 2017 \m87 data on April 5, 6, 10, and 11 (left to right). The heat map indicates the Stokes $\mathcal{I}$ structure, and the contours indicate the Stokes $\mathcal{V}$ structure. These fits are also presented in \citetalias{M87PaperIX}.}
    \label{fig:m87_stokesv}
\end{figure*}

\begin{figure*}
    \centering
    \includegraphics[width=\textwidth]{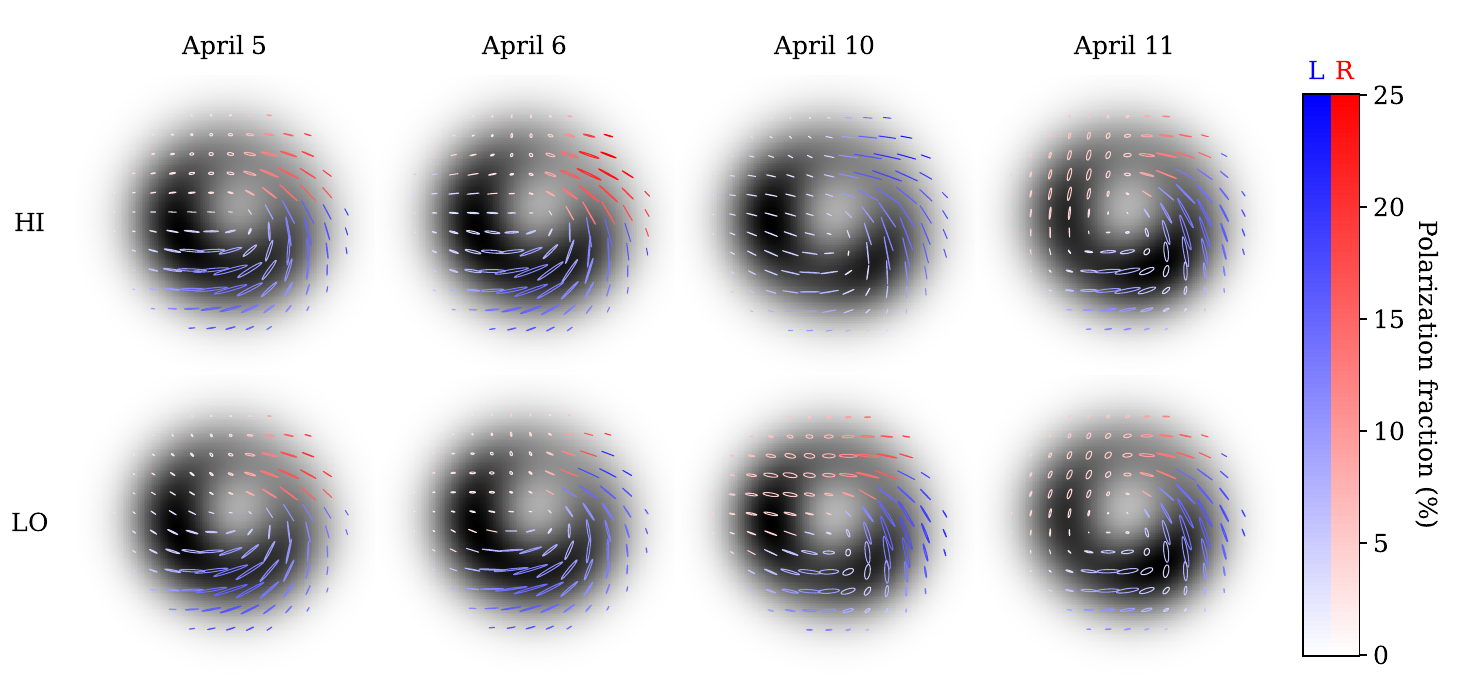}
   
    \caption{Polarization ellipse plots of our polarimetric m-ring model fits shown in \autoref{fig:m87_stokesv}, blurred with a 20 $\mu$as Gaussian beam. The total intensity structure is indicated in gray scale, and the ellipses indicate the total polarization  $\sqrt{\mathcal{Q}^2 + \mathcal{U}^2 + \mathcal{V}^2}/\mathcal{I}$. The ellipse sizes are proportional to the total polarized brightness, the orientations indicate the linear polarization EVPA, the ellipticities indicate the the circular polarization fraction, and the colors indicate the circular polarization sign and fraction. Posterior means for the April 11 fits are presented in the same format in \citetalias{M87PaperIX}.}
    \label{fig:m87_pe}
\end{figure*}

\autoref{fig:m87_stokesv} shows our circular polarization m-ring fits of EHT \m87 data on April 5, 6, 10, and 11. \autoref{fig:m87_pe} shows complete polarization ellipse plots for the same fits, which include total intensity, linear, and circular polarization information. Corresponding posterior ranges are reported in \autoref{tab:fitcomp}. These figures and table represent fits with an $m_{\mathcal{V}}=1$ m-ring to closure products; fits with varying Stokes $\mathcal{V}$ m-ring order and fits to RR/LL visibility ratios are explored in Figures~\ref{fig:m87_morder} and \ref{fig:m87_beta1}. Our fits identify a first-order (dipolar) circular polarization asymmetry that is broadly consistent across the four observing epochs spanning a six-day window, with more negative Stokes~$\mathcal{V}$ in the South and more positive Stokes~$\mathcal{V}$ in the North. The strength of the dipolar asymmetry slightly increases towards the later epochs, as indicated by the increase in $|\beta_{\mathcal{V},1}|$ and $\langle|\mathcal{V/I}|\rangle$. The closure-only fit to the April 10 high-band data is an outlier; the $\beta_{\mathcal{V},1}$ posterior is too broad for significant dipole structure to be detected (\autoref{tab:fitcomp}).

As shown in \autoref{fig:m87_morder}, the overall Stokes $\mathcal{V}$ morphology of the fits is remarkably consistent between fitting to different data products and fitting an $m_{\mathcal{V}}=1$ or $m_{\mathcal{V}}=2$ model, on all days. This consistency starts to break down for $m_{\mathcal{V}}=3$, although it persists for fits using data from April 10 and 11, where the asymmetry for the low m-order fits is most prominent.

The $\beta_{\mathcal{V},1}$ posteriors in \autoref{fig:m87_beta1} indicate that a significant dipolar signal is found by nearly all our fits, since $\beta_{\mathcal{V},1}$ is nonzero at a $>3\sigma$ level, especially for the $m_{\mathcal{V}}=1$ and $m_{\mathcal{V}}=2$ fits. The posteriors are generally broader for the closure only fits. The posteriors also indicate a preferred $\beta_{\mathcal{V},1}$ orientation close to the positive real axis, indeed corresponding to positive Stokes $\mathcal{V}$ in the North.

Considering the Bayesian evidence $\ln\mathcal{Z}$ and reduced $\chi^2$ as shown in \autoref{fig:m87_fitquality}, the preferred Stokes $\mathcal{V}$ m-ring order depends on the data product, day, and band that is fit. For the closure only fits, the Bayesian evidence mostly decreases and the goodness of fit remains approximately equal towards higher $m_{\mathcal{V}}$, indicating a preference for a low-order m-ring. Interestingly, $\ln\mathcal{Z}$ is larger for $m_{\mathcal{V}}=1$ than for $m_{\mathcal{V}}=0$ for the April 11 low-band fit ($m_{\mathcal{V}}=0$ corresponds to constant $\mathcal{V}$ along the ring), while the $m_{\mathcal{V}}=1$ values are slightly lower for the other data sets. Additionally, the $\beta_{\mathcal{V},1}$ posteriors (\autoref{fig:m87_beta1}) are furthest from zero compared to the other days. These trends indicate that the evidence for the presence of dipolar circular polarization structure is largest for April 11 (especially low band).

For the fits to visibility ratios, the Bayesian evidence mostly increases and the $\chi^2$ decreases towards higher $m_{\mathcal{V}}$, indicating a preference for higher-order m-rings. The difference in Bayesian evidence is especially large between $m_{\mathcal{V}}=0$ and $m_{\mathcal{V}}=1$ on April 6 and 11, which is a strong indicator of the presence of horizon-scale Stokes $\mathcal{V}$ structure. The Bayesian evidence peaks at $m_{\mathcal{V}}=1$ for April 6 and 11 (low band), indicating a preference for a dipole structure. Considering the increased inconsistency among the fit Stokes $\mathcal{V}$ structure for $m_{\mathcal{V}}=3$ (\autoref{fig:m87_morder}), the same trends occurring in our GRMHD fits, and the fact that the visibility ratio fits could be affected by R/L gain calibration uncertainties, we have presented the closure-only $m_{\mathcal{V}}=1$ fits as the main modeling result in \citetalias{M87PaperIX}. We do not deem any further fitted substructure trustworthy. The $m_{\mathcal{V}}=3$ fits to visibility ratios may be picking up on smaller-scale structure that the lower-order m-rings cannot account for, but in some of our GRMHD tests we have seen that these fits may present images that do not correctly reproduce this smaller-scale structure (\autoref{fig:grmhd_stokesv_maps}). Future EHT datasets with better $uv$-coverage and sensitivity will allow us to detect circular polarization structure more strongly and in more detail than these first-order structure results.

\autoref{fig:m87_vnet} shows that the basic structure of the posterior maxima is independent of the assumed $\mathcal{V_{\mathrm{net}}}$, within the range reported from ALMA-only measurements by \citet{Goddi2021}. 

\subsubsection{Sensitivity to R/L gain calibration strategy}
\label{sec:m87cpolcasa}
\begin{figure*}
    \centering
    \includegraphics[height=0.19\textwidth]{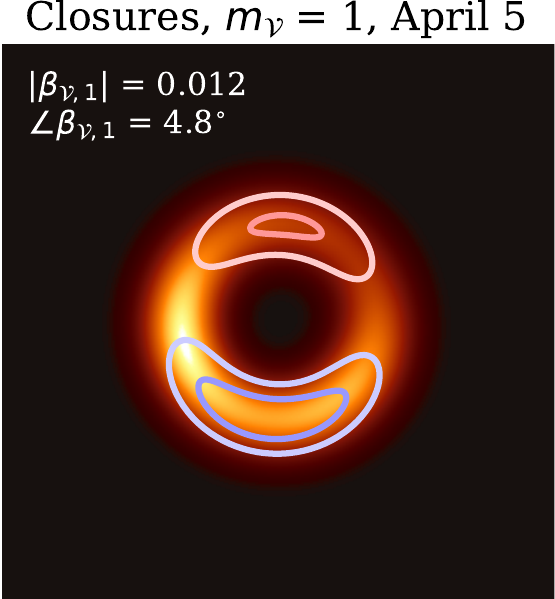}
    \includegraphics[height=0.19\textwidth]{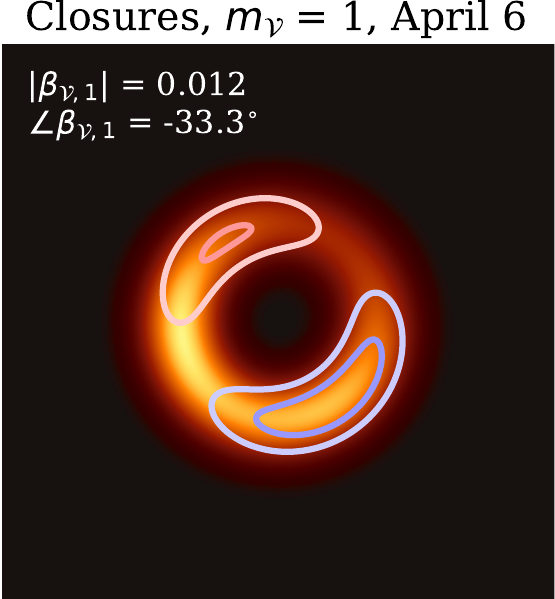}
    \includegraphics[height=0.19\textwidth]{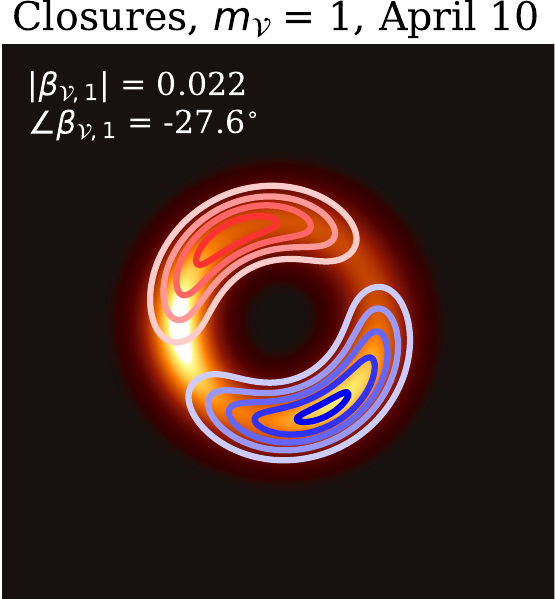}
    \includegraphics[height=0.19\textwidth]{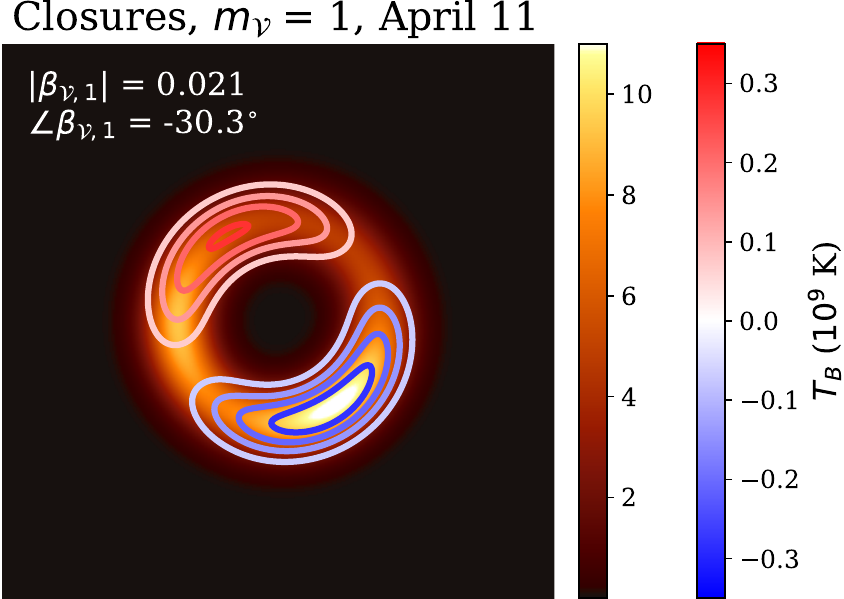} \\
    \includegraphics[height=0.19\textwidth]{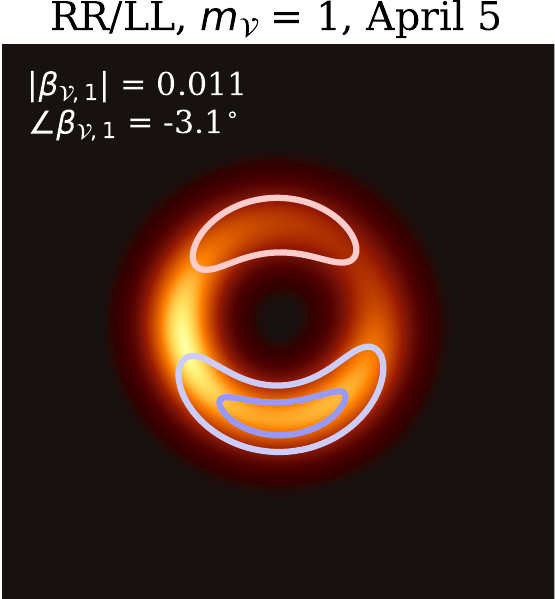}
    \includegraphics[height=0.19\textwidth]{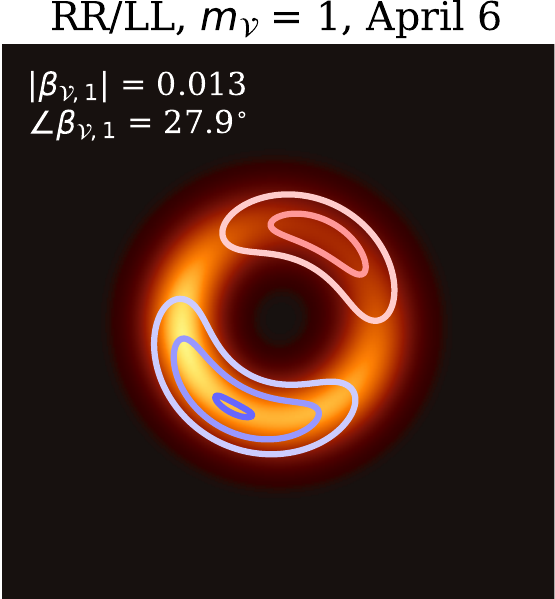}
    \includegraphics[height=0.19\textwidth]{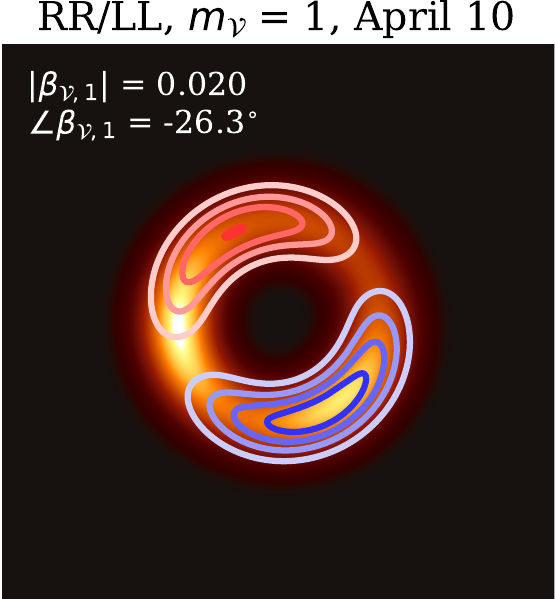}
    \includegraphics[height=0.19\textwidth]{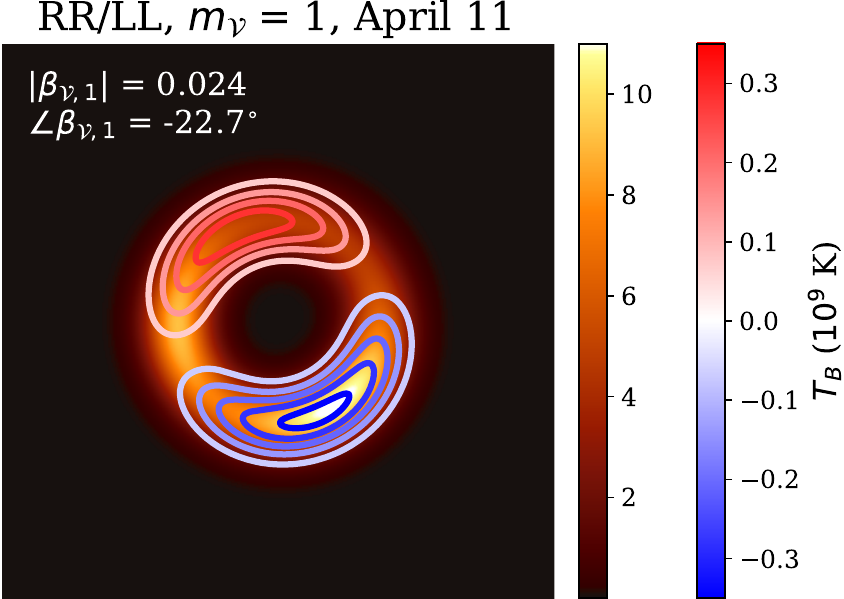} \\
    \vspace{6mm}
    \includegraphics[height=0.19\textwidth]{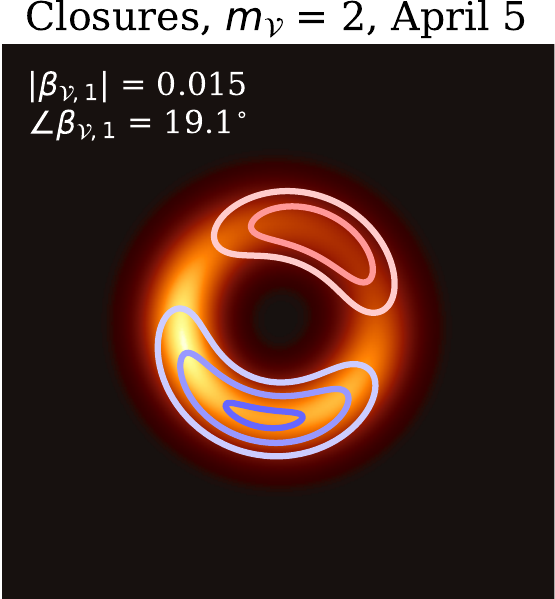}
    \includegraphics[height=0.19\textwidth]{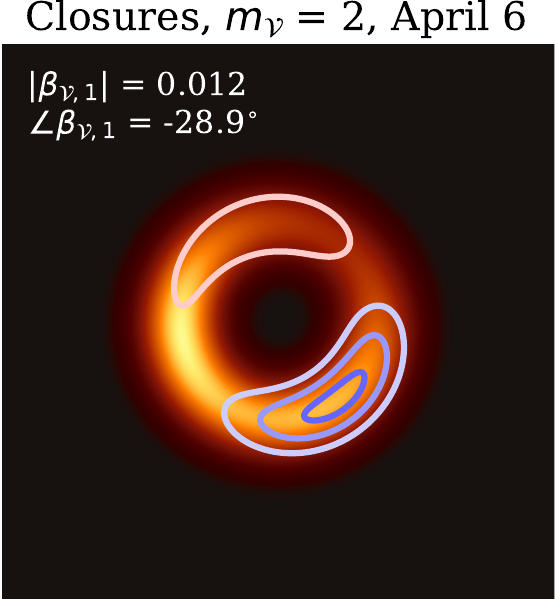}
    \includegraphics[height=0.19\textwidth]{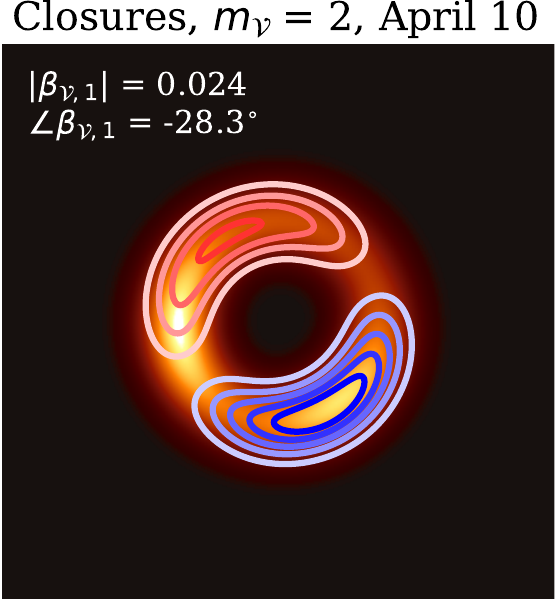}
    \includegraphics[height=0.19\textwidth]{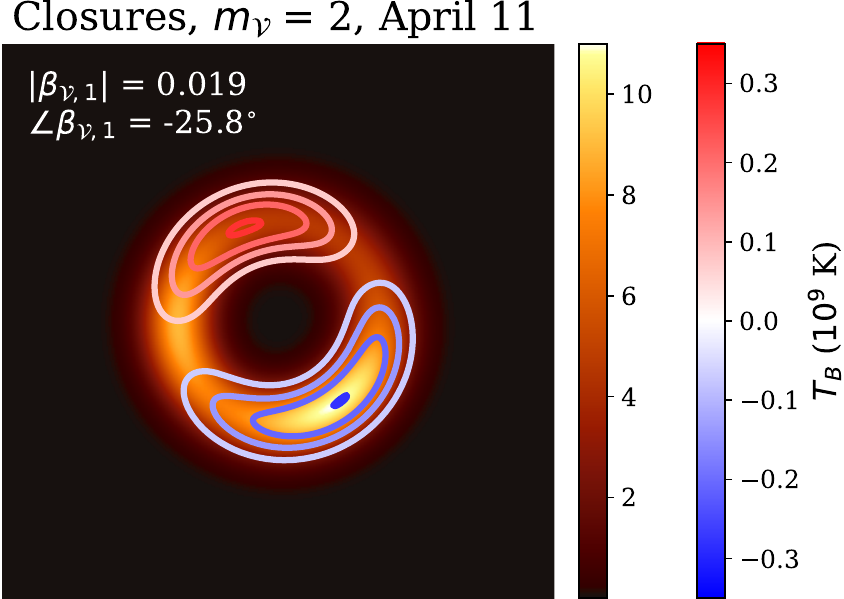} \\
    \includegraphics[height=0.19\textwidth]{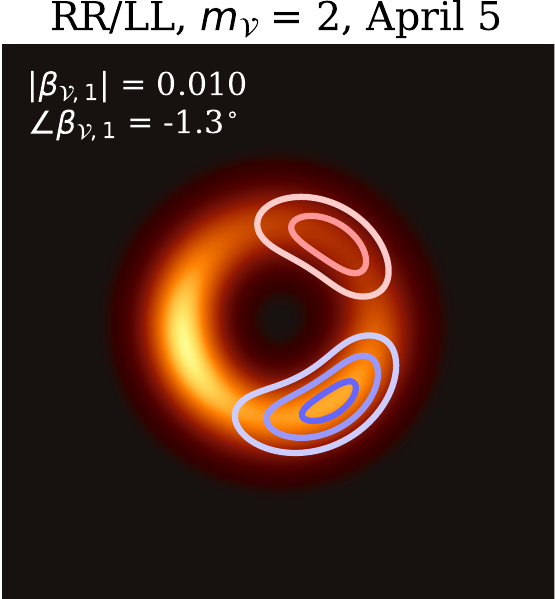}
    \includegraphics[height=0.19\textwidth]{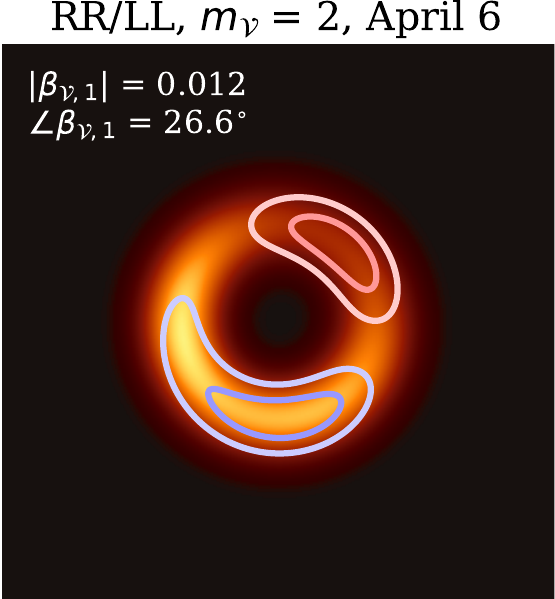}
    \includegraphics[height=0.19\textwidth]{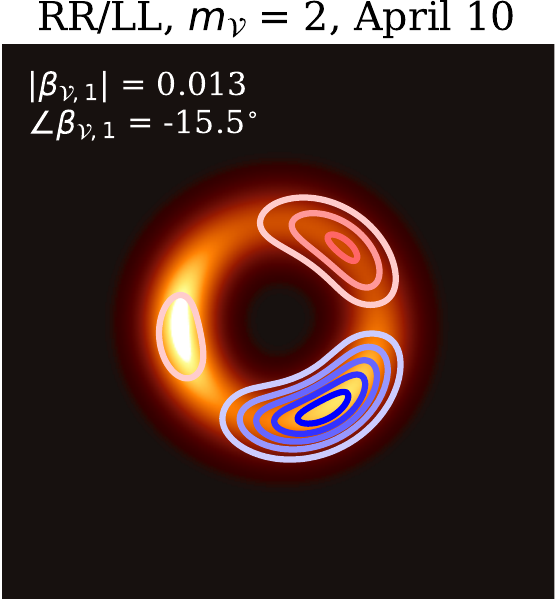}
    \includegraphics[height=0.19\textwidth]{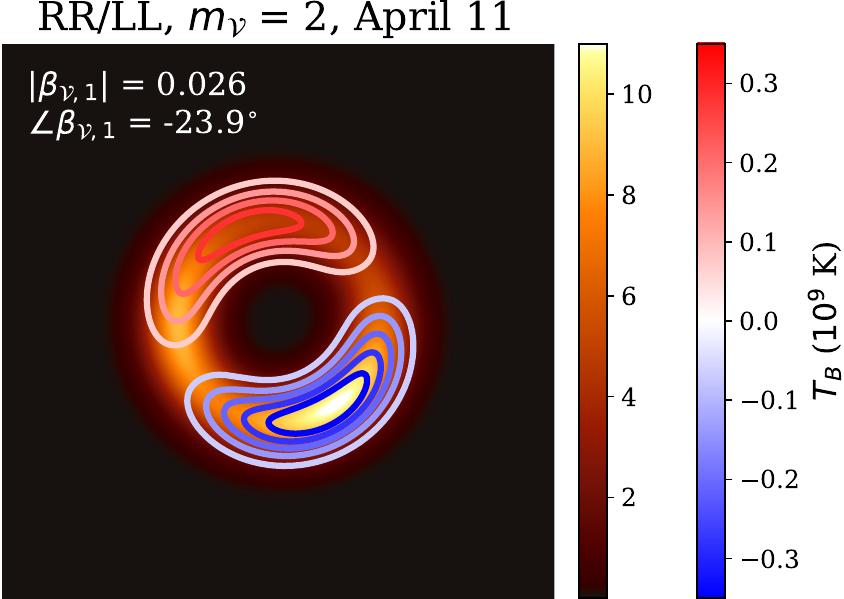} \\
    \vspace{6mm}
    \includegraphics[height=0.19\textwidth]{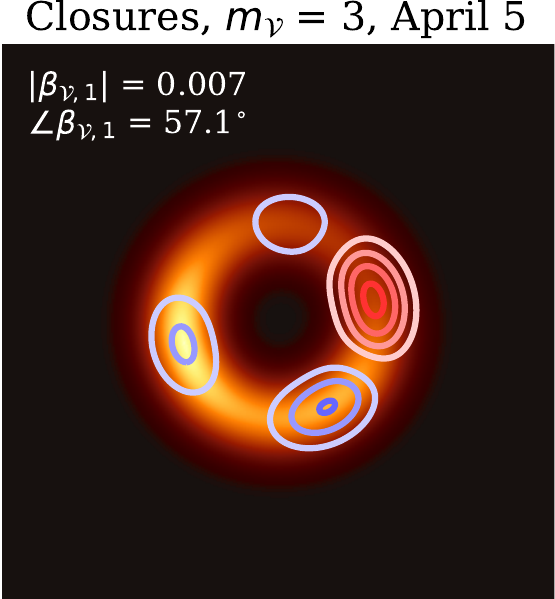}
    \includegraphics[height=0.19\textwidth]{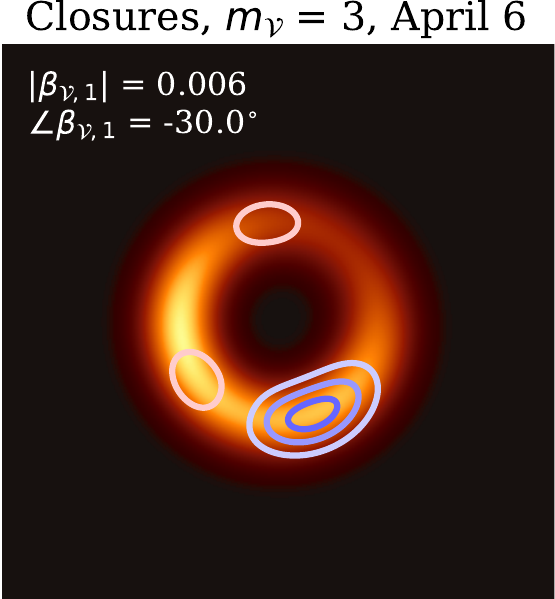}
    \includegraphics[height=0.19\textwidth]{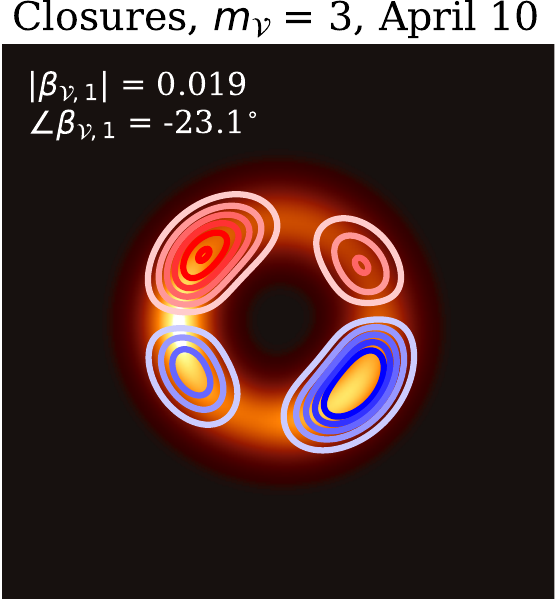}
    \includegraphics[height=0.19\textwidth]{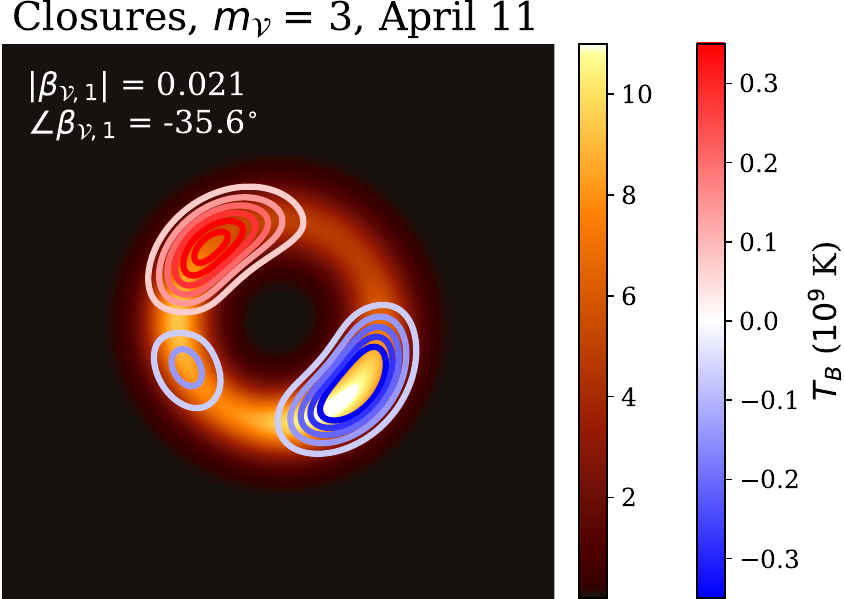} \\
    \includegraphics[height=0.19\textwidth]{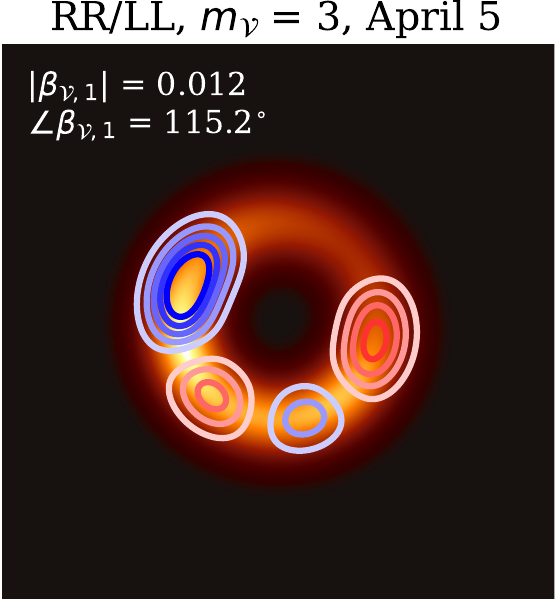}
    \includegraphics[height=0.19\textwidth]{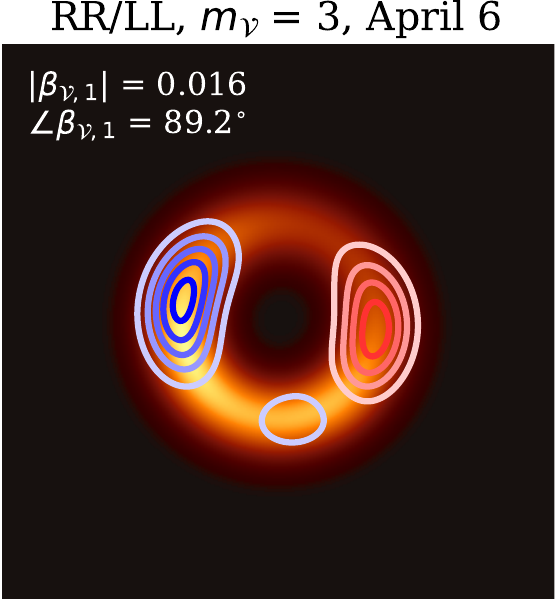}
    \includegraphics[height=0.19\textwidth]{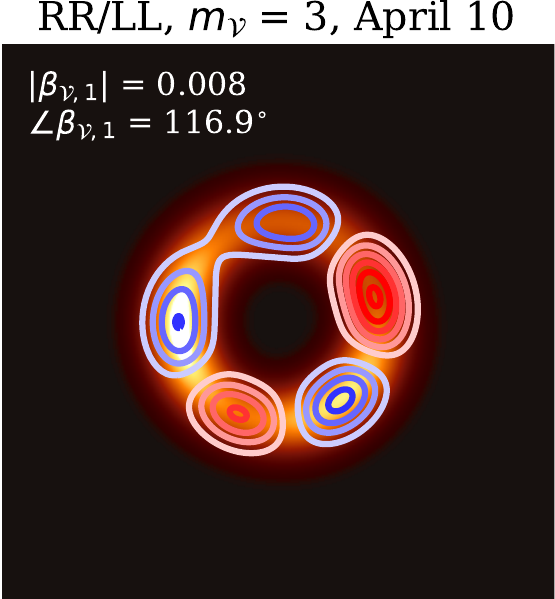}
    \includegraphics[height=0.19\textwidth]{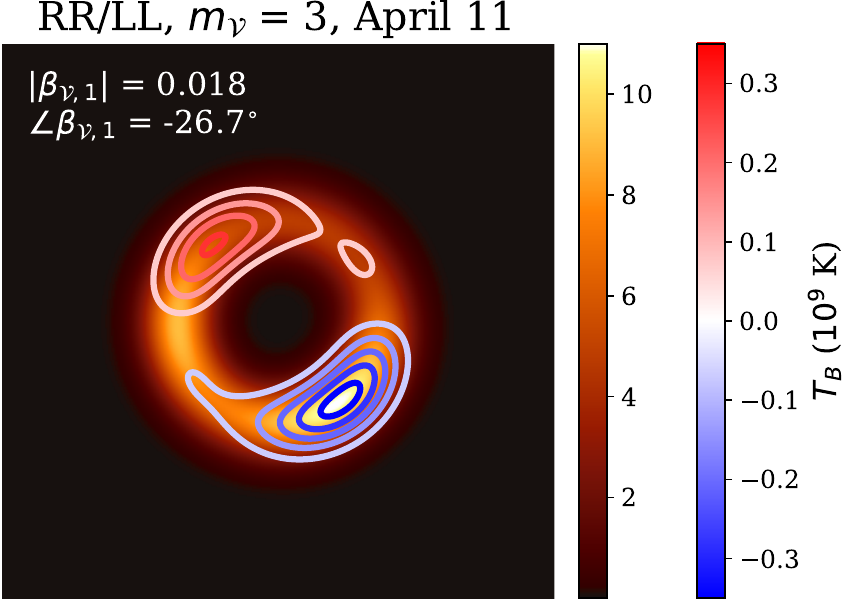} \\
    \caption{Circular polarization m-ring fits (posterior maxima) to low-band EHT 2017 \m87 data on April 5, 6, 10, and
11 (left to right). The three sets of two rows show fits with $m_{\mathcal{V}}=1, 2, 3$, fitting to closure quantities (upper panels for each $m$-order) and RR/LL visibility ratios (lower panels for each $m$-order). The top row panels correspond to the top row panels in \autoref{fig:m87_stokesv}.}
    \label{fig:m87_morder}
\end{figure*}

\begin{figure*}
    \centering
    \includegraphics[height=0.18\textwidth]{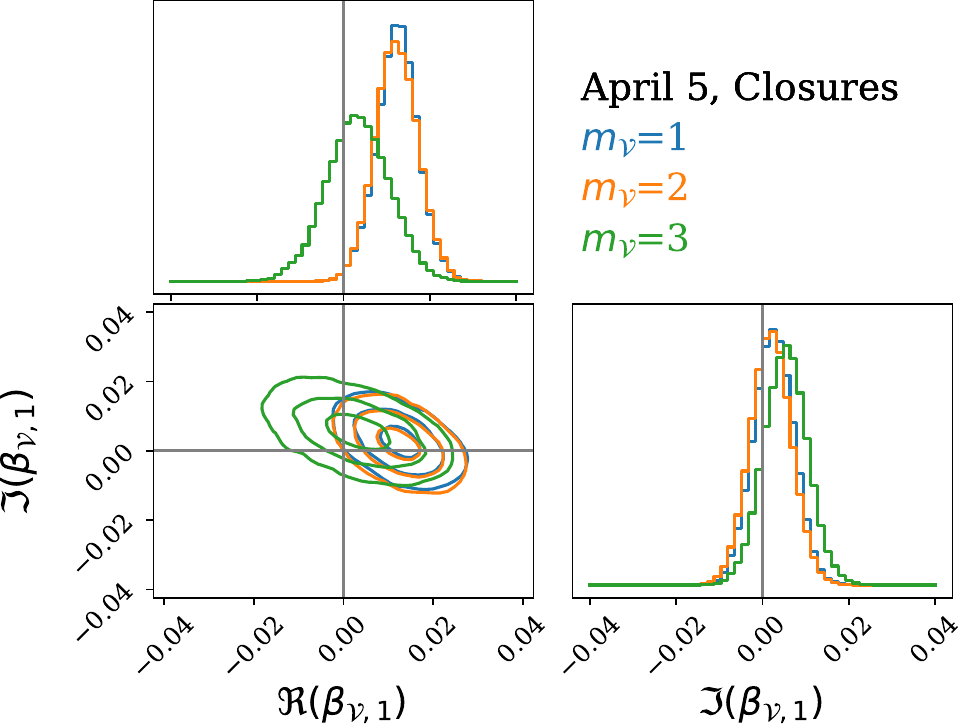}
    \includegraphics[height=0.18\textwidth]{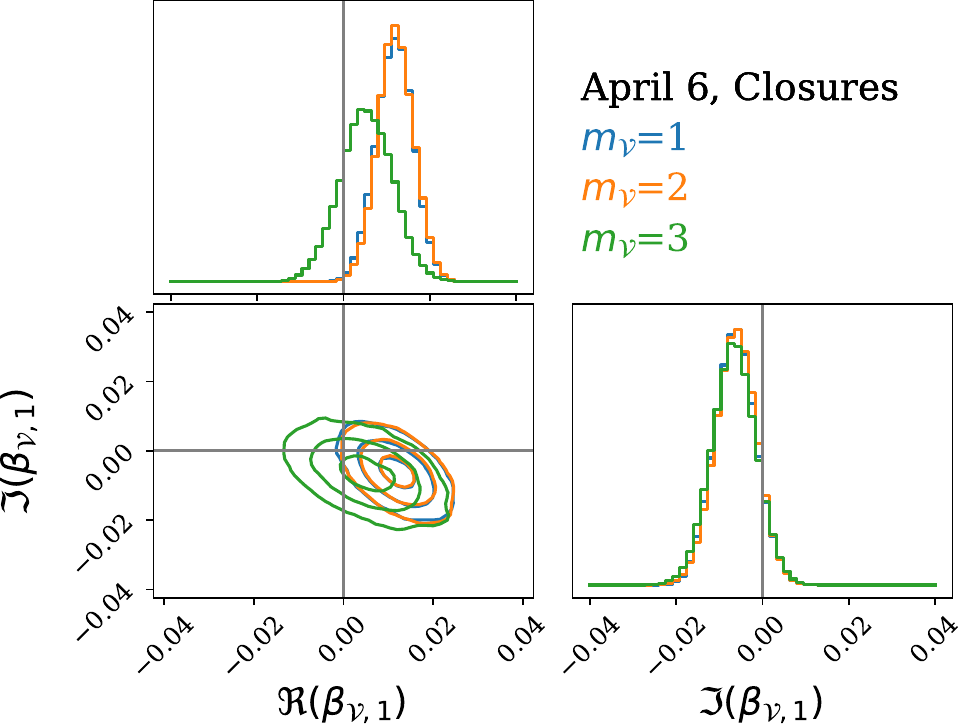}
    \includegraphics[height=0.18\textwidth]{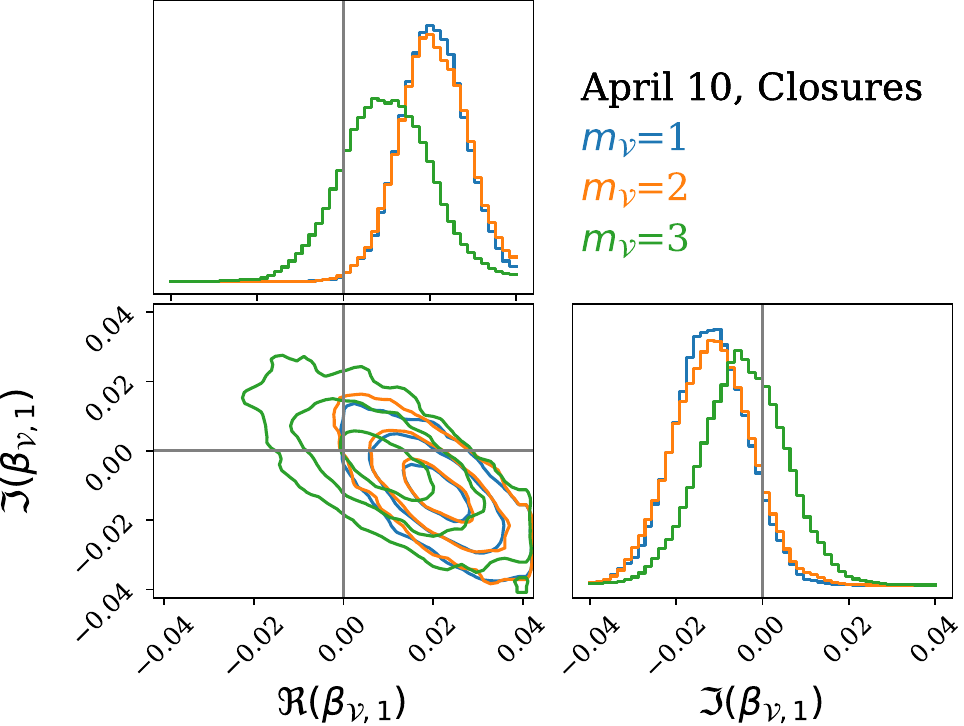}
    \includegraphics[height=0.18\textwidth]{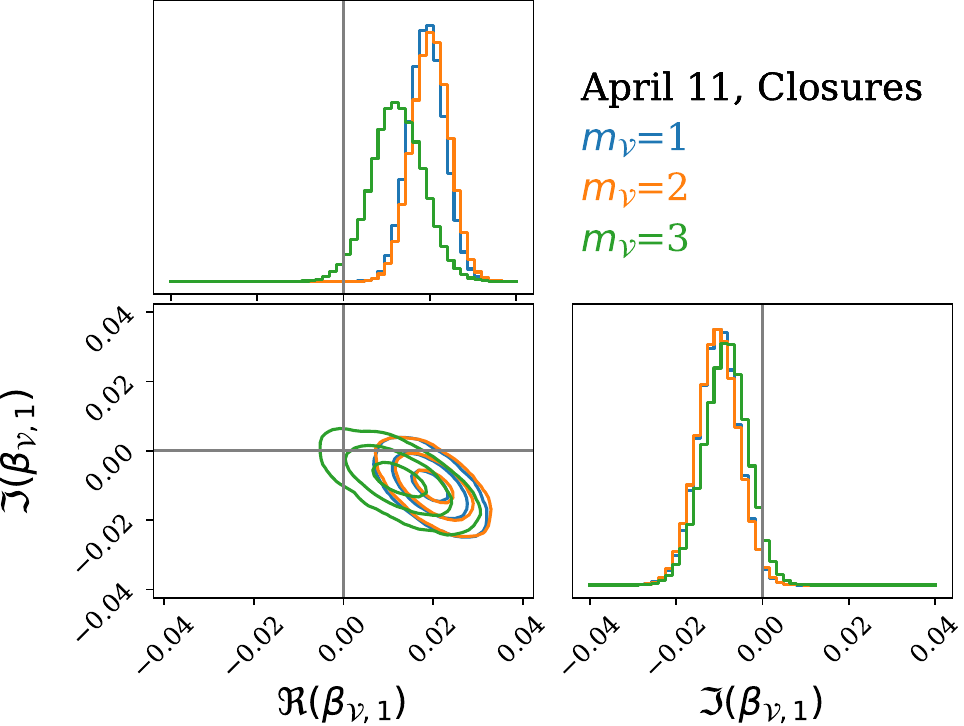} \\
    \includegraphics[height=0.18\textwidth]{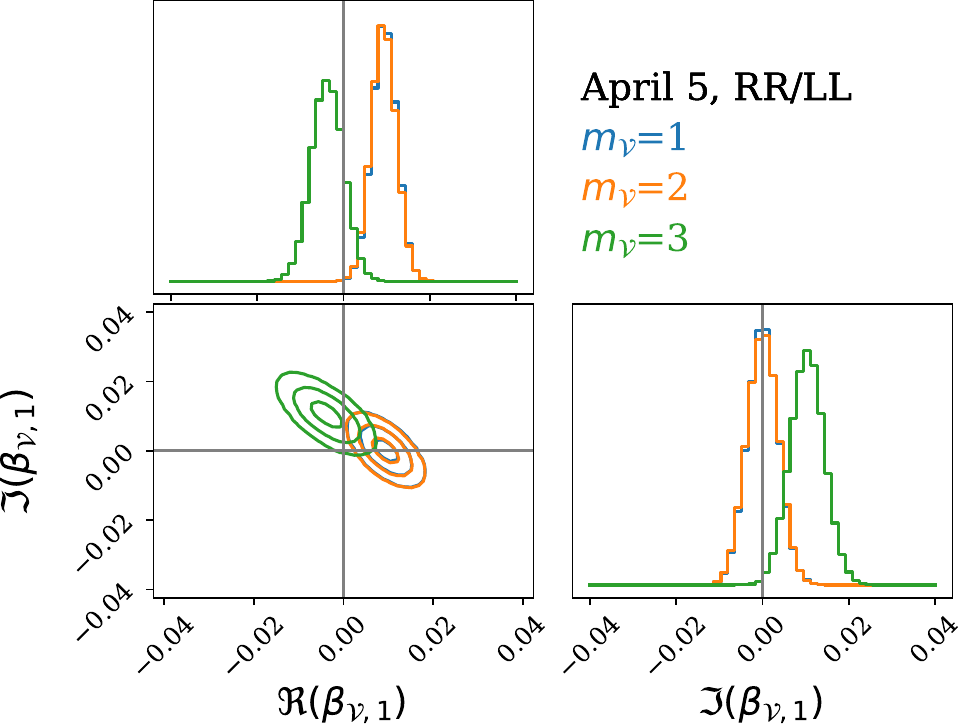}
    \includegraphics[height=0.18\textwidth]{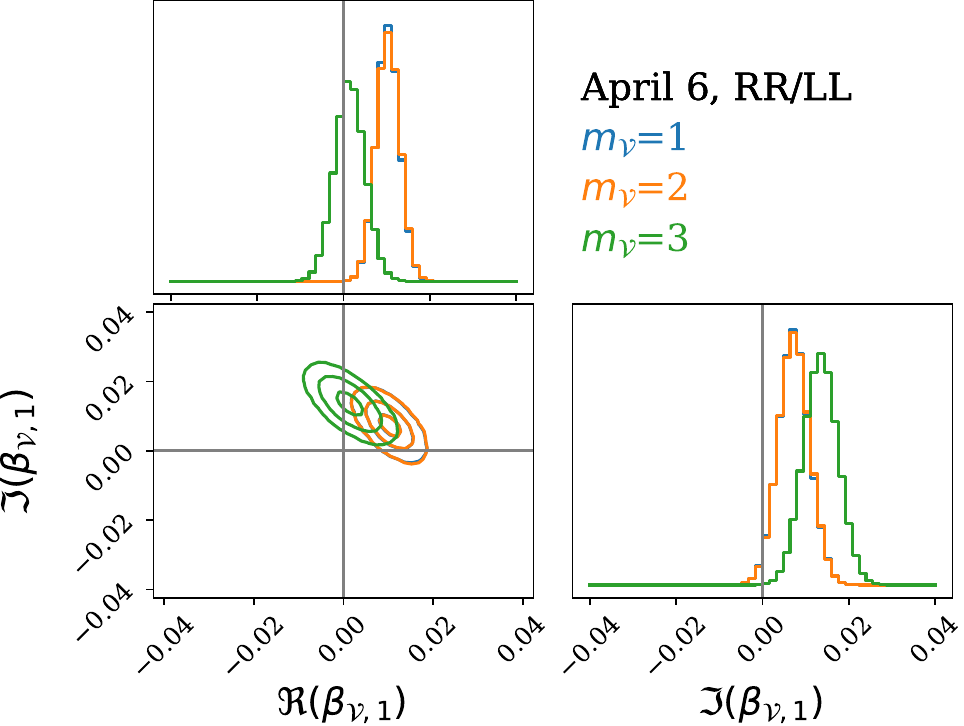}
    \includegraphics[height=0.18\textwidth]{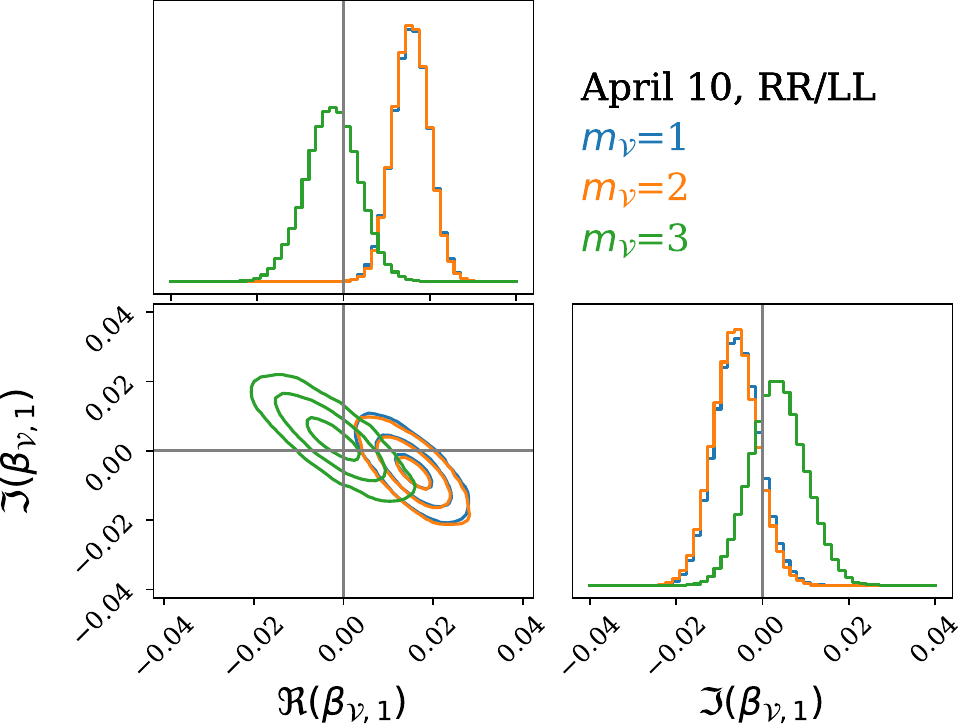}
    \includegraphics[height=0.18\textwidth]{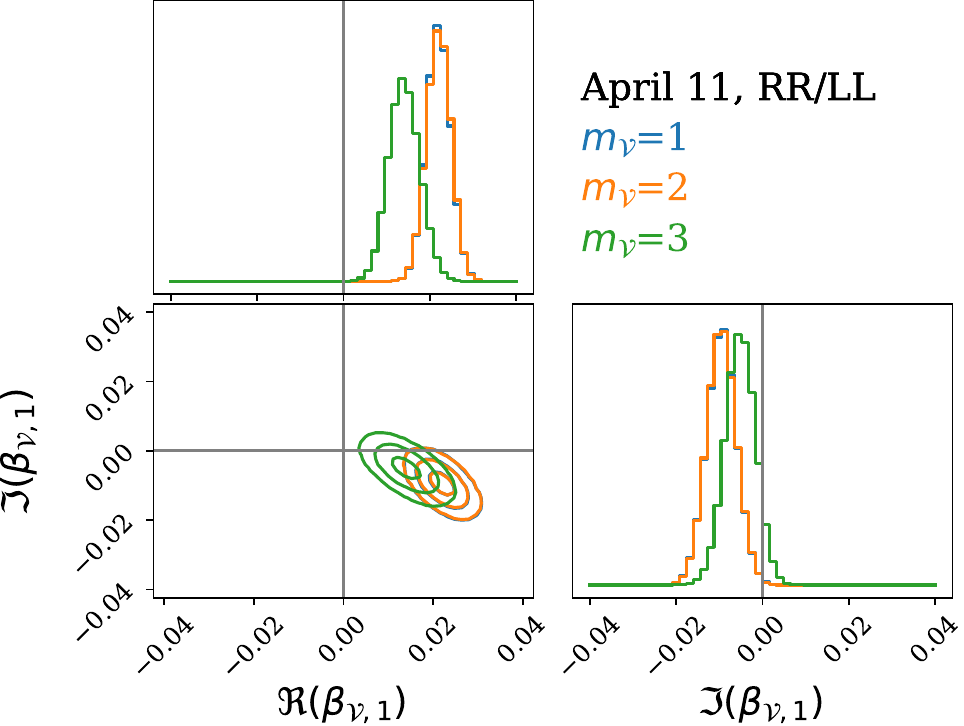}
    \caption{$\beta_{\mathcal{P},1}$ posteriors for the Stokes $\mathcal{V}$ m-ring fits to closure quantities (top row) and RR/LL visibility ratios (bottom row) from low-band EHT 2017 \m87 data on April 5, 6, 10, and
11 (left to right). The contours show 1$\sigma$, 2$\sigma$, and 3$\sigma$ levels.}
    \label{fig:m87_beta1}
\end{figure*}

\begin{figure*}
    \centering
    \includegraphics[width=0.4\textwidth]{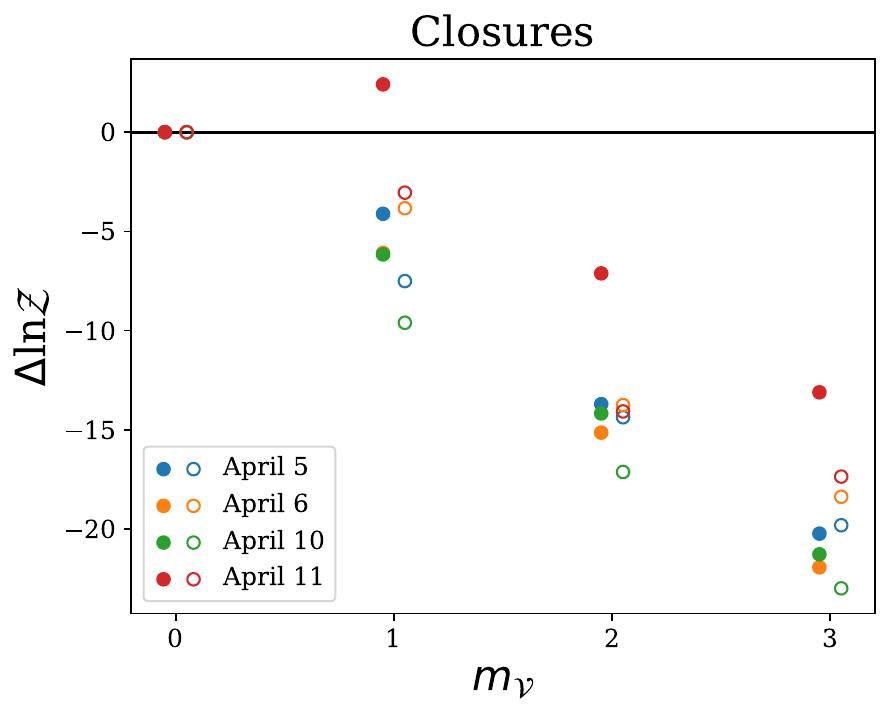}
    \includegraphics[width=0.4\textwidth]{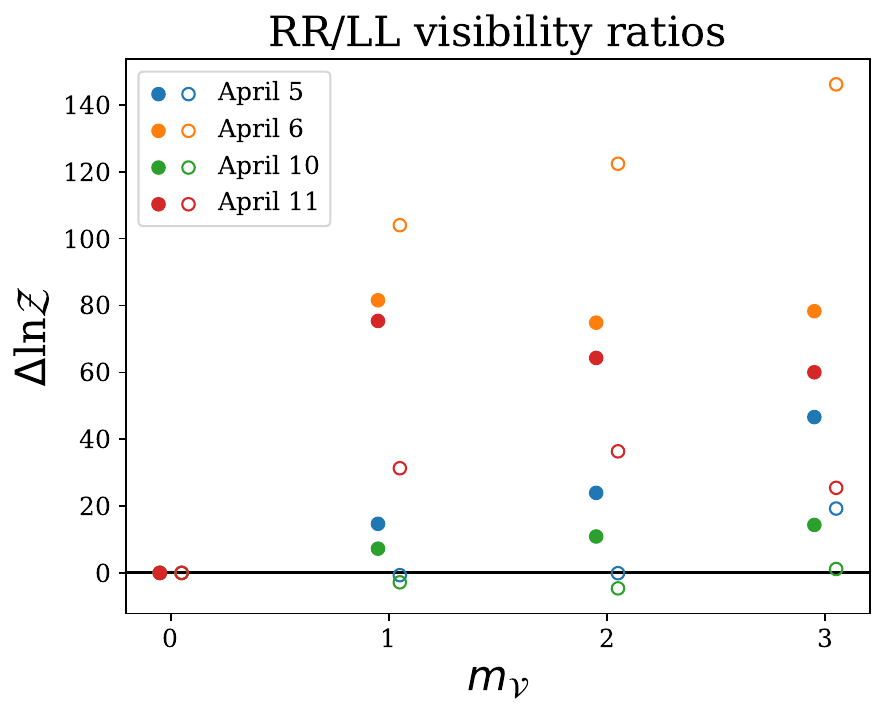}
    \includegraphics[width=0.4\textwidth]{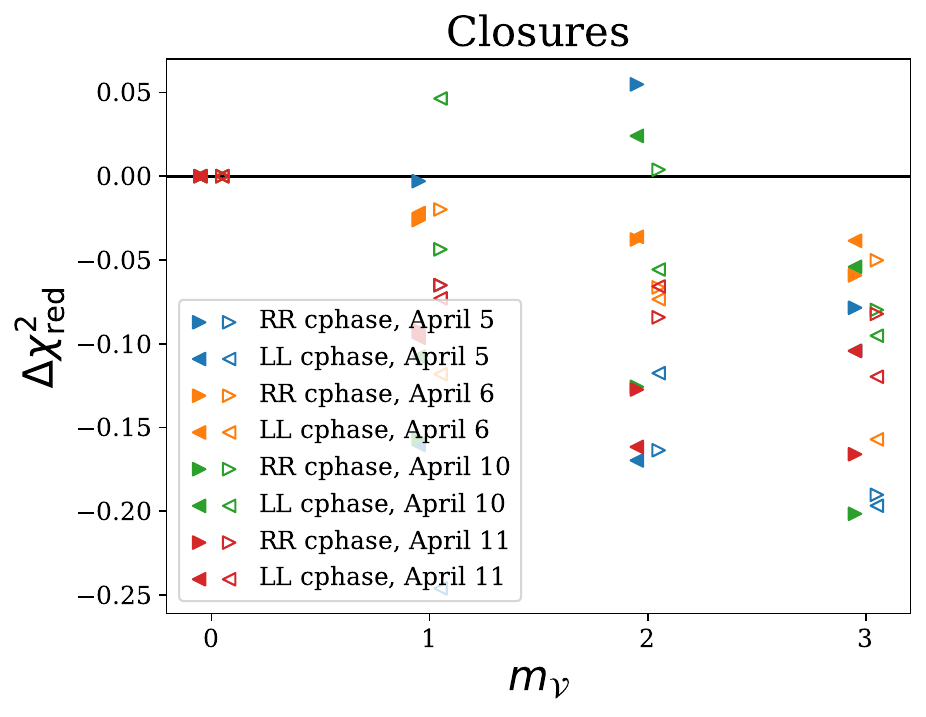}
    \includegraphics[width=0.4\textwidth]{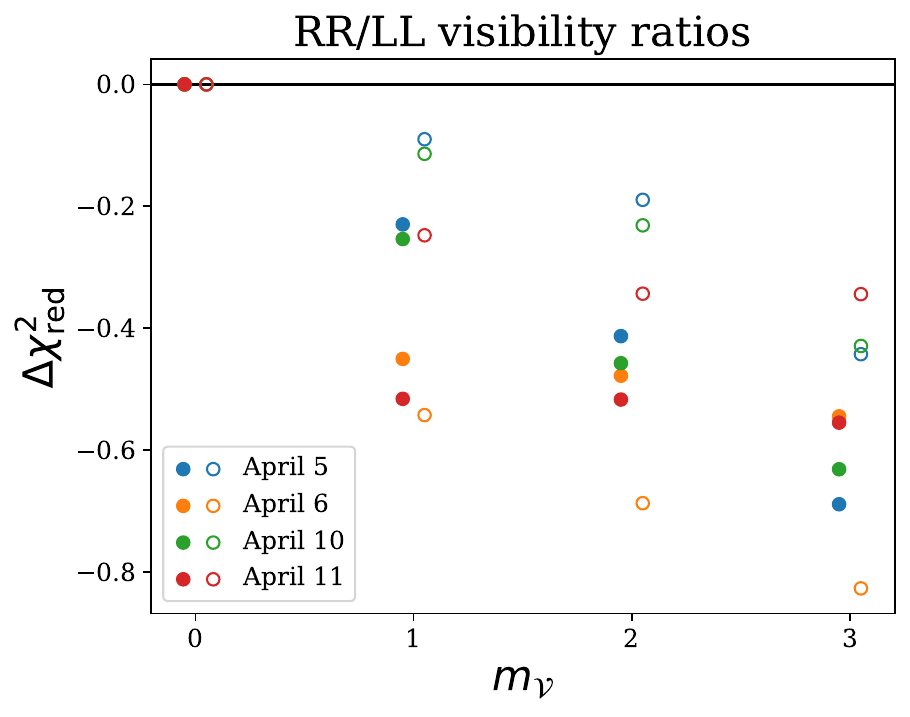}
    
    \caption{Relative Bayesian evidence (upper panels) and reduced $\chi^2$ (lower panels) as a function of Stokes $\mathcal{V}$ m-ring order $m_{\mathcal{V}}$, for the EHT \m87 m-ring fits shown in Figures~\ref{fig:m87_morder} and \ref{fig:m87_beta1}. Low-band fits are indicated with filled symbols, and high-band fits are indicated with open symbols. All values are relative to the $m_{\mathcal{V}}=0$ fit on each day and band.}
    \label{fig:m87_fitquality}
\end{figure*}

\begin{figure*}
    \centering
    \includegraphics[height=0.19\textwidth]{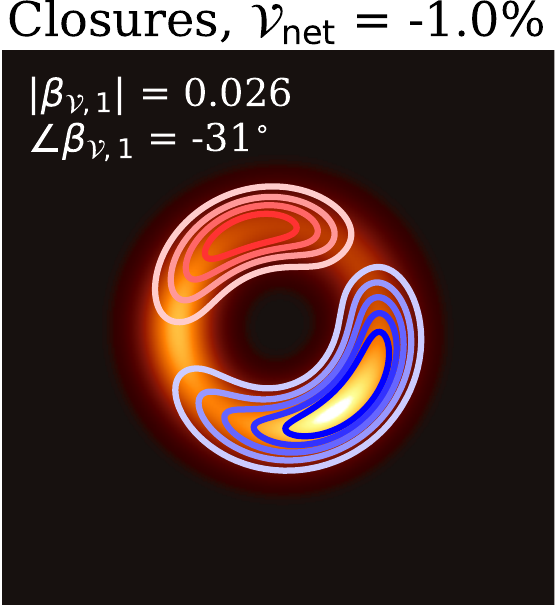}
    \includegraphics[height=0.19\textwidth]{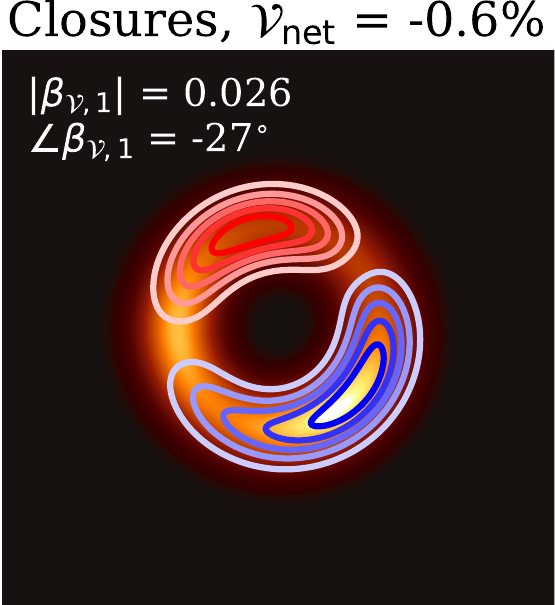}
    \includegraphics[height=0.19\textwidth]{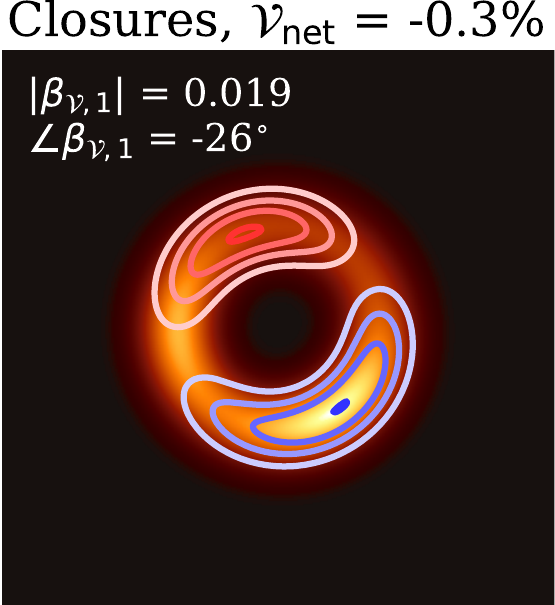}
    \includegraphics[height=0.19\textwidth]{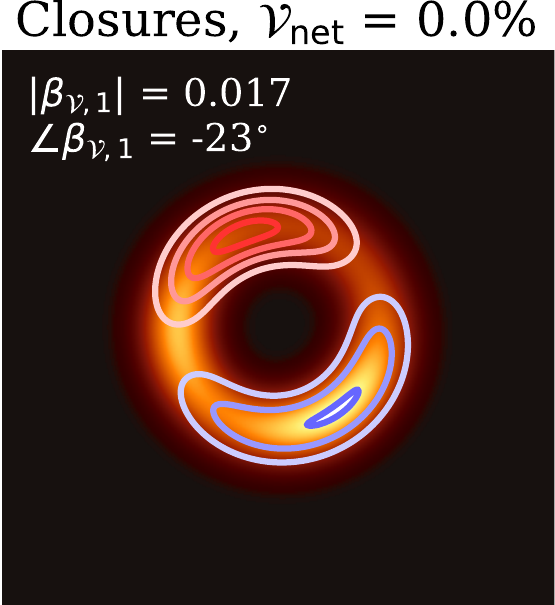}
    \includegraphics[height=0.19\textwidth]{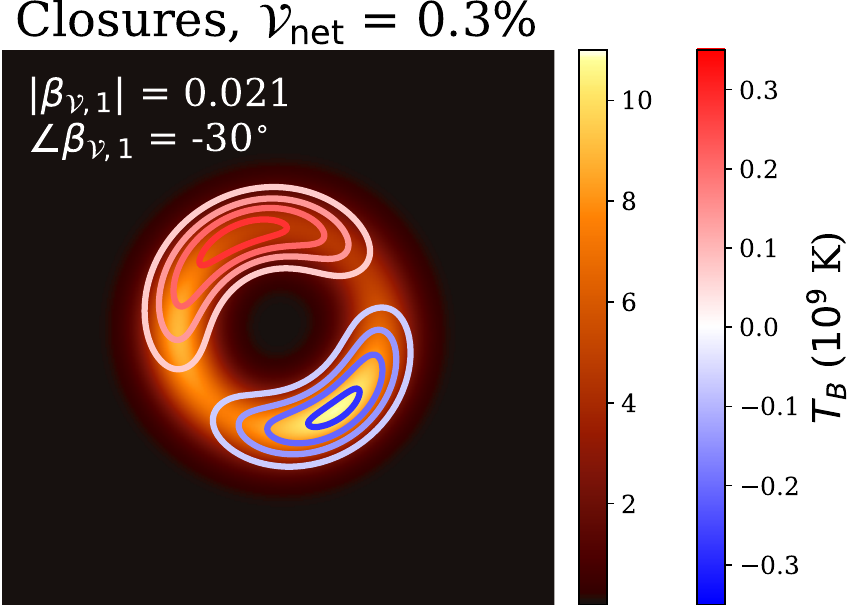} \\
    \includegraphics[height=0.19\textwidth]{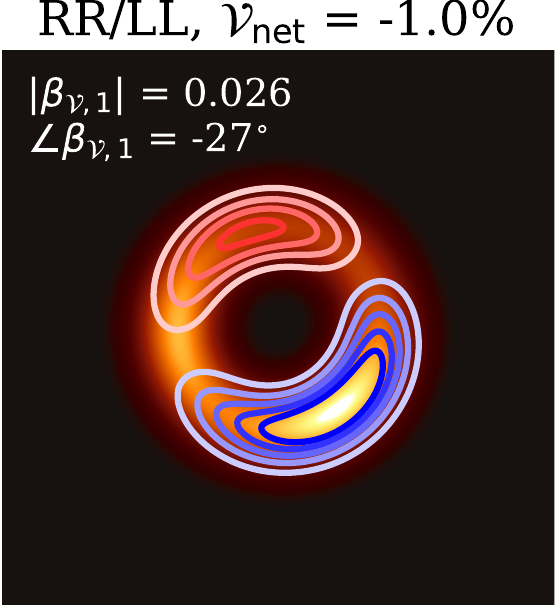}
    \includegraphics[height=0.19\textwidth]{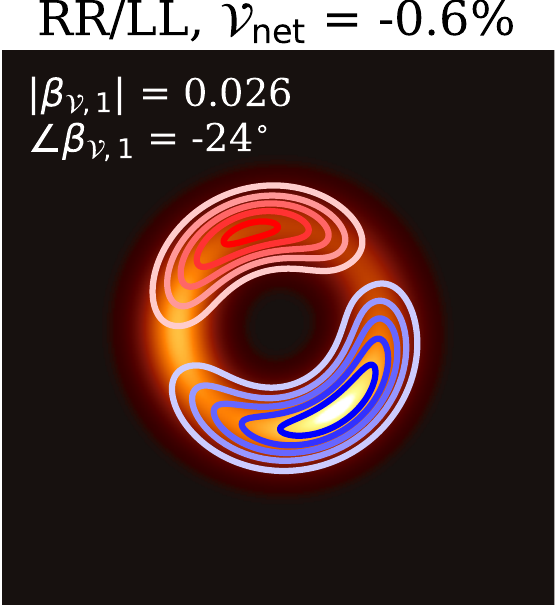}
    \includegraphics[height=0.19\textwidth]{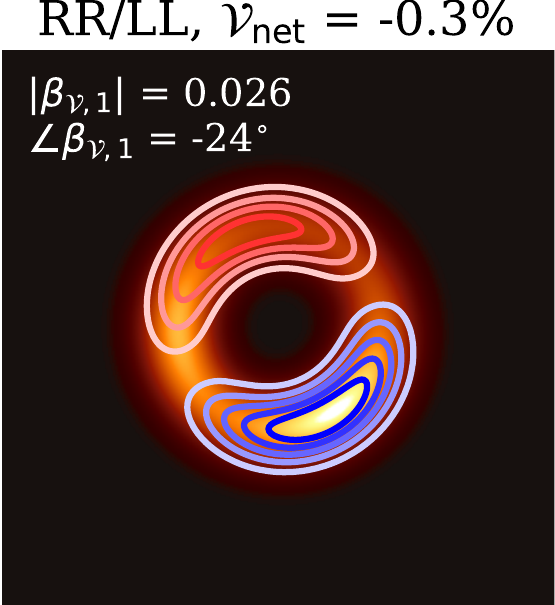}
    \includegraphics[height=0.19\textwidth]{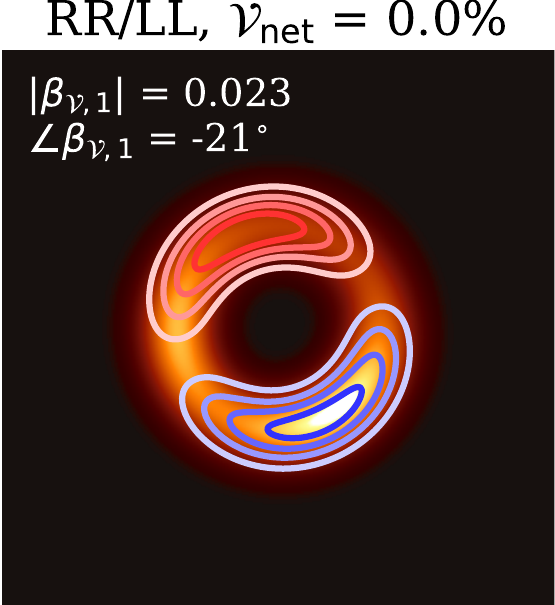}
    \includegraphics[height=0.19\textwidth]{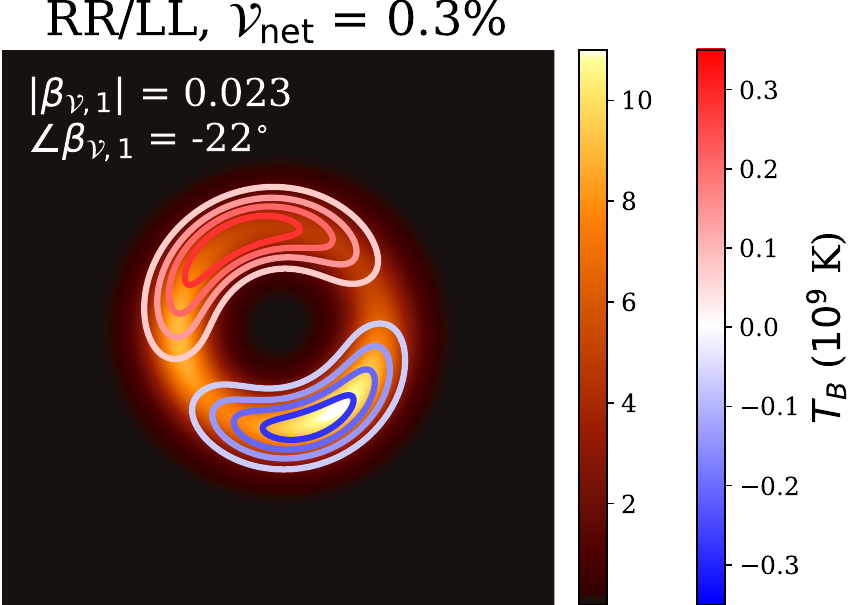}
    \caption{Circular polarization m-ring fits ($m_{\mathcal{V}}=2$, posterior maxima) to EHT 2017 \m87 data on April 11, assuming different total circular polarization fractions $\mathcal{V}_{\mathrm{net}}$ that span the range reported by \citet{Goddi2021} (left to right). The top row shows fits to closure quantities and the bottom row shows fits to RR/LL visibility ratios.}
    \label{fig:m87_vnet}
\end{figure*}

As noted in \autoref{sec:data_products}, the RR/LL visibility ratio data product is affected by non-unity R/L gain ratios, while the parallel-hand closure data products are not. In this section, we explore the sensitivity of our \m87 fit results to the gain calibration strategy. The R/L gain ratios in the \texttt{HOPS} data (used in \autoref{sec:m87linpol} and \autoref{sec:m87cpolhops}) were calibrated by fitting a polynomial to the RR and LL visibility offsets (amplitude ratios and phase differences) as a function of time for ten sources observed during the five days of the 2017 EHT campaign. By using data from multiple sources and days, circular polarization signatures of individual sources (assumed to be independent and thus averaging out) could be separated from instrumental R/L gain offsets (shared between the sources). The visibility differences could be fit with a constant complex gain ratio for all stations, except for APEX and SMA, which showed stronger time-dependence of the RR and LL visibility phase differences and hence required a time-dependent polynomial fit for the R/L gain phases. Some more details are given in \citetalias{M87PaperIX}, with an illustration of a R/L phase fit shown in Fig. 14 therein, indicating a well-behaving set of sources, without any strongly polarized outlier. The general similarity of our RR/LL fits to our closure-only fits (\autoref{fig:m87_morder}) can be taken as an indication that the effect of any residual R/L gain ratios on our fits is limited.

\begin{figure*}
    \centering
    \includegraphics[height=0.19\textwidth]{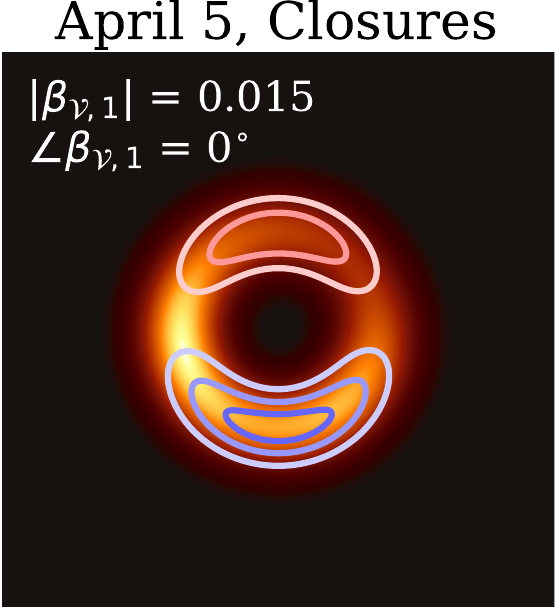}
    \includegraphics[height=0.19\textwidth]{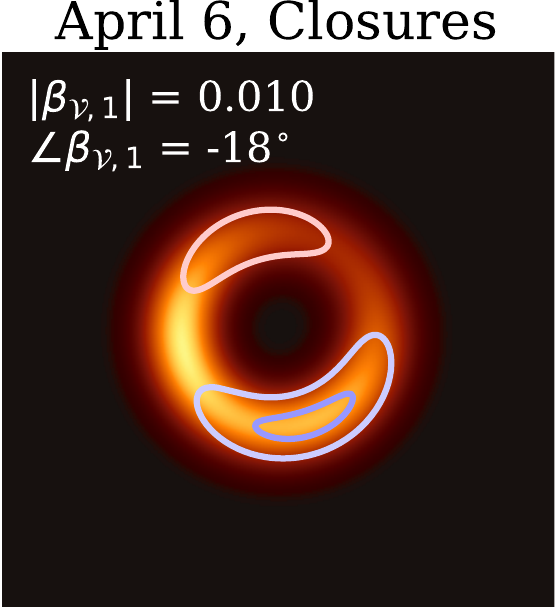}
    \includegraphics[height=0.19\textwidth]{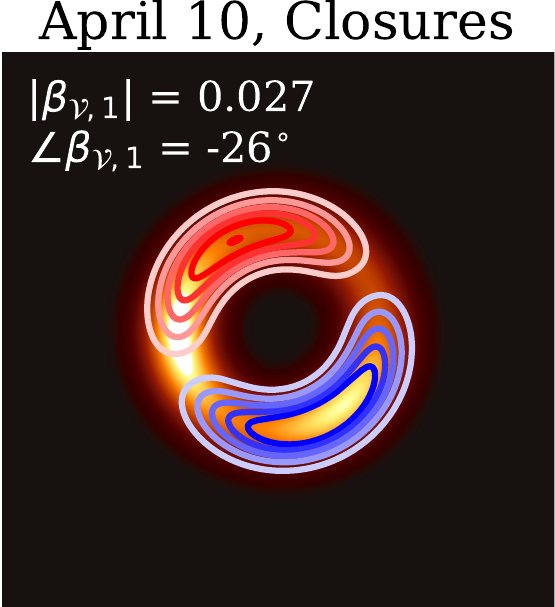}
    \includegraphics[height=0.19\textwidth]{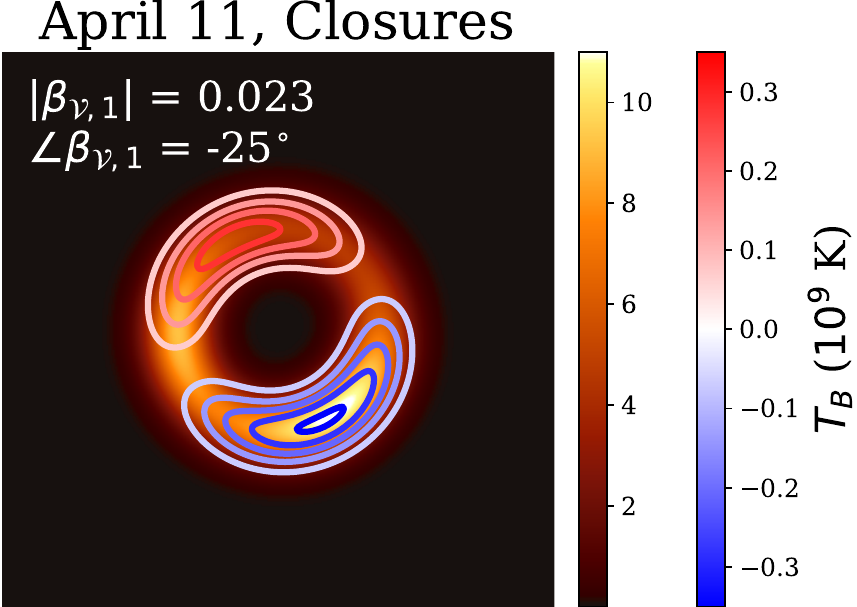} \\
    \includegraphics[height=0.19\textwidth]{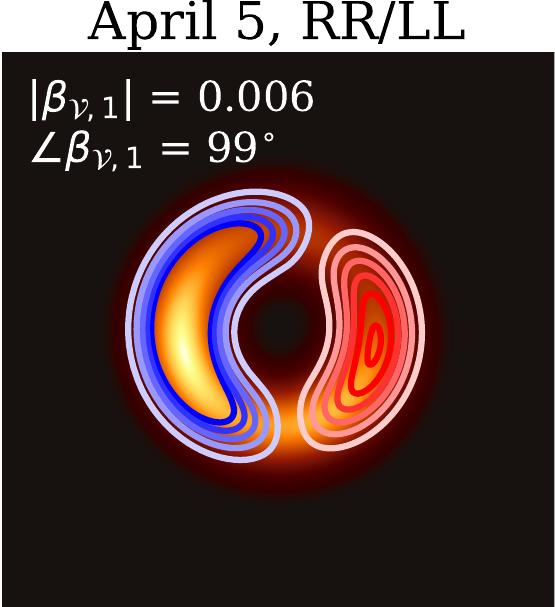}
    \includegraphics[height=0.19\textwidth]{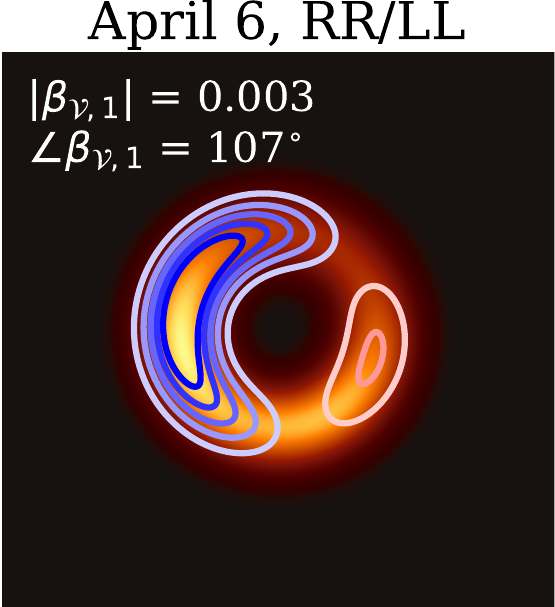}
    \includegraphics[height=0.19\textwidth]{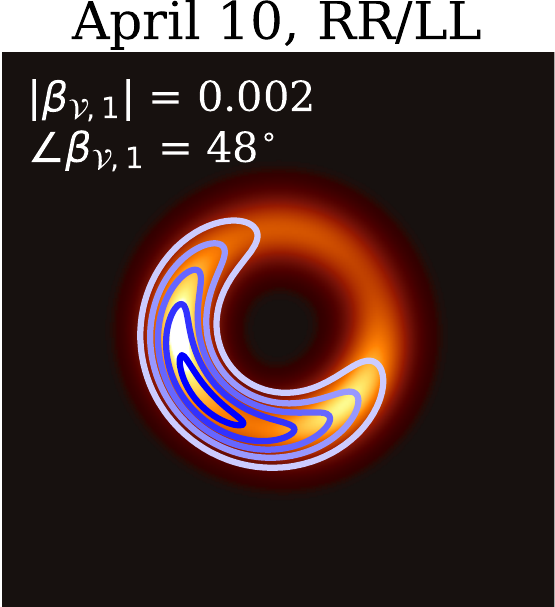}
    \includegraphics[height=0.19\textwidth]{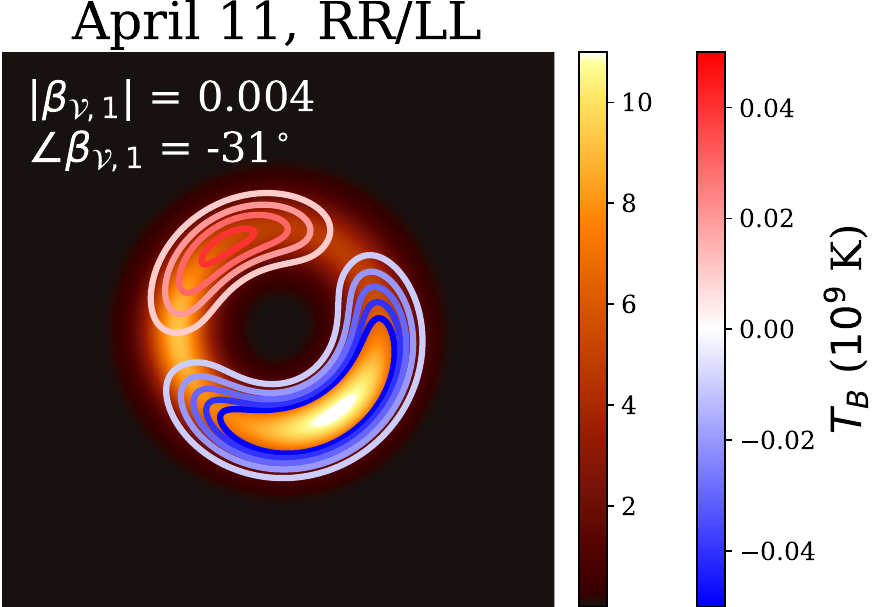} \\
    \caption{Circular polarization m-ring fits (low band, $m_{\mathcal{V}}=1$, posterior maxima) to EHT 2017 \m87 data, performed after self-calibrating the R and L gains to the total intensity fit assuming $\mathcal{V}=0$. The top row shows fits to closure quantities and the bottom row shows fits to RR/LL visibility ratios.}
    \label{fig:m87_zerovselfcal}
\end{figure*}

The robustness of our RR/LL fits in particular can also be tested in an exercise where we attempt to remove the circular polarization signal by self-calibrating the R and L gains separately to our total intensity model assuming $\mathcal{V}=0$, before fitting the Stokes $\mathcal{V}$ structure \citep[e.g.][]{Homan2004}. \autoref{fig:m87_zerovselfcal} shows the result of performing such an exercise on the EHT \texttt{HOPS} data. Here, we have set a solution interval of two hours for both the gain amplitudes and gain phases. As expected, the closure-only fits are unaffected by this operation, since the closure products are robust against station gains. The RR/LL fits are clearly affected, showing substantially weaker circular polarization structure (note the difference in scale between the two rows), and a different orientation on most days. The exception is April 11, where the zero-$\mathcal{V}$ self-calibration failed to remove the dipole structure with approximately North-South orientation, although it is substantially weaker.

To test the sensitivity of our model fitting procedure to upstream calibration choices, we also fit our polarimetric m-ring model to a new version of the 2017 EHT data, which has been calibrated in a slightly different way. We utilize the CASA-VLBI-based \citep{2022Bemmel} \texttt{rPICARD} \citep{Janssen2019} pipeline to solve for instrumental offsets \citep{PaperIII} and then combine the two LO plus HI frequency bands and all polarization correlation products to solve for fringes and atmospheric phases (Janssen et al. in prep.). In contrast to the multi-source polynomial fit R/L gain calibration described above, either no R/L amplitude gain ratios are applied to the CASA data used here, or an R/L gain amplitude ratio calibration was performed assuming zero circular polarization.

\begin{figure*}
    \centering
    \includegraphics[height=0.19\textwidth]{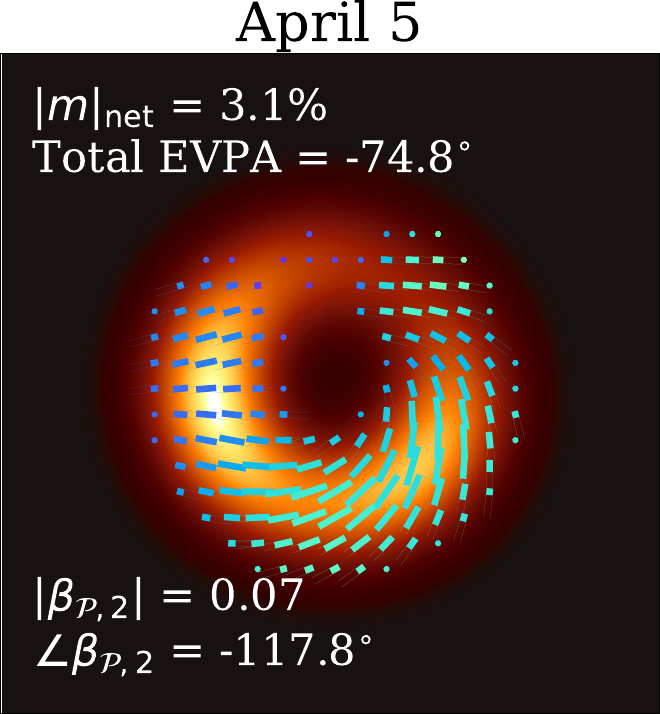}
    \includegraphics[height=0.19\textwidth]{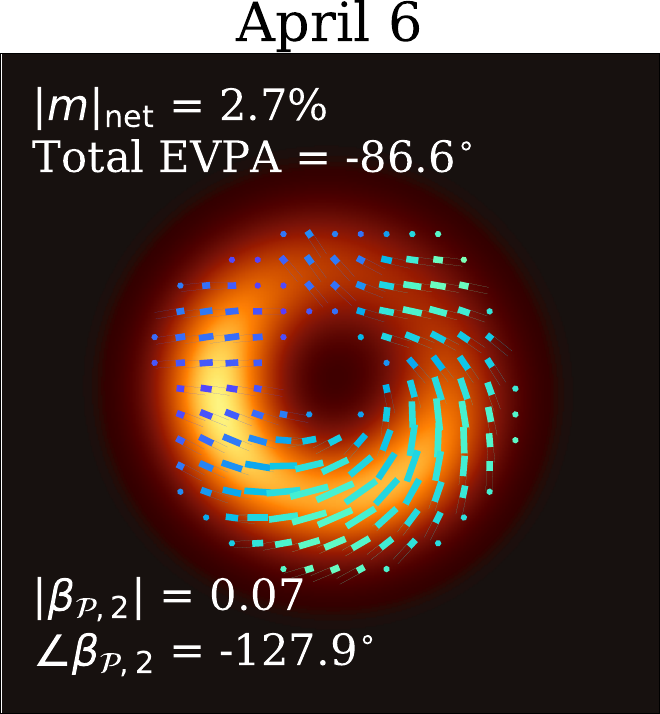}
    \includegraphics[height=0.19\textwidth]{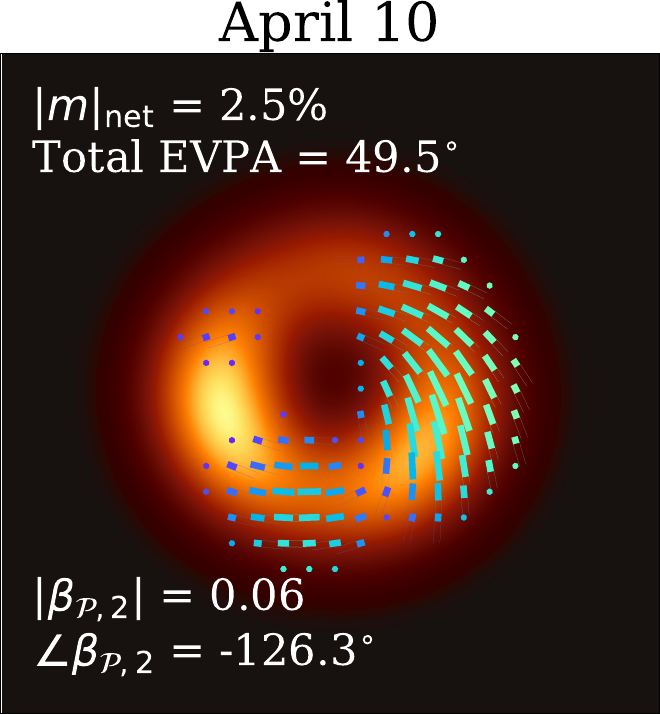}
    \includegraphics[height=0.19\textwidth]{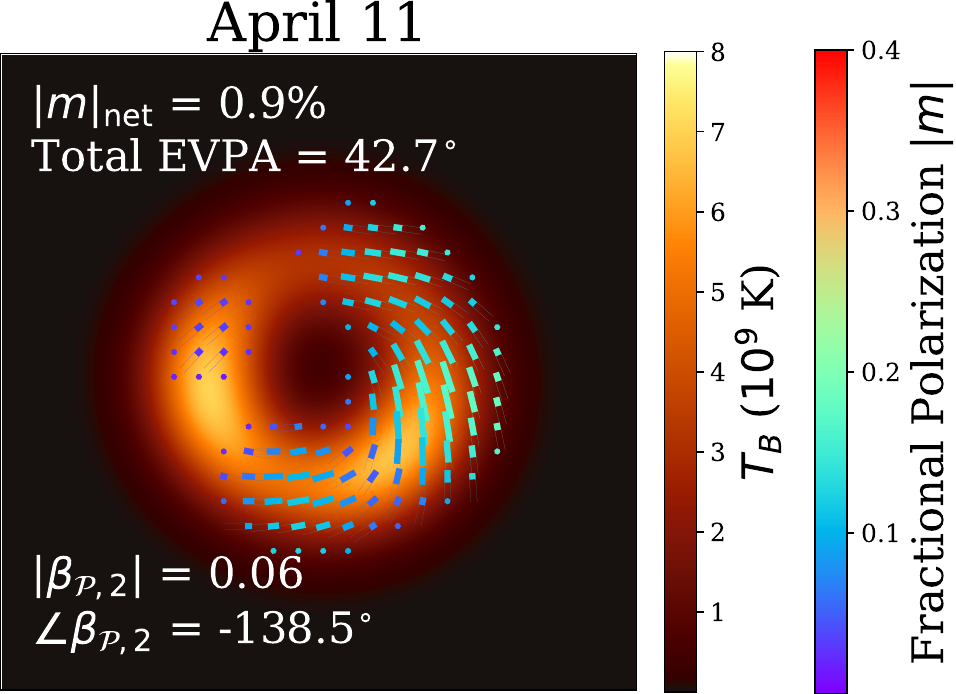} \\
    \includegraphics[height=0.19\textwidth]{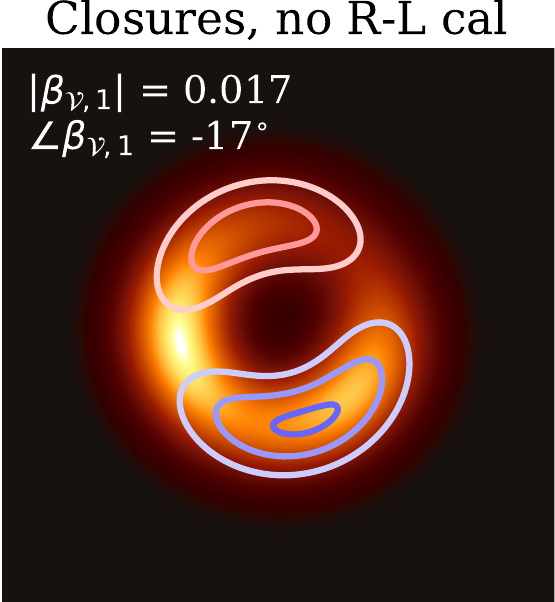}
    \includegraphics[height=0.19\textwidth]{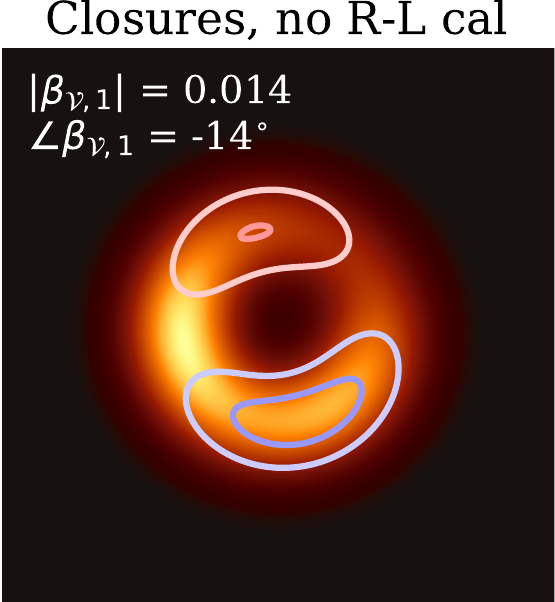}
    \includegraphics[height=0.19\textwidth]{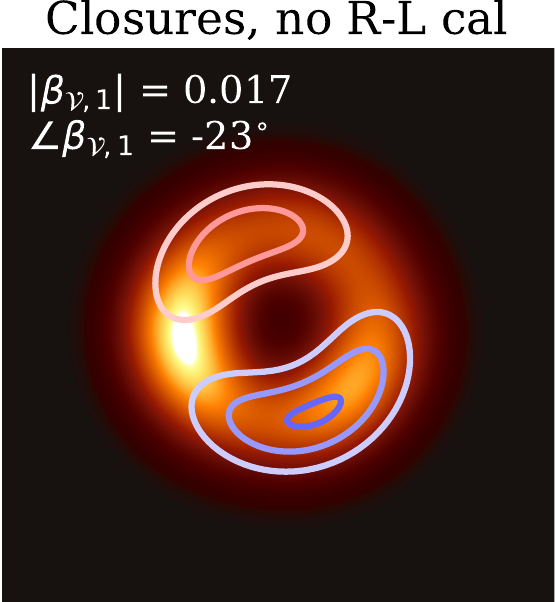}
    \includegraphics[height=0.19\textwidth]{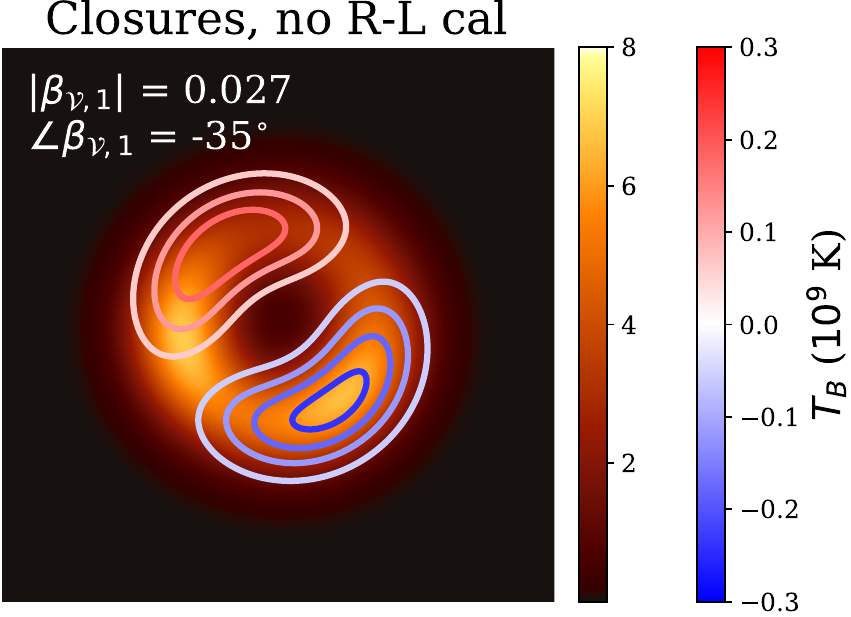} \\
    \includegraphics[height=0.19\textwidth]{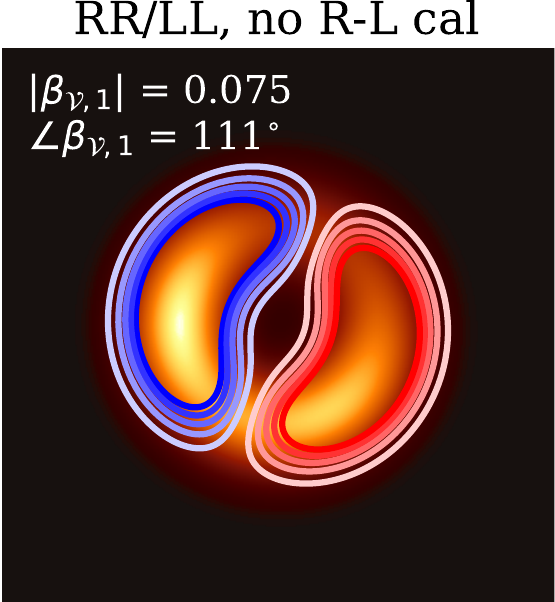}
    \includegraphics[height=0.19\textwidth]{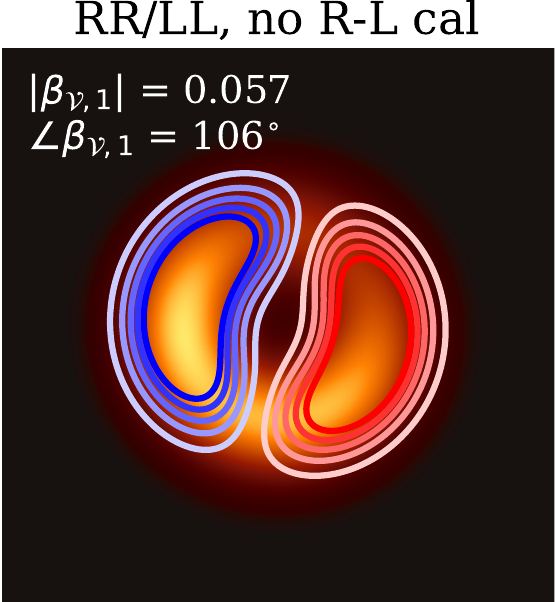}
    \includegraphics[height=0.19\textwidth]{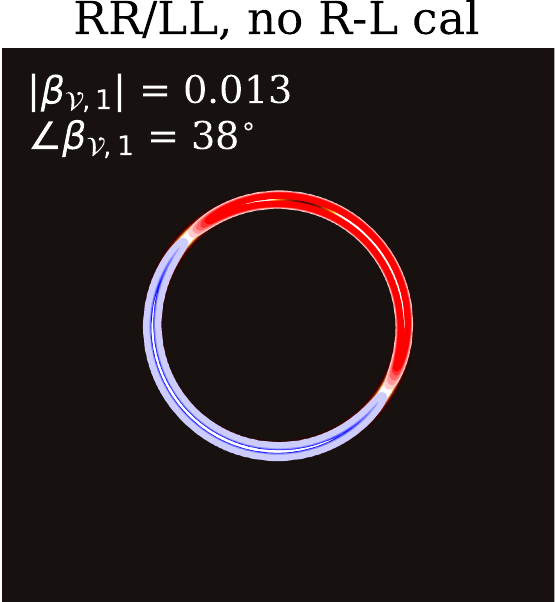}
    \includegraphics[height=0.19\textwidth]{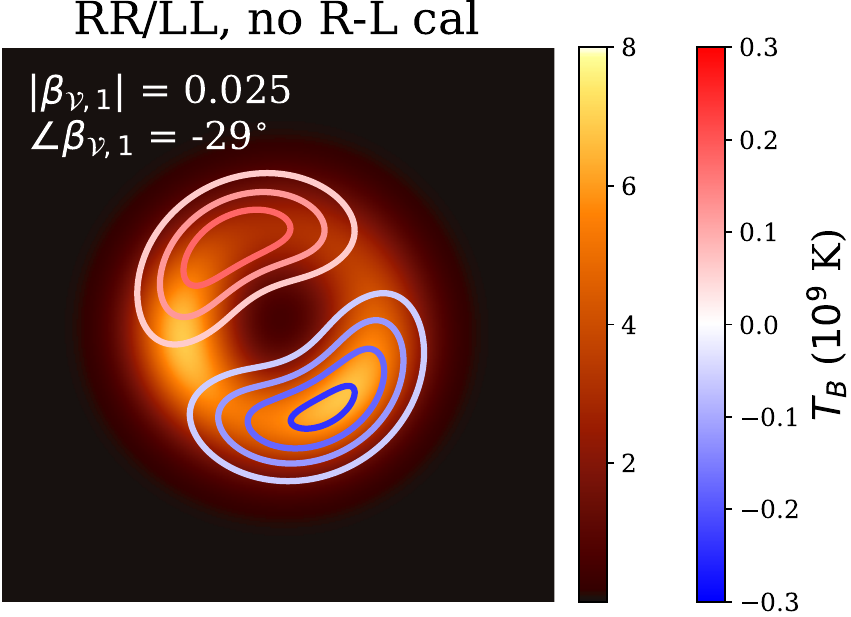} \\
    \includegraphics[height=0.19\textwidth]{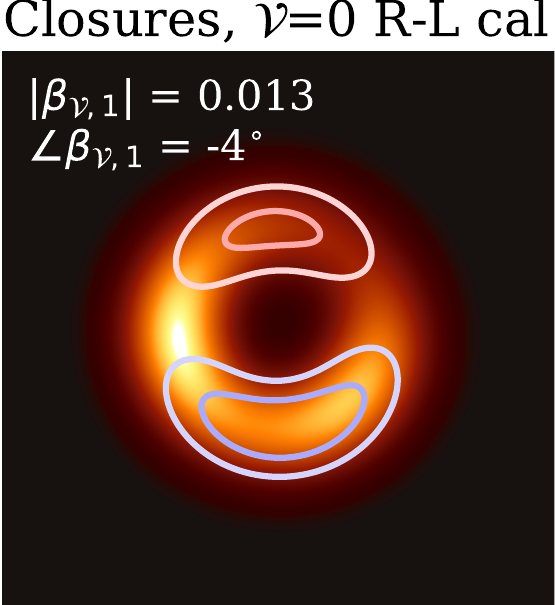}
    \includegraphics[height=0.19\textwidth]{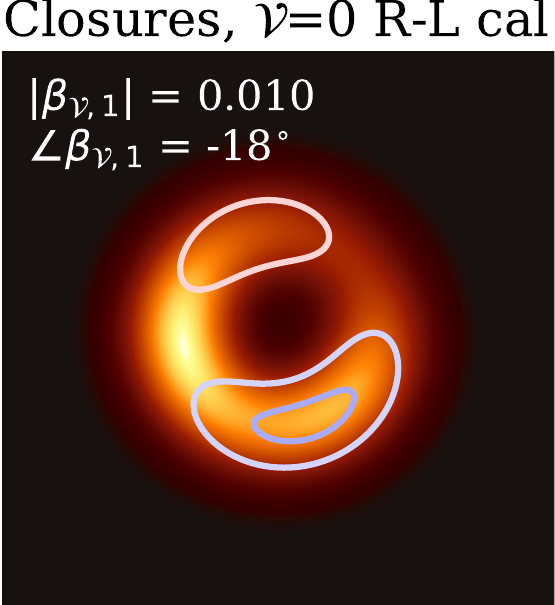}
    \includegraphics[height=0.19\textwidth]{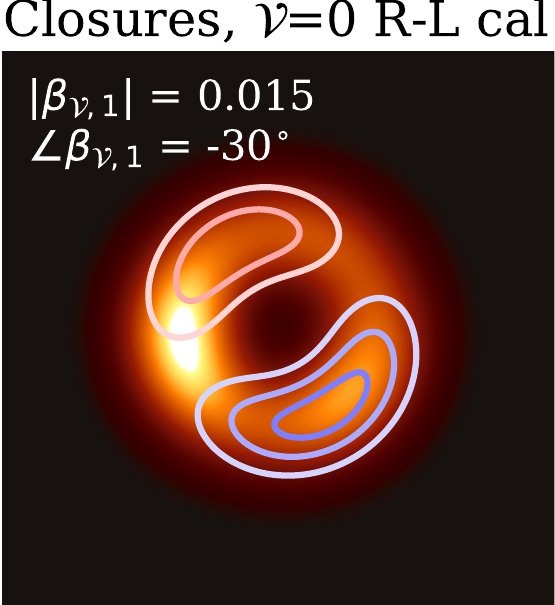}
    \includegraphics[height=0.19\textwidth]{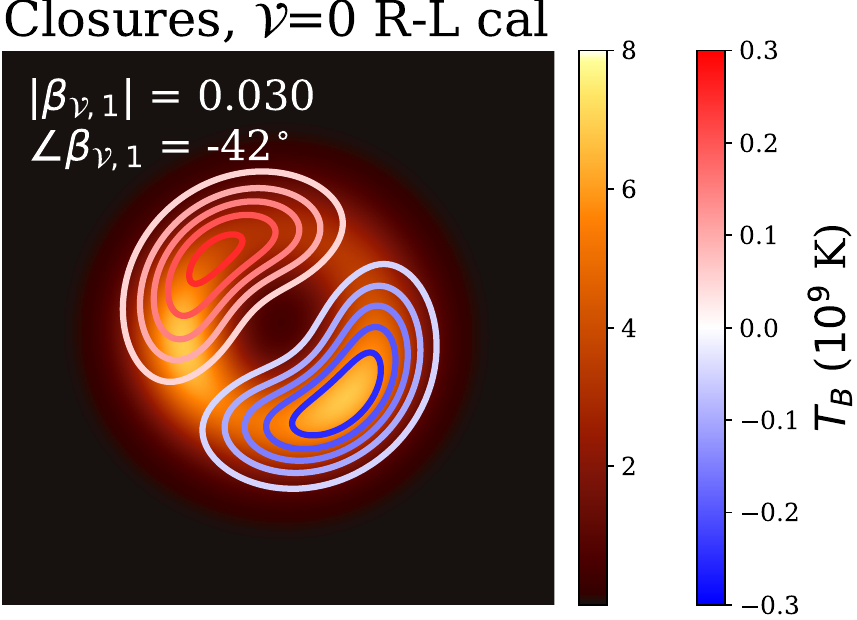} \\
    \includegraphics[height=0.19\textwidth]{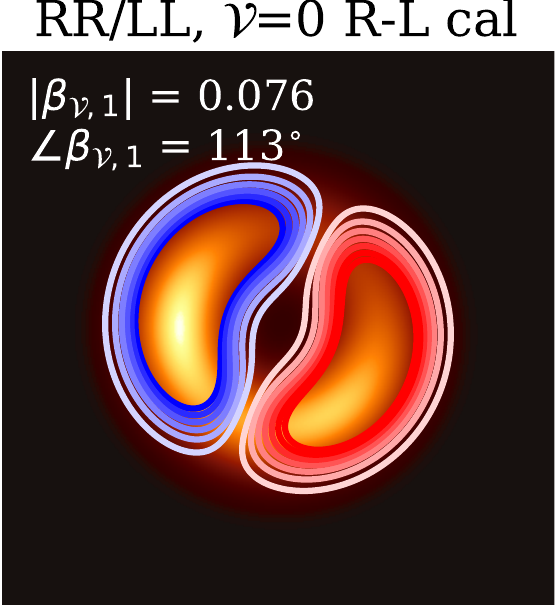}
    \includegraphics[height=0.19\textwidth]{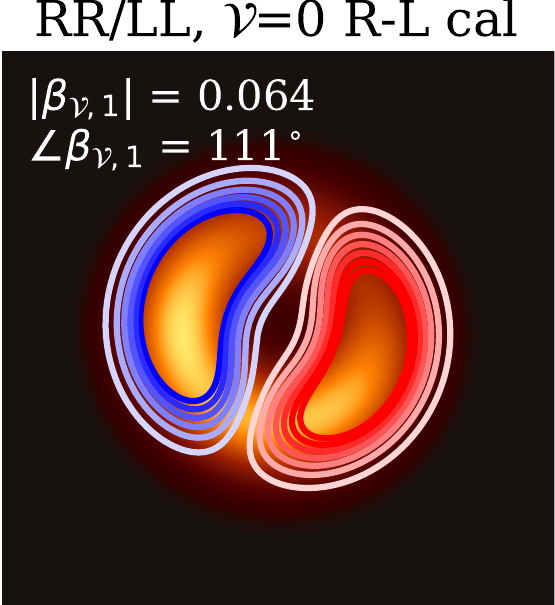}
    \includegraphics[height=0.19\textwidth]{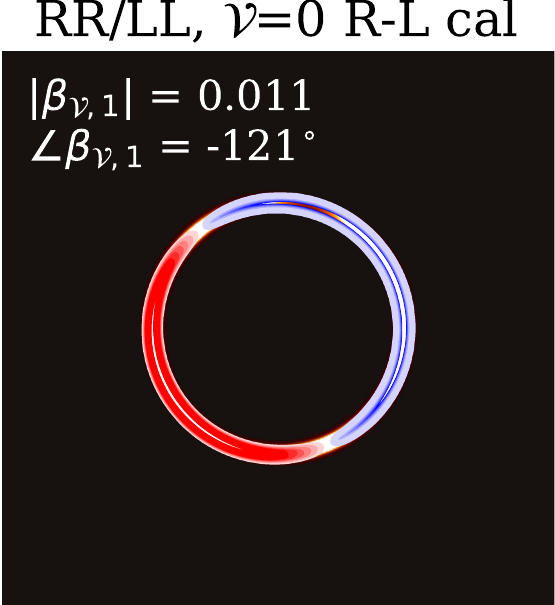}
    \includegraphics[height=0.19\textwidth]{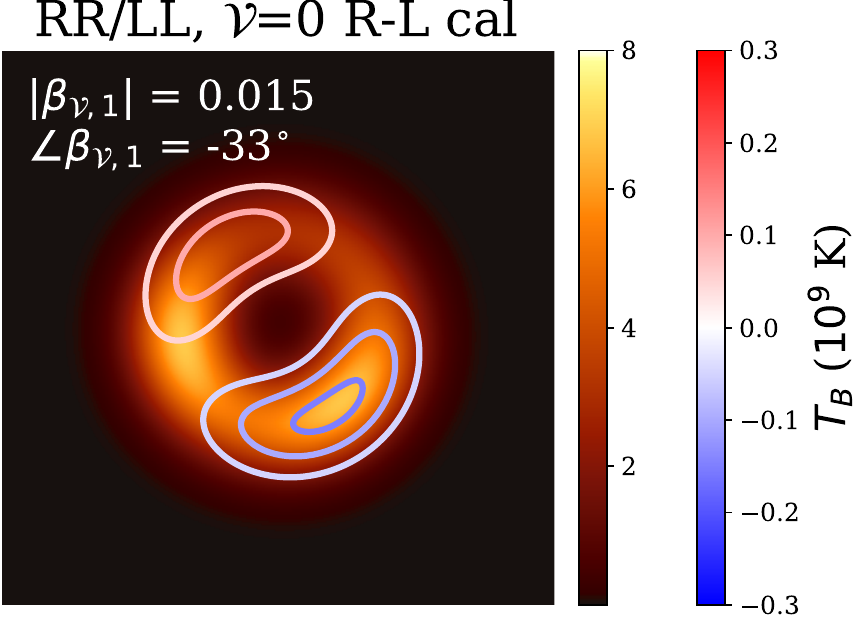}
    
    \caption{Full-Stokes m-ring fits (posterior maxima) to low-band 2017 EHT \m87 data, calibrated with the CASA/\texttt{rPICARD} EHT calibration pipeline \citep{Janssen2019, PaperIII}. Rows from top to bottom show linear polarization fits, circular polarization fits to closure quantities without any R/L gain ratio calibration, circular polarization fits to RR/LL visibility ratios without any R/L gain ratio calibration, and circular polarization fits to the same data with an R/L gain ratio calibration assuming $\mathcal{V}=0$, respectively. The fits shown in the leftmost and rightmost panels of the fourth row are also presented in \citetalias{M87PaperIX}.}
    \label{fig:m87_casafits}
\end{figure*}

\autoref{fig:m87_casafits} shows the polarimetric modeling results of the new CASA data. As expected, the R/L gain-insensitive closure-based fitting is in excellent agreement with the \texttt{HOPS} data results. Even without any R/L gain calibration or assuming $\mathcal{V}=0$, consistent circular polarization signals are retrieved here. However, for the RR/LL visibility ratio fits, the differences in Stokes $\mathcal{V}$ structures with the \texttt{HOPS} results and the inconsistency between observing epochs demonstrate a significant sensitivity to the R/L gain calibration strategy (since these are posterior maxima, the structure itself does not look more noisy than for the \texttt{HOPS} data). For April 11 data, on which the \texttt{HOPS} fits indicated the strongest evidence for the presence of dipolar Stokes $\mathcal{V}$ structure, the CASA RR/LL fits are in agreement with the closure and \texttt{HOPS} data fits regardless of the R/L gain calibration strategy, which may indicate a reduced sensitivity to R/L gains due to the stronger Stokes $\mathcal{V}$ signal on this day.

The consistency of the results with the previously used, well vetted, 2017 calibrated EHT data demonstrates the robustness of our method relative to different calibration assumptions and serves as a first validation of the updated CASA/\texttt{rPICARD} data reduction pathway. Our fits across different data sets, data products, and modeling assumptions support the presence of a persistent dipolar asymmetry in the circular polarization of \m87 especially on April 11, where this asymmetry persists in the RR/LL fits regardless of calibration strategy.

\section{Summary and outlook}
\label{sec:summary}
In this work, we have developed a novel method to reconstruct polarimetric image structure from VLBI observations, making use of data products with different levels of calibration-invariance and simple geometric models. Specifically, polarimetric m-ring fitting is a useful method to obtain information on the polarimetric structure of horizon-scale observations of supermassive black holes. We have shown that ground-truth polarization parameters can be recovered from synthetic EHT data from geometric m-ring models and GRMHD models, with accuracies depending on the level of model misspecification. Even with total and resolved circular polarization fractions as low as 0.5\%, the first-order circular polarization asymmetry can be recovered from GRMHD models to within a few to ${\sim}30-40$ degrees, depending on the model. Polarized structure is recovered most faithfully in image regions with high total intensity.

Application to EHT \m87 data has shown that the linear polarization structure imaged by \citet{PaperVII} is recovered well with our m-ring modeling framework. Our fits also indicate the presence of a persistent horizon-scale circular polarization asymmetry, with increased negative circular polarization in the South, near the total intensity maximum. This asymmetry persists across fits to different observing epochs, bands, and data products, across fits with circular polarization m-ring orders $m_{\mathcal{V}}$ of 1 and 2, for fits assuming different image-integrated circular polarization fractions, and for fits to data calibrated with different calibration pipelines and strategies. Support for the presence of the asymmetry is largest for April 11 data. For this day, a dipolar structure is favored by the Bayesian evidence, the structure persists in fits up to $m_{\mathcal{V}}=3$, and in R/L gain-sensitive RR/LL visibility ratio fits regardless of the R/L gain calibration strategy used. Some imaging methods reconstruct similar structure on this day \citepalias{M87PaperIX}. 

However, given the overall weakness of the circular polarization signal, the sensitivity of our RR/LL fit results to the R/L gain calibration strategy on three days, and the difficulty for imaging methods to reconstruct similarly consistent structure, caution should be exercised in interpreting this result and we are not reporting an unambiguous detection of dipolar structure with a specific orientation. While m-ring fitting reliably reconstructed the first-order circular polarization asymmetry in our GRMHD synthetic data tests (with an orientation offset of a few degrees up to $\sim$40 degrees), there is no guarantee that the underlying circular polarization structure of M87* has a strong dipolar component. Even for a strong asymmetry detection, GRMHD models show degeneracies in black hole and plasma parameters that may produce such asymmetry, and it may be short-lived due to plasma variability.

As more stations are added to the EHT and antenna sensitivity improves, the circular polarization structure of \m87 will become easier to detect and reconstruct. While the EHT 2017 data only provide upper limits on the resolved circular polarization fraction and tentative circular polarization images with large uncertainties, the future prospects for imaging and modeling \m87 in circular polarization are excellent.

Finally, our modeling methods have other applications as well. Apart from fitting horizon-scale structure with m-rings, polarimetric geometric modeling with different model prescriptions (e.g., a set of polarized Gaussians) may also be used to reconstruct the polarized structure of non-horizon active galactic nuclei. For \sgra, snapshot geometric modeling \citep[i.e., fitting to short time snippets of data and then combining the posteriors;][]{SgrAEHTCIV} can utilize our polarimetric m-ring model to mitigate the rapid source variability and constrain the polarimetric structure with only a small number of baselines. With future arrays with many more stations, such as the Next-Generation EHT \citep[ngEHT;][]{Doeleman2019, Johnson2023, Doeleman2023}, snapshot modeling may even be used to reconstruct real-time polarimetric black hole movies of \sgra.\\


The Event Horizon Telescope Collaboration thanks the following
organizations and programs: the Academia Sinica; the Academy
of Finland (projects 274477, 284495, 312496, 315721); the Agencia Nacional de Investigaci\'{o}n 
y Desarrollo (ANID), Chile via NCN$19\_058$ (TITANs), Fondecyt 1221421  and BASAL FB210003; the Alexander
von Humboldt Stiftung; an Alfred P. Sloan Research Fellowship;
Allegro, the European ALMA Regional Centre node in the Netherlands, the NL astronomy
research network NOVA and the astronomy institutes of the University of Amsterdam, Leiden University, and Radboud University;
the ALMA North America Development Fund; the Astrophysics and High Energy Physics programme by MCIN (with funding from European Union NextGenerationEU, PRTR-C17I1); the Black Hole Initiative, which is funded by grants from the John Templeton Foundation and the Gordon and Betty Moore Foundation (although the opinions expressed in this work are those of the author(s) 
and do not necessarily reflect the views of these Foundations); the Brinson Foundation; ``la Caixa'' Foundation (ID 100010434) through fellowship codes LCF/BQ/DI22/11940027 and LCF/BQ/DI22/11940030; 
Chandra DD7-18089X and TM6-17006X; the China Scholarship
Council; the China Postdoctoral Science Foundation fellowships (2020M671266, 2022M712084); Consejo Nacional de Ciencia y Tecnolog\'{\i}a (CONACYT,
Mexico, projects  U0004-246083, U0004-259839, F0003-272050, M0037-279006, F0003-281692,
104497, 275201, 263356);  the Colfuturo Scholarship; 
the Consejer\'{i}a de Econom\'{i}a, Conocimiento, 
Empresas y Universidad 
of the Junta de Andaluc\'{i}a (grant P18-FR-1769), the Consejo Superior de Investigaciones 
Cient\'{i}ficas (grant 2019AEP112);
the Delaney Family via the Delaney Family John A.
Wheeler Chair at Perimeter Institute; Direcci\'{o}n General
de Asuntos del Personal Acad\'{e}mico-Universidad
Nacional Aut\'{o}noma de M\'{e}xico (DGAPA-UNAM,
projects IN112417 and IN112820); 
the Dutch Organization for Scientific Research (NWO) for the VICI award (grant 639.043.513), the grant OCENW.KLEIN.113, and the Dutch Black Hole Consortium (with project No. NWA 1292.19.202) of the research programme the National Science Agenda; the Dutch National Supercomputers, Cartesius and Snellius  
(NWO grant 2021.013); 
the EACOA Fellowship awarded by the East Asia Core
Observatories Association, which consists of the Academia Sinica Institute of Astronomy and
Astrophysics, the National Astronomical Observatory of Japan, Center for Astronomical Mega-Science,
Chinese Academy of Sciences, and the Korea Astronomy and Space Science Institute; 
the European Research Council (ERC) Synergy
Grant ``BlackHoleCam: Imaging the Event Horizon
of Black Holes" (grant 610058); 
the European Union Horizon 2020
research and innovation programme under grant agreements
RadioNet (No. 730562) and 
M2FINDERS (No. 101018682); the Horizon ERC Grants 2021 programme under grant agreement No. 101040021; 
the Generalitat
Valenciana (grants APOSTD/2018/177 and  ASFAE/2022/018) and
GenT Program (project CIDEGENT/2018/021); MICINN Research Project PID2019-108995GB-C22;
the European Research Council for advanced grant `JETSET: Launching, propagation and 
emission of relativistic jets from binary mergers and across mass scales' (grant No. 884631); the FAPESP (Funda\c{c}\~ao de Amparo \'a Pesquisa do Estado de S\~ao Paulo) under grant 2021/01183-8; 
the Institute for Advanced Study; the Istituto Nazionale di Fisica
Nucleare (INFN) sezione di Napoli, iniziative specifiche
TEONGRAV; 
the International Max Planck Research
School for Astronomy and Astrophysics at the
Universities of Bonn and Cologne; 
DFG research grant ``Jet physics on horizon scales and beyond'' (grant No. FR 4069/2-1);
Joint Columbia/Flatiron Postdoctoral Fellowship (research at the Flatiron Institute is supported by the Simons Foundation); 
the Japan Ministry of Education, Culture, Sports, Science and Technology (MEXT; grant JPMXP1020200109); 
the Japan Society for the Promotion of Science (JSPS) Grant-in-Aid for JSPS
Research Fellowship (JP17J08829); the Joint Institute for Computational Fundamental Science, Japan; the Key Research
Program of Frontier Sciences, Chinese Academy of
Sciences (CAS, grants QYZDJ-SSW-SLH057, QYZDJSSW-SYS008, ZDBS-LY-SLH011); 
the Leverhulme Trust Early Career Research
Fellowship; the Max-Planck-Gesellschaft (MPG);
the Max Planck Partner Group of the MPG and the
CAS; the MEXT/JSPS KAKENHI (grants 18KK0090, JP21H01137,
JP18H03721, JP18K13594, 18K03709, JP19K14761, 18H01245, 25120007, 23K03453); the Malaysian Fundamental Research Grant Scheme (FRGS) FRGS/1/2019/STG02/UM/02/6; the MIT International Science
and Technology Initiatives (MISTI) Funds; 
the Ministry of Science and Technology (MOST) of Taiwan (103-2119-M-001-010-MY2, 105-2112-M-001-025-MY3, 105-2119-M-001-042, 106-2112-M-001-011, 106-2119-M-001-013, 106-2119-M-001-027, 106-2923-M-001-005, 107-2119-M-001-017, 107-2119-M-001-020, 107-2119-M-001-041, 107-2119-M-110-005, 107-2923-M-001-009, 108-2112-M-001-048, 108-2112-M-001-051, 108-2923-M-001-002, 109-2112-M-001-025, 109-2124-M-001-005, 109-2923-M-001-001, 110-2112-M-003-007-MY2, 110-2112-M-001-033, 110-2124-M-001-007, and 110-2923-M-001-001);
the Ministry of Education (MoE) of Taiwan Yushan Young Scholar Program;
the Physics Division, National Center for Theoretical Sciences of Taiwan;
the National Aeronautics and
Space Administration (NASA, Fermi Guest Investigator
grant 80NSSC20K1567, NASA Astrophysics Theory Program grant 80NSSC20K0527, NASA NuSTAR award 
80NSSC20K0645); 
NASA Hubble Fellowship 
grants HST-HF2-51431.001-A, HST-HF2-51482.001-A awarded 
by the Space Telescope Science Institute, which is operated by the Association of Universities for 
Research in Astronomy, Inc., for NASA, under contract NAS5-26555; 
the National Institute of Natural Sciences (NINS) of Japan; the National
Key Research and Development Program of China
(grant 2016YFA0400704, 2017YFA0402703, 2016YFA0400702); the National
Science Foundation (NSF, grants AST-0096454,
AST-0352953, AST-0521233, AST-0705062, AST-0905844, AST-0922984, AST-1126433, AST-1140030,
DGE-1144085, AST-1207704, AST-1207730, AST-1207752, MRI-1228509, OPP-1248097, AST-1310896, AST-1440254, 
AST-1555365, AST-1614868, AST-1615796, AST-1715061, AST-1716327,  
OISE-1743747, AST-1816420, AST-1935980, AST-2034306, AST-2307887); 
NSF Astronomy and Astrophysics Postdoctoral Fellowship (AST-1903847); 
the Natural Science Foundation of China (grants 11650110427, 10625314, 11721303, 11725312, 11873028, 11933007, 11991052, 11991053, 12192220, 12192223, 12273022); 
the Natural Sciences and Engineering Research Council of
Canada (NSERC, including a Discovery Grant and
the NSERC Alexander Graham Bell Canada Graduate
Scholarships-Doctoral Program); the National Youth
Thousand Talents Program of China; the National Research
Foundation of Korea (the Global PhD Fellowship
Grant: grants NRF-2015H1A2A1033752, the Korea Research Fellowship Program:
NRF-2015H1D3A1066561, Brain Pool Program: 2019H1D3A1A01102564, 
Basic Research Support Grant 2019R1F1A1059721, 2021R1A6A3A01086420, 2022R1C1C1005255); 
Netherlands Research School for Astronomy (NOVA) Virtual Institute of Accretion (VIA) postdoctoral fellowships; 
Onsala Space Observatory (OSO) national infrastructure, for the provisioning
of its facilities/observational support (OSO receives
funding through the Swedish Research Council under
grant 2017-00648);  the Perimeter Institute for Theoretical
Physics (research at Perimeter Institute is supported
by the Government of Canada through the Department
of Innovation, Science and Economic Development
and by the Province of Ontario through the
Ministry of Research, Innovation and Science); the Princeton Gravity Initiative; the Spanish Ministerio de Ciencia e Innovaci\'{o}n (grants PGC2018-098915-B-C21, AYA2016-80889-P,
PID2019-108995GB-C21, PID2020-117404GB-C21); 
the University of Pretoria for financial aid in the provision of the new 
Cluster Server nodes and SuperMicro (USA) for a SEEDING GRANT approved toward these 
nodes in 2020; the Shanghai Municipality orientation program of basic research for international scientists (grant no. 22JC1410600);
the Shanghai Pilot Program for Basic Research, Chinese Academy of Science, 
Shanghai Branch (JCYJ-SHFY-2021-013);
the State Agency for Research of the Spanish MCIU through
the ``Center of Excellence Severo Ochoa'' award for
the Instituto de Astrof\'{i}sica de Andaluc\'{i}a (SEV-2017-
0709); the Spanish Ministry for Science and Innovation grant CEX2021-001131-S funded by MCIN/AEI/10.13039/501100011033; the Spinoza Prize SPI 78-409; the South African Research Chairs Initiative, through the 
South African Radio Astronomy Observatory (SARAO, grant ID 77948),  which is a facility of the National 
Research Foundation (NRF), an agency of the Department of Science and Innovation (DSI) of South Africa; 
the Toray Science Foundation; the Swedish Research Council (VR); 
the US Department
of Energy (USDOE) through the Los Alamos National
Laboratory (operated by Triad National Security,
LLC, for the National Nuclear Security Administration
of the USDOE, contract 89233218CNA000001); and the YCAA Prize Postdoctoral Fellowship.

We thank
the staff at the participating observatories, correlation
centers, and institutions for their enthusiastic support.
This paper makes use of the following ALMA data:
ADS/JAO.ALMA\#2016.1.01154.V. ALMA is a partnership
of the European Southern Observatory (ESO;
Europe, representing its member states), NSF, and
National Institutes of Natural Sciences of Japan, together
with National Research Council (Canada), Ministry
of Science and Technology (MOST; Taiwan),
Academia Sinica Institute of Astronomy and Astrophysics
(ASIAA; Taiwan), and Korea Astronomy and
Space Science Institute (KASI; Republic of Korea), in
cooperation with the Republic of Chile. The Joint
ALMA Observatory is operated by ESO, Associated
Universities, Inc. (AUI)/NRAO, and the National Astronomical
Observatory of Japan (NAOJ). The NRAO
is a facility of the NSF operated under cooperative agreement
by AUI.
This research used resources of the Oak Ridge Leadership Computing Facility at the Oak Ridge National
Laboratory, which is supported by the Office of Science of the U.S. Department of Energy under contract
No. DE-AC05-00OR22725; the ASTROVIVES FEDER infrastructure, with project code IDIFEDER-2021-086; the computing cluster of Shanghai VLBI correlator supported by the Special Fund 
for Astronomy from the Ministry of Finance in China;  
We also thank the Center for Computational Astrophysics, National Astronomical Observatory of Japan. This work was supported by FAPESP (Fundacao de Amparo a Pesquisa do Estado de Sao Paulo) under grant 2021/01183-8.

APEX is a collaboration between the
Max-Planck-Institut f{\"u}r Radioastronomie (Germany),
ESO, and the Onsala Space Observatory (Sweden). The
SMA is a joint project between the SAO and ASIAA
and is funded by the Smithsonian Institution and the
Academia Sinica. The JCMT is operated by the East
Asian Observatory on behalf of the NAOJ, ASIAA, and
KASI, as well as the Ministry of Finance of China, Chinese
Academy of Sciences, and the National Key Research and Development
Program (No. 2017YFA0402700) of China
and Natural Science Foundation of China grant 11873028.
Additional funding support for the JCMT is provided by the Science
and Technologies Facility Council (UK) and participating
universities in the UK and Canada. 
The LMT is a project operated by the Instituto Nacional
de Astr\'{o}fisica, \'{O}ptica, y Electr\'{o}nica (Mexico) and the
University of Massachusetts at Amherst (USA). The
IRAM 30-m telescope on Pico Veleta, Spain is operated
by IRAM and supported by CNRS (Centre National de
la Recherche Scientifique, France), MPG (Max-Planck-Gesellschaft, Germany), 
and IGN (Instituto Geogr\'{a}fico
Nacional, Spain). The SMT is operated by the Arizona
Radio Observatory, a part of the Steward Observatory
of the University of Arizona, with financial support of
operations from the State of Arizona and financial support
for instrumentation development from the NSF.
Support for SPT participation in the EHT is provided by the National Science Foundation through award OPP-1852617 
to the University of Chicago. Partial support is also 
provided by the Kavli Institute of Cosmological Physics at the University of Chicago. The SPT hydrogen maser was 
provided on loan from the GLT, courtesy of ASIAA.

This work used the
Extreme Science and Engineering Discovery Environment
(XSEDE), supported by NSF grant ACI-1548562,
and CyVerse, supported by NSF grants DBI-0735191,
DBI-1265383, and DBI-1743442. XSEDE Stampede2 resource
at TACC was allocated through TG-AST170024
and TG-AST080026N. XSEDE JetStream resource at
PTI and TACC was allocated through AST170028.
This research is part of the Frontera computing project at the Texas Advanced 
Computing Center through the Frontera Large-Scale Community Partnerships allocation
AST20023. Frontera is made possible by National Science Foundation award OAC-1818253.
This research was done using services provided by the OSG Consortium~\citep{osg07,osg09}, which is supported by the National Science Foundation award Nos. 2030508 and 1836650.
Additional work used ABACUS2.0, which is part of the eScience center at Southern Denmark University. 
Simulations were also performed on the SuperMUC cluster at the LRZ in Garching, 
on the LOEWE cluster in CSC in Frankfurt, on the HazelHen cluster at the HLRS in Stuttgart, 
and on the Pi2.0 and Siyuan Mark-I at Shanghai Jiao Tong University.
The computer resources of the Finnish IT Center for Science (CSC) and the Finnish Computing 
Competence Infrastructure (FCCI) project are acknowledged. This
research was enabled in part by support provided
by Compute Ontario (http://computeontario.ca), Calcul
Quebec (http://www.calculquebec.ca), and Compute
Canada (http://www.computecanada.ca). 

The EHTC has
received generous donations of FPGA chips from Xilinx
Inc., under the Xilinx University Program. The EHTC
has benefited from technology shared under open-source
license by the Collaboration for Astronomy Signal Processing
and Electronics Research (CASPER). The EHT
project is grateful to T4Science and Microsemi for their
assistance with hydrogen masers. This research has
made use of NASA's Astrophysics Data System. We
gratefully acknowledge the support provided by the extended
staff of the ALMA, from the inception of
the ALMA Phasing Project through the observational
campaigns of 2017 and 2018. We would like to thank
A. Deller and W. Brisken for EHT-specific support with
the use of DiFX. We thank Martin Shepherd for the addition of extra features in the Difmap software 
that were used for the CLEAN imaging results presented in this paper.
We acknowledge the significance that
Maunakea, where the SMA and JCMT EHT stations
are located, has for the indigenous Hawaiian people.

\bibliography{ms}{}
\bibliographystyle{aasjournal}



\end{document}